\documentclass[twocolumn]{aastex631}
\usepackage{verbatim}
\usepackage{xspace}
\usepackage[utf8]{inputenc}
\usepackage{epsfig}
\usepackage{graphicx}
\usepackage{amssymb}
\usepackage{amsmath}
\usepackage{times}
\usepackage{hyperref}
\usepackage{natbib}
\usepackage{blindtext}
\usepackage{color}
\usepackage{mathrsfs}
\usepackage{cancel}
\usepackage{bm}
\usepackage{soul}
\usepackage{outlines}
\usepackage[inline]{enumitem}
    \setlist{nosep}
\newcommand{\sSun}{ {\scriptscriptstyle{\rm \odot}} }

\newcommand{\snull}{ {\scriptscriptstyle\text{0}} }
\newcommand{\sB}{ {\scriptscriptstyle\text{B}} }

\newcount\doepsf
\newcommand{\ie}{{i.e., }}
\newcommand{\cf}{{cf., }}
\newcommand{\eg}{{e.g., }}

\newcommand{\half}{{\tiny{\frac{1}{2}}}}
\newcommand{\oneforth}{{\tiny{\frac{1}{4}}}}

\newcommand{\threeforth}{{\tiny{\frac{3}{4}}}}
\newcommand{\oneeigths}{{\tiny{\frac{1}{8}}}}
\newcommand{\be}{\begin{equation}}
\newcommand{\ee}{\end{equation}}
\newcommand{\bea}{\begin{eqnarray}}
\newcommand{\eea}{\end{eqnarray}}
\newcommand{\bean}{\begin{eqnarray*}}
\newcommand{\eean}{\end{eqnarray*}}

\newcommand{\alf}{{Alfv{\'e}n}}
\newcommand{\amp}{{Amp{\`e}re}}
\newcommand{\lame}{{Lam{\'e} }}

\begin{document}

\title{A Titov-D{\'e}moulin Type Eruptive Event Generator for $\beta>0$ Plasmas}
\author[0000-0002-6118-0469]{Igor V. Sokolov}
\author[0000-0001-9360-4951]{Tamas I Gombosi}
\affiliation{Department of Climate and Space Sciences and Engineering\\
University of Michigan, Ann Arbor, MI 48109, USA}

\correspondingauthor{Igor Sokolov}
\email{igorsok@umich.edu}

\shorttitle{Generalized Titov-D{\'e}moulin Eruption Generator}
\shortauthors{Sokolov and Gombosi}

\begin{abstract}
We provide exact analytical solutions for the magnetic field produced by prescribed current distributions located inside a toroidal filament of finite thickness. The solutions are expressed in terms of toroidal functions which are modifications of the Legendre functions. In application to the MHD equilibrium of a twisted toroidal current loop in the solar corona, the Grad-Shafranov equation is decomposed into an analytic solution describing an equilibrium configuration against the pinch-effect from its own current and an approximate solution for an external strapping field to balance the hoop force.  

Our solutions can be employed in numerical simulations of coronal mass ejections. When superimposed on the background solar coronal magnetic field, the excess magnetic energy of the twisted current loop configuration can be made unstable by applying flux cancellation to reduce the strapping field. Such loss of stability accompanied by the formation of an expanding flux rope is typical for the \cite{titov99} eruptive event generator. The main new features of the proposed model are:
\begin{enumerate*}[label=(\roman*)]
\item
The filament is filled with finite $\beta$ plasma 
with finite mass and energy,
\item
The model describes an equilibrium solution that will spontaneously erupt due to magnetic reconnection of the strapping magnetic field arcade, and
\item
There are analytic expressions connecting the model parameters to the asymptotic velocity and total mass of the resulting CME, providing a way to connect the simulated CME properties to multipoint coronograph observations.
\end{enumerate*}

\end{abstract}


\section{Introduction}
\label{sec:intro}
Solar eruptions, including Coronal Mass Ejections (CMEs), are associated with a major restructuring of the coronal magnetic field and the ejection of solar material ($\sim$~$10^{12}-10^{13}$~kg) and magnetic flux ($\sim$~$10^{13}-10^{15}$~Wb) into interplanetary space \citep[\eg][]{Roussev2006}. Among many aspects of CMEs which justify the heliophysics community's interest in numerical simulations of CMEs is their contribution to the acceleration of Solar Energetic Particles (SEPs). To explain the observed signatures of CME-SEP events, global models of solar eruptions need to incorporate the realistic \textit{background solution} for the solar corona (SC) and  magnetic field driven by observed magnetograms \cite[\cf][]{Roussev2004}. 

The fundamental process producing a CME is the conversion of magnetic free energy to the kinetic energy of the ejecta; that is why magnetically-driven CME models are the most promising. A simple, but well working, way to drive a CME in a global simulation is to \textit{superimpose} a \cite{gibson98} (GL) or \cite{titov99} (TD) magnetic flux-tube configuration onto the background state of SC. These magnetic configurations describe an erupting magnetic filament. That filament becomes an expanding flux rope (magnetic cloud) in the ambient solar wind while evolving and propagating outward from the Sun, thus allowing the simulation of the propagation of a magnetically driven CME. 

Our recent work on  the GL model allowed us to significantly simplify the process of triggering CMEs. The product of the effort is the Eruptive Event Generator based on Gibson-Low magnetic configuration (EEGGL) \citep{Jin:2017b}, which is described in more details in \cite{Bor:2017} in terms of an analytical solution of the Grad-Shafranov \cite[][herafter GS]{grad58, shafranov66} equation. While the GL model represents a significant progress in physics-based CME initiation modeling, it also has important limitations. When superimposed  on the external field of the active region the GL flux-rope is already out of equilibrium and it is expanding in a self-similar manner, therefore it sidesteps the CME initiation problem. More importantly, the analysis based on the GS equation in \cite{Bor:2017} demonstrated that the GL flux rope has regions of negative plasma $\beta$ (the ratio of thermal to magnetic pressures), a clearly unphysical regime. The TD model inserts a toroidal loop (filament) carrying an electric current, $I^{\rm tot}$, on top of the active region in a way that only part of the current loop is above the photosphere. The superposed magnetic configuration is stabilized by the effect of a strapping magnetic field, $B^{\rm (s)}$ in the active region, such that the action of this field on the loop current, $\propto I^{\rm tot}B^{\rm (s)}$, balances the hoop force, $\propto \left(I^{\rm tot}\right)^2$ \cite[see][]{titov14}, which allows derivation of the current, $I^{\rm tot}$, in terms of the observed magnetic field in the active region. If the equilibrium  breaks,  the filament immediately starts to expand, initiating an eruption. Recently, the model was generalized for inserting non-toroidal current loops \cite{Titov:2021}, as well as for producing near-critical current loops using a helicity condensation method \cite{Titov:2022}.   The force-free TD model also has its own important limitation: the assumption of no mass (pressure) inside the filament is part of the equilibrium analysis ($\beta=0$). 

In spite of its limitations, the original TD flux rope model has been used in a number of studies \cite[\eg][]{Roussev2003a, manchester08, Manchester2012, Jin2013}. Starting with the work of \cite{Linker2016}, numerous impressive results were obtained with the modified TD configuration \citep{titov14}, simulating historic CME events with unprecedented clarity and completeness.

In this paper we describe an alternative TD approach, which addresses major limitations of the \cite{titov99} and \cite{titov14} models: our \textit{equilibrium} analysis of the filament superposed with the potential field of the active region allows for finite mass and pressure ($\beta\!\!>\!\!0$) inside the filament.

Similarly to the approach by \cite{Bor:2017}, this work is based on an analytical solution of the \textit{scalar} GS equation. With the GS equation one can describe a toroidal filament of twisted magnetic field lines filled with finite density plasma (ejecta). The solutions are expressed in terms of toroidal functions (see Appendix~\ref{app:toroidal}) which are straightforward modifications of Legendre functions. 

We will apply this method to describe a twisted toroidal current loop in the solar corona that is in MHD equilibrium. In order to combine this filament with the active region magnetic field, an external strapping field must be accounted for in the force balance that balances the hoop force and thus assures equilibrium. We provide an approximate analytic solution to describe this combined configuration.

\paragraph{Note about notations.} 
This paper is highly mathematical and some of the notations are easy to confuse. Here we briefly summarize our guiding philosophy concerning notations.

In general (dimensional) physical quantities described by functions of cylindrical coordinates $z,r$ will be denoted by upper case:
\begin{outline}
\1 $\mathbf{J}(z,r)$ -- current density, 
\1 $\Psi(z,r)$ -- flux function, 
\1 $P(z,r)$ -- gas-kinetic pressure.
\end{outline}
Quantities denoted by lower case letters are reduced functions (or representative functions) of toroidal coordinates, $u,v$:
\begin{outline}
\1 $\Psi(z,r) = \mu_0\sqrt{R_\infty r}\  \psi(u.v)$, 
\1 $J_\varphi(z,r) = \sqrt{R_\infty / r^5}\  j(u,v)$, 
\1 $P(z,r)= (R_\infty/r)^3 \ p(u,v)$,
\1 $B_\varphi(z,r)= (R_\infty/r)^{\frac32} \ b(u,v)$.
\end{outline}
Finally, quantities denoted by $\sim$ represent normalized (dimensionless) quantities:

\begin{outline}
\1 $\tilde{I}_n(u) = I_n(u) / I_{n_0}$,
\1 $\tilde{j}_n = j_n / I _{n_0}$
\1 $\tilde{\psi}_n = \psi_n / I_{n_0}$.   
\end{outline}

\section{Magnetostatics in toroidal coordinates}
\label{sec:statics}
Equilibrium confinement of a toroidal plasma filament with a finite gas-kinetic pressure is controlled by a steady-state toroidal electric current, which produces an axially symmetric magnetic field that is independent of the toroidal angle, $\varphi$. The magnetostatics of such  fields can be formulated in arbitrary orthogonal coordinates, $\left(u(r,z),v(r,z), \varphi\right)$. The meridional plane coordinates, $\left(u,v\right)$, may or may not differ from cylindrical ones, $\left(r,z\right)$, $r$ being a distance from the axis of symmetry.  

In the 3-D vector of magnetic field, $\mathbf{B}=\mathbf{B}_2+B_\varphi\mathbf{e}_\varphi$ the poloidal components in the $(z,r)$ plane,  $\mathbf{B}_2$, can be expressed via the toroidal component of a vector potential, $A_\varphi$, using the \lame coefficients, $h_u$, $h_v$ (which describe the length element in terms of infinitesimal coordinate increments:  $\mathrm{d}s^2=h_u^2\mathrm{d}u^2+h_v^2\mathrm{d}v^2+r^2\mathrm{d}\varphi^2$):
\be\label{eq:BviaPsi}
    \mathbf{B}_2 = \frac{\nabla_2\Psi}{r} \times \mathbf{e}_\varphi,
\ee
where the two-dimensional (2D) differential operator,
\be\label{eq:nabla2uv}
\nabla_2\Psi=\frac{1}{h_u}\frac{\partial\Psi}{\partial u}\mathbf{e}_u+\frac{1}{h_v}\frac{\partial\Psi}{\partial v}\mathbf{e}_v,
\ee
is applied to the flux function, $\Psi=r A_\varphi$. Instead of the full flux function we will use (everywhere except Section~\ref{sec:GS}) the ``reduced'' flux function, $\psi\left(u,v\right)$, that is defined in $u,v$ coordinates:
\be\label{eq:tildepsi}
\Psi\left(z,r\right)=\mu_\snull \sqrt{R_\infty r} \, \psi\left(u,v\right),
\ee
where $R_\infty$ is a characteristic scale to be specified later. Using \amp's law, $\nabla_2\times\mathbf{B}_2=\mu_\snull J_\varphi\mathbf{e}_\varphi${,} in $u,v$ coordinates the expression for the toroidal current density, $J_\varphi$, can be simplified with the reduced flux function:
\bea\label{eq:jphi1}
&&J_\varphi\left(z,r\right) =\sqrt{\frac{R_\infty}{r}}\,
\frac{j(u,v)}{r^2},\\
&&j=\frac{3\psi}{4}-\frac{r^2}{h_u h_v}\left[\frac{\partial}{\partial u}\left(
\frac{h_v}{h_u}
\frac{\partial \psi}{\partial u}\right)+\frac{\partial}{\partial v}
\left(\frac{h_u}{h_v}\frac{\partial \psi}{\partial v}\right)\right],\nonumber
\eea
where we introduced a representative function for the toridal current density, $j(u,v)$, that only depends on the generalized coordinates.

\begin{figure}[thb]
\centering
\includegraphics[width=1\linewidth]{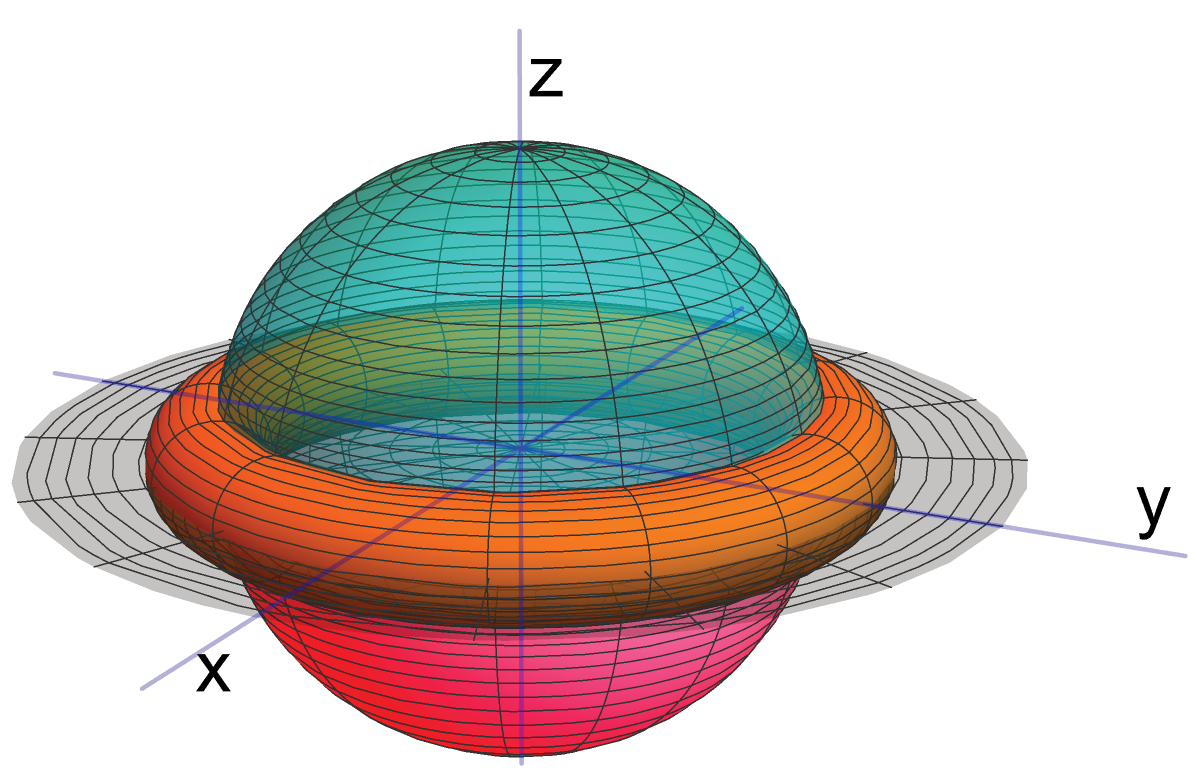}
\caption{Toroidal coordinate surfaces for $R_\infty=1$: $\sinh u =5$  ($\kappa^\prime\approx0.1$)  - orange torus;   $v=0$ - gray part of plane $z=0$; $v=\pi/2$ - blue hemisphere; $v=\pi$ - invisible part of plane $z=0$;  $v=3\pi/2$ - red hemisphere. Coordinate $u$ decreases outward the torus and increases inward,  turning to infinity at the (invisible) circumference, $x^2+y^2=1, z=0$.}
\label{fig:ToroidalCoords}
\end{figure}

Next, we define the \textit{toroidal} coordinates in the meridional plane $\left(u,v\right)$ the following way \cite[see][and Fig.~\ref{fig:ToroidalCoords}]{Mors53}:
\bea\label{eq:lame}
r&=&\frac{R_\infty\sinh u}{\cosh u-\cos v},
\qquad z=\frac{R_\infty\sin v}{\cosh u-\cos v},\nonumber\\
h_u&=&h_v=\frac{R_\infty}{\cosh u-\cos v}=\frac{r}{\sinh u}.
\eea
From these definitions, we get the following relation for the normalized radius vector to the ($r,z$) point:
\bea
\left(\frac{r}{R_\infty}\right)^2+\left(\frac{z}{R_\infty}\right)^2=\frac{\cosh u +\cos v}{\cosh u-\cos v}=\nonumber\\
=1+\frac{2z}{R_\infty}\cot v=-1+\frac{2r}{R_\infty}\coth u.
\eea
This means that surfaces of constant $v=v_\snull $ are \textit{spheres} with centers  at $r=0$, $z=R_\infty\cot v_\snull $, and radii of $R_\infty/|\sin v_\snull |$. Surfaces of constant $u=u_\snull $ are \textit{tori} with major radii $R_\infty\coth  u_\snull $ and minor radii $R_\infty/\sinh u_\snull $.  Specifically, when $u\rightarrow\infty$, the major and minor radii become $R_\infty$ and $0$, respectively. This is a degenerated toroidal surface of zero minor radius (\textit{toroidal magnetic axis}).

The inverse transformations determine the toroidal coordinates, $u,v$, and the \lame coefficients  in terms of $r,z$:
\bea\label{eq:toroidalcoords}
\sin v&=&\frac{2R_\infty z}{R_+R_-},\qquad\cos v=\frac{R^2-R_\infty^2}{R_+R_-},\nonumber\\
\sinh u&=&\frac{2R_\infty r}{R_+R_-},\qquad h_{u,v}=\frac{R_+R_-}{2R_\infty},\label{eq:4}
\eea
where $R=\sqrt{\mathbf{R}^2}$, $\mathbf{R} = r\mathbf{e}_{r} + z\mathbf{e}_{z}$ is the radius vector pointing from the center, $z=0,r=0$ to a given point,
\be
\label{eq:Rpm}
R_\pm = \sqrt{(r\pm R_\infty)^2+z^2}
\ee
are the maximum (+) and minimum (-) distances from the given point to the toroidal magnetic axis. 

The magnetic field can be expressed in terms of \textit{toroidal} special functions of the toroidal coordinate, $u$. Note, that the presence of $u,u_\snull$ in the equations is quite formal and actually they are not calculated, since  in effect the special functions can be expressed and efficiently calculated as hypergeometric power series of either $\kappa(u)$ or $\kappa^\prime(u)$, which can in turn be expressed in terms of $R_\pm$:
\bea\label{eq:kappas}
    \kappa^2(u) &=& 1 - e^{-2u} = \frac{4R_\infty r}{R^2_+}, \nonumber\\ \kappa^\prime(u) &=& \sqrt{1-\kappa^2} = e^{-u} = \frac{R_-}{R_+}.
\eea 

Using these notations,  the toroidal  coordinate surface, $u=\mathrm{const}$, has minor radius,  $a=2\kappa^\prime R_\infty/ (1 - \kappa^{\prime^2})$, and major radius, $R_\snull =\sqrt{R_\infty^2+a^2}$, determined by the constant value of $\kappa^\prime$, at the surface. Any such  surface can be taken as the boundary of a toroidal current filament.  Conversely, the field of a toroidal current filament with known minor and major radii of $a,R_\snull $, can be described using toroidal coordinates with a characteristic length scale of $R_\infty = \sqrt{R_\snull ^2 - a^2}$, so that the filament boundary is a $u = u_\snull  = \mathrm{const}$ surface at which $\kappa^\prime_\snull  = \kappa^\prime(u_\snull ) = a/\left(R_\snull  + R_\infty\right)$. This surface separates the filament interior ($u_\snull \le u<\infty$) from its exterior ($0<u<u_\snull $).  Note, that for $u\to0$ $\lim_{u\to0}\kappa = 0$, while for $u\to\infty$ (at   the toroidal magnetic axis)  $\lim_{u\to\infty}\kappa^\prime = 0$.

The coordinate unit vectors are:
\bea\label{eq:ebeta}
    \mathbf{e}_v &=& \frac{\left(R^2-R_\infty^2\right) \mathbf{e}_{z} - 2\mathbf{ R} \left(\mathbf{R} \cdot \mathbf{e}_{z}\right)}{R_- R_+}\equiv\nonumber\\
    &\equiv&\frac{\left(\cosh u\cos v-1\right)\mathbf{e}_{z}-\sinh u\sin v\,\mathbf{e}_{r}}{\cosh u -\cos v},\\ \mathbf{e}_u &=& \left[\mathbf{ e}_v \times \mathbf{ e}_\varphi\right]=\nonumber\\
    &=&\frac{-\sinh u\sin v\,\mathbf{e}_{z}-\left(\cosh u\cos v-1\right)\mathbf{e}_{r}}{\cosh u -\cos v}.\label{eq:ebeta1}
\eea
With the help of the \lame coefficients (Eq.~\ref{eq:lame}) one can express the (poloidal) magnetic field (Eq.~\ref{eq:BviaPsi}), 
\bea\label{eq:Bfield}
\mathbf{B}_2 = \frac{\mu_\snull R_\infty^{\half}}{2r^{\frac32}} 
\left[\psi\mathbf{ e}_z + \frac{ \kappa^2}{\kappa^\prime} \left(\frac{\partial \psi}{\partial v} \mathbf{ e}_u - \frac{\partial\psi}{\partial u} \mathbf{ e}_v\right)\right],
\eea
and the toroidal current density (Eq.~\ref{eq:jphi1}):
\be\label{eq:itopsi}
    j= \sinh^2u \left[-\frac{\partial^2\psi}{\partial u^2} - \frac{\partial^2\psi}{\partial v^2} + \frac{3\psi}{ 4\sinh^2u}\right].
\ee
Any solution of \textit{scalar} equation Eq.(\ref{eq:itopsi}) which relates the form-factor of the toroidal current to the reduced flux function, allows expressing the \textit{vector} poloidal magnetic field via Eq.~(\ref{eq:Bfield}). In particular, we will present \textit{manufactured} solutions, which, for some special choices of the current form-factor, lead to analytic expressions for the magnetic field.      
\section{Deriving the reduced flux function with the Fourier Method}
\label{sec:fourier}
Eq.~(\ref{eq:itopsi}) can be solved because the Laplacian in toroidal coordinates allows the \textit{separation of variables} within the framework of the \textit{Fourier method}. Both currents and fields are expressed in terms of products of eigenfunctions of a single variable depending either on $u$ or on $v$. This way a variety of solutions can be derived expressing the field analytically in terms of special \textit{toroidal} functions. Indeed, the scalar function appearing in the poloidal 
magnetic field (see Eq.~\ref{eq:Bfield}) can be expressed as a complex series: 
\be\label{eq:pcalseries}
    \psi = \sum_{n=-\infty}^\infty e^{\mathrm{i}nv} \psi_n (u), \quad \psi_n=\psi^*_{-n},
\ee
where $\mathrm{i}^2=-1$ and the superscript asterisk means complex conjugation.

To find the magnetic field harmonics,  the current, $j(u,v)$, in Eq.~(\ref{eq:jphi1}) is also expanded into a Fourier series:
\be\label{eq:jphiseries}
    j(u,v) = \sum_{n=-\infty}^\infty e^{\mathrm{i} nv} j_n(u), \qquad j_n=j^*_{-n}.
\ee
Equations relating the magnetic field and current harmonics can be derived from Eqs.~ (\ref{eq:itopsi}, \ref{eq:pcalseries} and \ref{eq:jphiseries}):
\be\label{eq:pcalviaj}
    -\frac{d^2\psi_n}{du^2}+\left(n^2+\frac3{4\sinh^2u}\right)\psi_n=\frac{ j_n(u)}{\sinh^2u}.
\ee
The substitution, $\psi_n = \sqrt{2\sinh u}{\mathcal P}_n(u)$, reduces this equation  with zero RHS to the equation for Legendre functions of semi-integer index, $P^{-1}_{n-\half}(\cosh u)$ and $Q^{-1}_{n-\half}(\cosh u)$. The eigenfunctions of Eq.~(\ref{eq:pcalviaj}) are
\bea\label{eq:toroidalfunctions}
    \bar{P}^{-1}_{n-\half}(u) &=& \sqrt{2\sinh u} P^{-1}_{n-\half}(\cosh u), \nonumber \\
    \bar{Q}^{-1}_{n-\half}(u) &=& \sqrt{2\sinh u} Q^{-1}_{n-\half}(\cosh u),
\eea
(see Eqs.~\ref{eq:Pminus1} and \ref{eq:qminus1}), which are referred to as \textit{toroidal functions}. Since the toroidal current is assumed to vanish  \textit{outside} the torus, i.e., $J_\varphi=0$ for $0<u<u_\snull $, the field in this region can be expressed as a series of toroidal functions of the first kind: $\psi_n\propto \bar{P}^{-1}_{n-\half}( u)$, because the Legendre functions of the second  kind, $\bar{Q}^{-1}_{n-\half}( u)$, are singular at $u\to0$.

The current, $I(u)$, through a contour of constant $u$, can be expressed as a series of harmonics:
\bea\label{eq:Iu}
    I(u) &=& \int\limits_{u}^\infty \int\limits_\snull ^{2\pi}J_\varphi h_u h_v  \, \mathrm{d}v \mathrm{d}u_1 = \sum_{n=-\infty}^\infty I_n(u),\nonumber \\
    I_n(u)&=& \int\limits_{u}^\infty \left(\int\limits_\snull ^{2\pi} e^{\mathrm{i} nv} \sqrt{\frac{R_\infty}{r}} \mathrm{d}v\right) \frac{j_n(u_1)}{\sinh^2u_1} \mathrm{d}u_1  \quad
\eea
where the inner integral over $v$ can be evaluated using Eq.~(\ref{eq:qminus1series}), yielding:
\be\label{eq:In}
    I_n(u) = \int_{u}^\infty \bar{Q}^{-1}_{n-\half}( u_1) \frac{j_n(u_1)}{\sinh^2u_1} \mathrm{d}u_1.
\ee
The $n$-th harmonics of the total current through the current loop, $I_n(u_\snull )$, will be denoted as $I_{n_\snull }$:
\be\label{eq:In0}
I^\mathrm{tot}=\sum\limits_{n=-\infty}^{\infty}{I_{n_\snull}}=\sum\limits_{n=-\infty}^{\infty}{I_n(u_\snull)}.
\ee
Similarly, with the help of Eq.~(\ref{eq:q1series}) the harmonics of magnetic moment defined as the volume integral, $\mathcal{M}=\frac12\int{rJ_\varphi\mathrm{d}V}$, can be obtained in terms of quantities introduced above:
\bea\label{eq:magmomentharm}
&&\mathcal{M}=\pi\int\limits_{u_\snull}^\infty \int\limits_\snull ^{2\pi}J_\varphi r^2h_u h_v  \, \mathrm{d}v \mathrm{d}u_1=\nonumber\\
&&=\pi R^2_\infty\int\limits_{u}^\infty \left(\int\limits_\snull ^{2\pi} e^{\mathrm{i} nv}
\sqrt{\frac{r^3}{R^3_\infty}}
\mathrm{d}v\right)\frac{j_n(u_1)}{\sinh^2u_1} \mathrm{d}u_1=\nonumber\\
&&=\pi R^2_\infty\sum\limits_{n=-\infty}^{\infty}{\left(1-4n^2\right)I_{n_\snull}}
\eea

The quantities, $I_n(u)$, $j_n(u)$ as well as $\psi_n(u)$ all have dimensions of current, therefore, it is convenient to characterize the distributions of current and reduced flux function with the dimensionless quantities normalized by the appropriate harmonics of total current:
\bea\label{eq:normalized}
&&\tilde{I}_n(u)=\frac{I_n(u)}{I_{n_\snull}},\quad \tilde{I}_n(u_0)=1,\nonumber\\ 
&&\tilde{j}_n(u)=\frac{j_n(u)}{I_{n_\snull}},\quad \int_{u_0}^\infty \bar{Q}^{-1}_{n-\half}( u_1) \frac{\tilde{j}_n(u_1)}{\sinh^2u_1} \mathrm{d}u_1=1,\nonumber\\ &&\tilde{\psi}_n(u)=\frac{\psi_n(u)}{I_{n_\snull}}.
\eea

One can express the magnetic field harmonics in terms of the current harmonics using a convolution integral (see Eq.~\ref{eq:wronskian}):
\be\label{eq:convol}
    \tilde{\psi}_n = \int_{u_\snull }^\infty {G_n(u,u_1) \frac{\tilde{j}_n(u_1)}{\sinh^2u_1} \mathrm{d}u_1}
\ee
where we introduced the \textit{Green function},
\bea\label{eq:Green}
    G_n(u,u_1) &=& \left(\oneeigths-\frac{n^2}{2}\right) \bar{P}^{-1}_{n-\half}(\min(u,u_1)) \times \nonumber \\ & \times & \bar{Q}^{-1}_{n-\half}(\max(u,u_1)).
\eea
Following general rules, the Green function is constructed from the eigenfunctions of Eq.~(\ref{eq:pcalviaj}) satisfying the proper boundary conditions. While it is continuous, its derivative, $\partial G_n/\partial u=1$, has a discontinuity at $u=u_1$ in a way that the second derivative equals to the negative of the Dirac $\delta$-function. This is why Eq.~(\ref{eq:convol}) provides a solution to Eq.~(\ref{eq:pcalviaj}) for a given current in the right hand side (RHS).

\textit{Inside} the current filament, ($u>u_\snull $), the integration of the Green function (Eq.~\ref{eq:Green}) gives:
\bea\label{eq:fluxfunint}
    & \tilde{\psi}_n &(u > u_\snull ) = \left(\oneeigths-\frac{n^2}{2}\right) \left[\bar{P}^{-1}_{n-\half}(u) \tilde{I}_n(u)  \right.\nonumber \\
    & & \left. + \bar{Q}^{-1}_{n-\half}(u) \int\limits_{u_\snull }^u {\frac{\bar{P}^{-1}_{n-\half}(u_1) \tilde{j}_n(u_1) \mathrm{d}u_1}{\sinh^2u_1}}\right] .
\eea
\textit{Outside} the current loop where $u \le u_\snull \le u_1$, Eqs.~(\ref{eq:convol} and \ref{eq:Green}) give:
\be\label{eq:fluxfuncext}
    \tilde{\psi}_n(u\le u_\snull ) = \left(\oneeigths-\frac{n^2}{2}\right) \bar{P}^{-1}_{n-\half}(u).
\ee
Even though the reduced flux functions in Eqs.~(\ref{eq:fluxfunint} and \ref{eq:fluxfuncext}) are continuous at the filament surface ($u=u_\snull$), the derivatives might be discontinuous when the finite surface current is concentrated at the filament boundary. 

To conclude this Section, we provide equation for the total reduced flux function for the case when the current distribution is symmetric with respect to the $z=0$ plane, so that the flux is an even function of $v$, the current amplitudes are real functions, and one can use $e^{\mathrm{i}nv}\equiv\cos(nv)$. With these simplifications, the reduced flux function ibecomes (see Eqs.~\ref{eq:pcalseries} and \ref{eq:fluxfuncext}):
\bea\label{eq:extfieldseries}
\psi(u\le u_\snull ) &=& 
\sum_{n=-\infty}^\infty
{\left(\oneeigths-\frac{n^2}{2}\right)
I_{n_\snull} \bar{P}^{-1}_{n-\half}(u) 
\cos\left(nv\right)}, \nonumber\\
I_{n_\snull}&=&I_{-n_\snull}.
\eea
\section{Constructing Magnetic Field Configurations for the Zeroth Harmonic}
\label{sec:n=0}
Our objective is to construct simple, analytic expressions for a twisted toroidal magnetic flux rope that can be superimposed on observed solar active region magnetic fields. Such a configuration can be obtained using the lowest order harmonics of the Fourier series solution discussed in Sect.~\ref{sec:fourier}.

Let us assume that there is only the $n=0$ Fourier harmonic in the current distribution, $\tilde{j}(u)\equiv \tilde{j}_\snull(u)$,
which only depends on $u$. The subscript ``0'' that denotes quantities related to the $n=0$ harmonic is omitted herewith. The only contribution to the total current comes from this harmonic, $I^\mathrm{tot}=I_{0_\snull}=I_\snull(u_\snull)$. Now, we   consider the $n=0$ harmonic of the magnetic field, in which the reduced flux function, $\tilde{\psi}\equiv\tilde{\psi}_\snull (u)$, also only depends on $u$, so that Eq.~(\ref{eq:Bfield}) becomes:
\be
\label{eq:Bn=0a}
\mathbf{B}_\snull =
B_c\left(\frac{R_\infty}{r}\right)^{\frac32}
\left[B^{(z)}(u)\mathbf{e}_z
- B^\text{(p)}(u)\mathbf{e}_v\right],\quad
\ee
where
\be\label{eq:Bc}
    B_c = \frac{\mu_\snull  I^\mathrm{tot}}{2R_\infty},
\ee
is the magnetic field at  the origin, $R=0$. The dimensionless amplitudes,
\be\label{eq:BAmplitudes}
    B^{(z)}(u) \equiv \tilde{\psi},
    \quad
    B^\text{(p)}(u) \equiv \frac{\kappa^2}
    {\kappa^\prime} \frac{\mathrm{d}
    \tilde{\psi} }{\mathrm{d}u},
\ee
describe the axial and  poloidal fields, respectively. It is convenient to eliminate the false singularity in $\mathbf{e}_v$, by transforming the denominator in Eq.~(\ref{eq:ebeta}) using the definitions of $R_\pm$ and $\kappa^\prime$ (Eqs.~\ref{eq:Rpm} and ~\ref{eq:kappas}): $R_-R_+=\kappa^\prime R_+^2
$, so that Eq.~(\ref{eq:Bn=0a}) can be written as:
\be
\label{eq:Bn=0}
\mathbf{B}_\snull =
\left(\frac{R_\infty}{r}\right)^{\frac32}B_c
\left[B^{(z)}(u)\mathbf{e}_z
- \frac{B^\text{(p)}(u)}{\kappa^\prime}\left(\kappa^\prime\mathbf{e}_v\right)\right],
\ee
where the singularity is eliminated:
\be
\label{eq:kappaev}
\left(\kappa^\prime \mathbf{e}_v\right) = \frac{\left(R^2-R_\infty^2\right) \mathbf{e}_{z} - 2\mathbf{ R} \left(\mathbf{R} \cdot \mathbf{e}_{z}\right)}{ R_+^2}.
\ee

In order to eliminate another false singularity in Eq.~(\ref{eq:Bn=0}) \textit{outside} the torus  ($u < u_\snull$) we use the definition of $\kappa$ igiven by Eq.~(\ref{eq:kappas}):
\be
\left(\frac{R_\infty}{r}\right)^{\frac32}
=\frac{8 R_\infty^3}{\kappa^3R_+^3}.
\ee
Substituting this expression into Eq.~(\ref{eq:Bn=0}) yields
\be
\label{eq:Bn=0Calc}
\mathbf{B}_\snull =
\frac{8 R_\infty^3}{R_+^3} B_c
\left[\frac{B^{(z)}(u)}{\kappa^3}\mathbf{e}_z
- \frac{B^\text{(p)}(u)}{\kappa^3\kappa^\prime} \left(\kappa^\prime \mathbf{e}_v\right) \right].
\ee

For $n=0$ the field amplitude and its derivative appearing in Eq.~(\ref{eq:BAmplitudes}) can be obtained outside the filament from the reduced flux function, Eq.~(\ref{eq:fluxfuncext}):
\bea\label{eq:extfieldamplitudes}
&& \tilde{\psi}(u\le u_\snull ) = \frac{1}{8}\bar{P}^{-1}_{-\half}(u),
\\
&&\frac{\kappa^2}{\kappa^\prime} \frac{\tilde{\psi}(u\le u_\snull)}{\mathrm{d}u} = \tiny{\frac{3}{8}} \bar{P}^{-1}_{\half}(u), \nonumber
\eea
so that Eq.~(\ref{eq:Bn=0Calc}) reads:
\bea
\label{eq:BExternal}
&&\mathbf{B}_\snull (u \le u_\snull) = \frac{R_\infty^3 }{R_+^3}B_c\times\nonumber\\
&&\times \left\{\left[\frac{\bar{P}^{-1}_{-\half}(u)}{\kappa^3}\right] \mathbf{e}_z - 3\left[\frac{\bar{P}^{-1}_{\half}(u)}{\kappa^3\kappa^\prime}\right] \left(\kappa^\prime \mathbf{e}_v\right)\right\}.
\eea

The  ratio,
$\bar{P}^{-1}_{n-\half}(u)/ \left[\kappa^3 \left(\kappa^\prime\right)^n \right]$, that appears twice in Eq.~(\ref{eq:BExternal}) can be expressed  in terms of a hypergeometric series of powers of $\kappa^2$ (see Eq.~\ref{eq:Pminus1}). For $\kappa\to0$ it approaches $1/4$. Specifically, at the center where, according to Eqs.~(\ref{eq:Rpm},\ref{eq:kappas}, and \ref{eq:ebeta}), $\kappa=0$, $R=0$, $R_+=R_\infty$ and $\left(\kappa^\prime\mathbf{e}_v\right)\to-\mathbf{e}_z$, this  approximation of the toroidal functions in Eq.~(\ref{eq:BExternal}) gives $\lim_{R\to0}\mathbf{B}_\snull = B_c\mathbf{e}_z$, as required. 

At large distances from the filament, $R\gg R_\infty$, Eq. (\ref{eq:BExternal}) approaches the magnetic field of a dipole with the magnetic moment of the $n=0$ harmonic given by Eq.~(\ref{eq:magmomentharm}):
\be\label{eq:magmoment}
\bm{\mathcal{M}} = \pi R^2_\infty I^\mathrm{tot}\mathbf{e}_z.
\ee
Close to the current loop, where $\kappa\approx1$ and Eqs.~(\ref{eq:ebeta} and \ref{eq:ebeta1}) at $u\to\infty$ can be approximated as follows,
\bea\label{eq:evapprox}
\mathbf{e}_v&\approx&\cos v\,\mathbf{e}_z-\sin v\,\mathbf{e}_r,\\
\mathbf{e}_u&\approx&-\sin v\,\mathbf{e}_z-\cos v\,\mathbf{e}_r,\nonumber
\eea
one can approximate functions $\bar{P}^{-1}_{\pm\half}(u)$ using Eqs.~(\ref{eq:PTransformed}) and demonstrate that the external poloidal field dominates: 
\be
    \mathbf{B}_\snull \approx \frac{\mu_\snull I^\mathrm{tot}}{2\pi R_-} \left(\mathbf{e}_r \sin v - \mathbf{e}_z \cos v\right).
\ee

In effect, Eq.~(\ref{eq:BExternal}), describes the magnetic field of an infinitely thin ring current with major radius of $R_\infty$, even though it is derived as the magnetic field of an arbitrary $u-$dependent current distribution. Furthermore, it is not assumed that the ratio, $a/R_\snull$, is infinitesimal. In addition, the major radius of the current filament differs from that of the infinitely thin ring, $R_\snull \ne  R_\infty$.

\begin{figure*}[htb]
\centering
\includegraphics[width=0.33\textwidth]{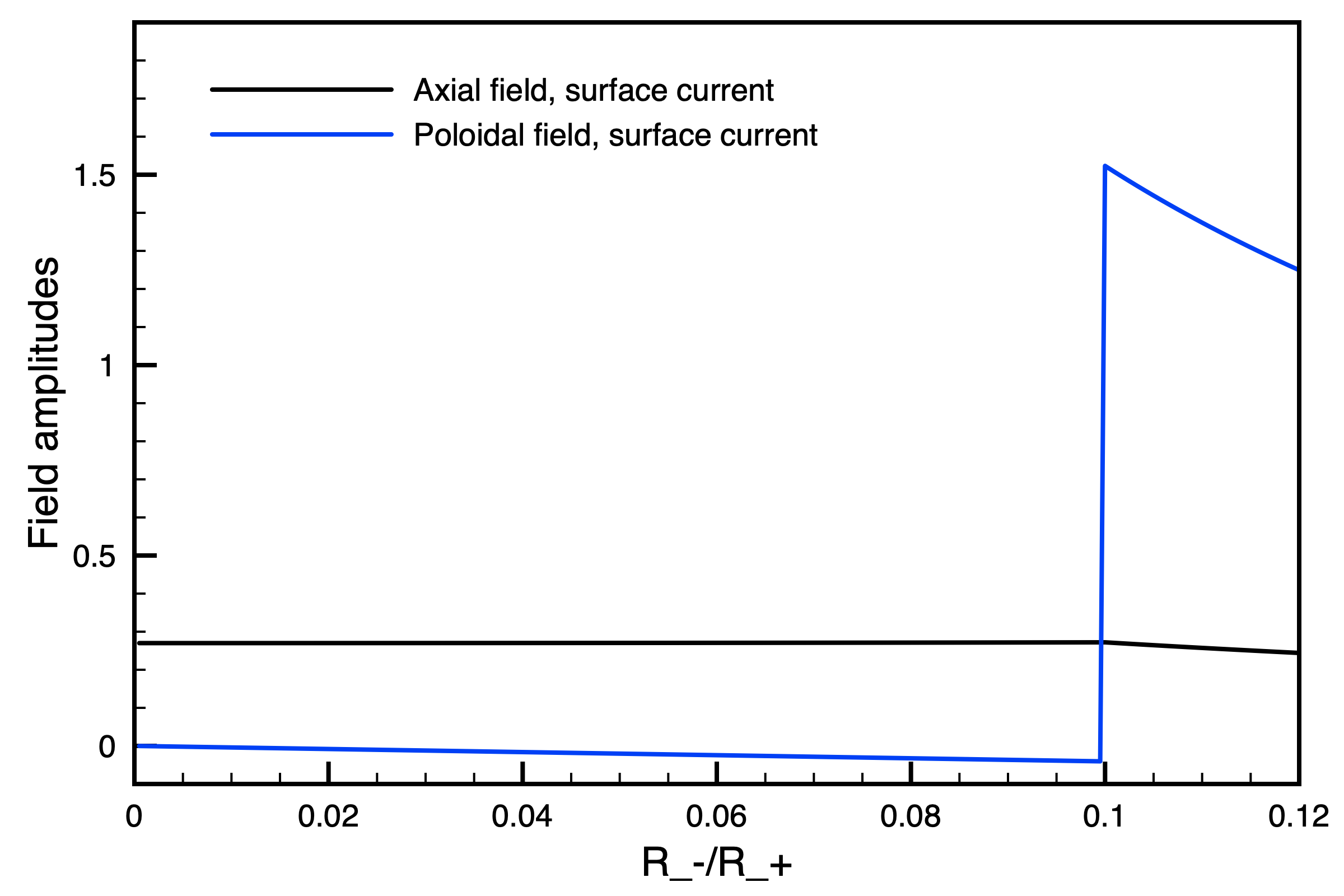}
\includegraphics[width=0.33\textwidth]{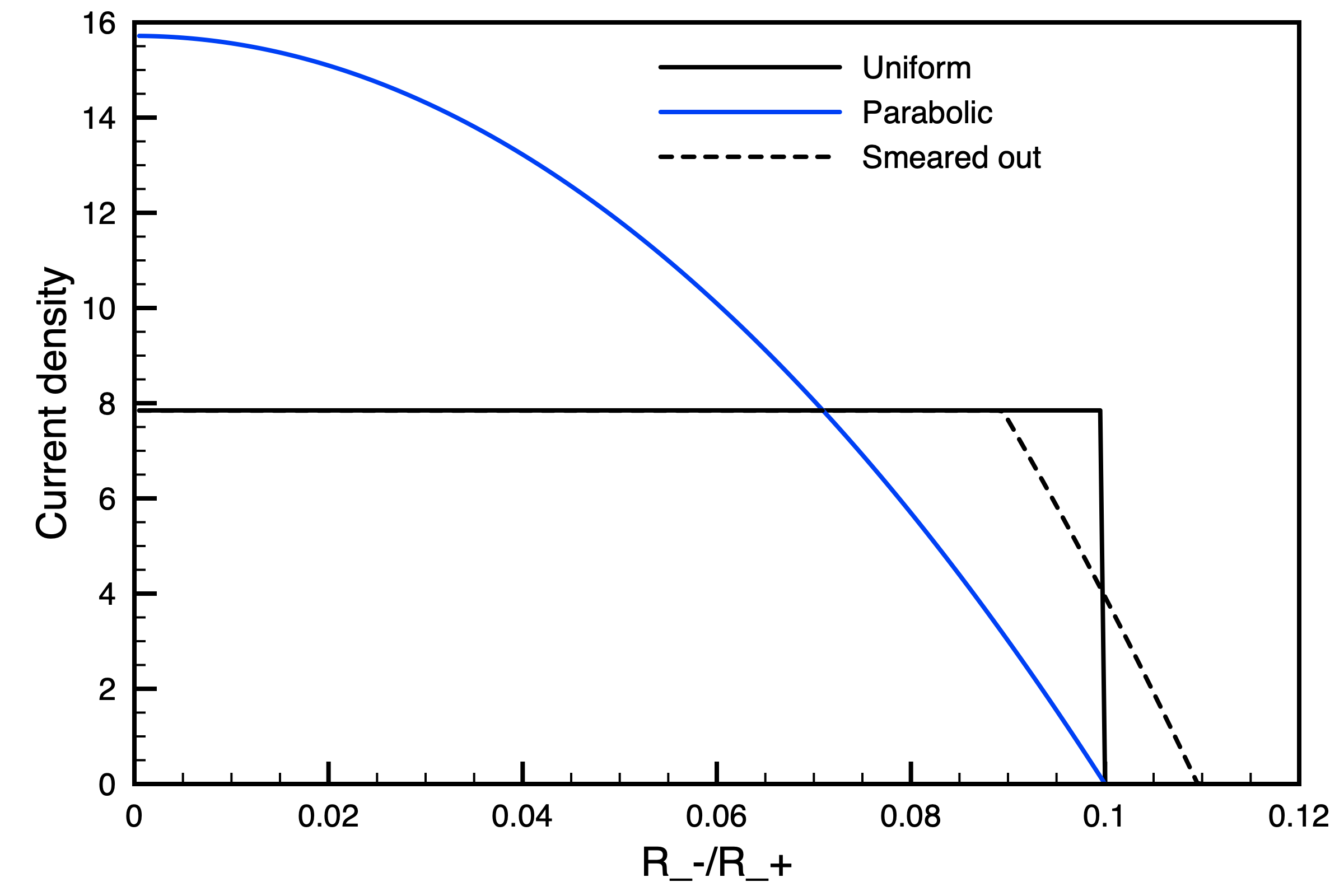}
\includegraphics[width=0.33\textwidth]{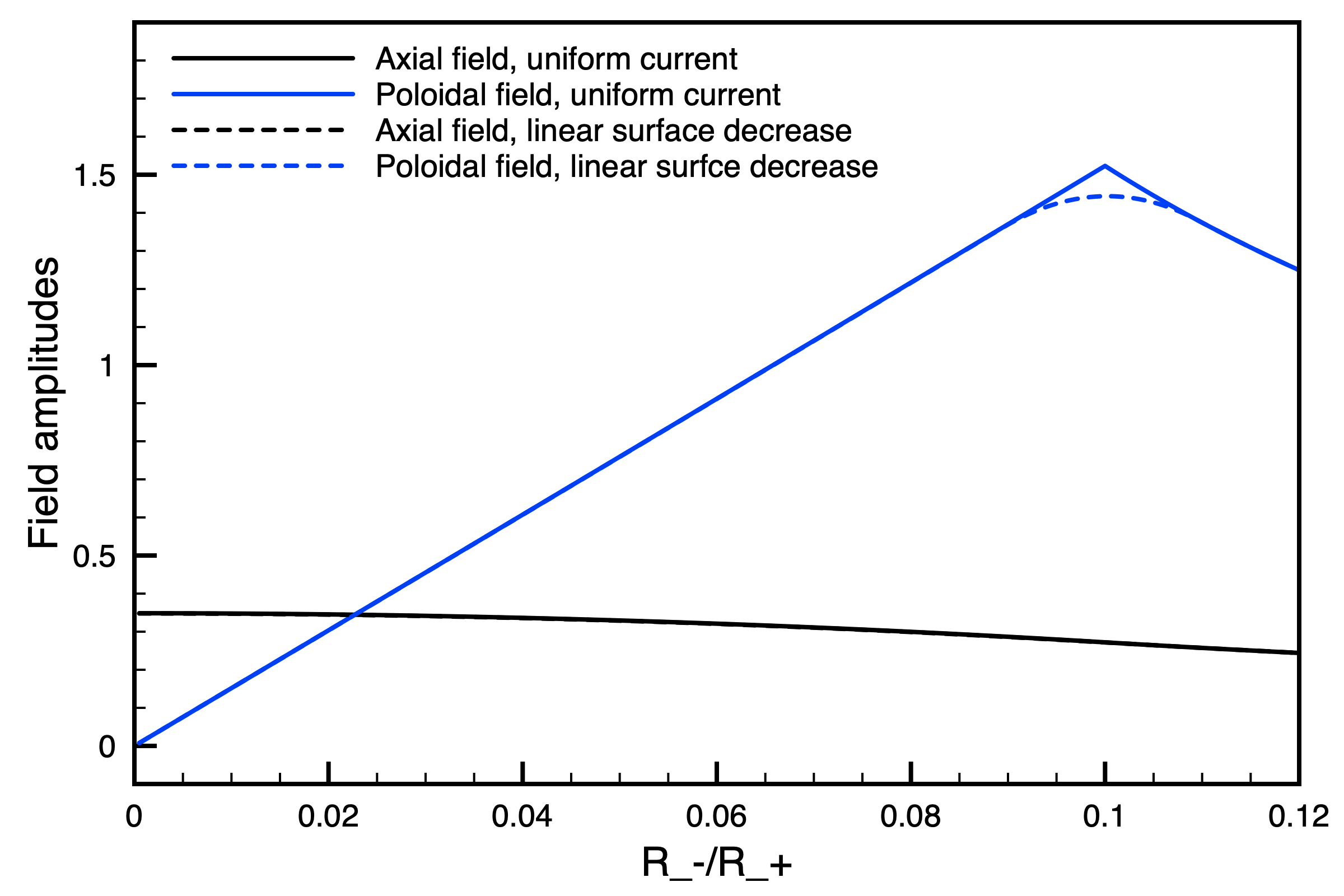}
\caption{\textbf{Left panel}: Amplitudes of poloidal (blue line) and axial (black line) magnetic field components for the current concentrated at the filament boundary. \textbf{Middle panel:} Normalized current distributions for uniform (black solid line), parabolic (blue line) and linear surface decrease current (black dashed line for $\varepsilon=0.1$. \textbf{Right panel:} Amplitudes of poloidal (blue line) and axial magnetic field (black line) components for uniform current (solid lines) and for linear surface decrease current (dashed lines).  For $\kappa^\prime_\snull =0.1$ ($a/R_\snull \approx0.2$) the argument, $\kappa^\prime=R_-/R_+$, ranges from $\kappa^\prime=0$ (at the toroidal magnetic axis) to $\kappa^\prime=0.1$ at the filament boundary; while $\kappa^\prime>0.1$ values correspond to the loop exterior.}
\label{fig:Inner}
\end{figure*}

A simple example for the field \textit{inside} the filament can be found if the current is concentrated at the filament surface: 
\bea\label{eq:innerscr}
\tilde{\psi}(u\ge u_\snull ) &=& \ell^{(\text{s})} (u_\snull ) \bar{Q}^{-1}_{-\half}(u),\\
\frac{\kappa^2}{\kappa^\prime} 
\frac{\tilde{\psi}(u\le u_\snull)}{\mathrm{d}u}&=&3\ell^{(\text{s})} (u_\snull ) \bar{Q}^{-1}_{\half}(u),\nonumber 
\eea
where
\be\label{eq:q_0}
    \ell^{(\text{s})} (u_\snull ) = \frac{\bar{P}^{-1}_{-\half}( u_\snull )}{8\bar{Q}^{-1}_{-\half}(u_\snull )}.
\ee
Here $l^{(\text{s})}$ is a dimensionless induction coefficient (proportional to the flux-to-current ratio) for the surface current, which is-- as demonstrated below (see Section~\ref{Section:Energy}) -- closely related to the energy of the external poloidal magnetic field produced by this current.
Eqs.~(\ref{eq:extfieldamplitudes}) and (\ref{eq:innerscr}) can be combined and written in terms of the Green function, Eq.~(\ref{eq:Green}):
\be\label{eq:viagreen}
\tilde{\psi}=\frac{G_\snull\left(u,u_\snull\right)}{\bar{Q}^{-1}_{-\half}\left(u_\snull\right)}. 
\ee

Below we consider several specific situations and express $\tilde{\psi}$ in terms of the Green function(s), while the field amplitudes, $B^{(z,\text{p})}$, are expressed in terms of the normalized reduced flux function, $\tilde{\psi}$.
The amplitudes of poloidal (blue line) and axial (black line) magnetic fields  are shown in the left panel of Fig.~\ref{fig:Inner}.

A variety of more realistic solutions for the magnetic field inside a plasma can be constructed by approximating the current profile as a linear combination of specially chosen current profiles (``form-factors''), $j^{(m)}$ with constant coefficients, $c_m$,
\be\label{eq:imanufactured1}
    j (u)=\sum\limits_m c_m j^{(m)}(u).
\ee
The specially chosen form-factors satisfy the equation:
\bea\label{eq:imanufactured2}
   && \left(-\sinh^2u \frac{d^2}{du^2} + \frac{3}{4}
    \right) j^{(m)} = E^{(m)} j^{(m)},\\
    &&j^{(0)}(u)=1,\quad  j^{(1)}(u)=\coth{u}, \dots\nonumber\\
    &&E^{(0)} = \frac{3}{4},\quad E^{(1)}=- \frac{5}{4}, \dots\nonumber
\eea
With this choice, the integral in Eq.~(\ref{eq:In}) can be evaluated analytically:
\bea
\label{eq:I0u}
&&I(u) = \bar{Q}^{-1}_{-\half}(u) \frac{dj_E(u)}{du} - j_E(u) \frac{d\bar{Q}^{-1}_{-\half}(u)}{du}, \nonumber\\ 
&&j_E(u) = \sum_m{\frac{c_mj_\snull^{(m)}(u)}{E^{(m)}}},
\eea
(see Sect.~\ref{sec:integral} for more details). Eq.~(\ref{eq:I0u}) provides a simple normalization recipe: (1) for a current profile given by Eq.~(\ref{eq:imanufactured1}) the modified current distribution, $j_E(u)$, should be constructed according to Eq.~(\ref{eq:I0u}); (2) using $j_E(u)$ the \textit{normalization integral},
\be\label{eq:normalizationint}
N = \bar{Q}^{-1}_{-\half}(u_\snull) \frac{dj_E(u_\snull)}{du_\snull} - j_E(u_\snull) \frac{d\bar{Q}^{-1}_{-\half}(u_\snull)}{du_\snull}
\ee
should be calculated; and (3) the normalized current distributions are calculated then as  $\tilde{j}(u)=j(u)/N$ and $\tilde{j}_E(u)=j_E(u)/N$. The normalized current satisfies the identity:
\be\label{eq:identity}
\bar{Q}^{-1}_{-\half}(u_\snull) \frac{d\tilde{j}_E(u_\snull)}{du_\snull} - \tilde{j}_E(u_\snull) \frac{d\bar{Q}^{-1}_{-\half}(u_\snull)}{du_\snull}=1.
\ee
Specifically, for ``uniform'' current when $j=\mathrm{const}$:
\bea\label{eq:currentuniform}
&&\tilde{j}=\frac{1}{N^\text{uni}(u_\snull)},\quad \tilde{j}_E=\frac{1}{E^{(0)}N^\text{uni}(u_\snull)},\nonumber\\
&&N^\text{uni}(u_\snull)=-\frac1{E^{(0)}}\frac{d\bar{Q}^{-1}_{-\half}(u_\snull)}{du_\snull}.
\eea

For a ``parabolic'' current profile we get:
\bea\label{eq:currentpar}
&&\tilde{j}=\frac{\coth(u_\snull)-\coth u}{N^\text{par}(u_\snull,u_\snull)},\nonumber\\
&&\tilde{j}_E=\frac{\coth(u_\snull)}{E^{(0)}N^\text{par}(u_\snull,u_\snull)}-\frac{\coth u}{E^{(1)}N^\text{par}(u_\snull)},\nonumber\\
&&N^\text{par}(u_\snull,u)=\frac{\bar{Q}^{-1}_{-\half}(u)}{E^{(1)}\sinh^2(u)}-\nonumber\\
&&\quad -\left(\frac{\coth(u_0)}{E^{(0)}}-\frac{\coth(u)}{E^{(1)}}\right)\frac{d\bar{Q}^{-1}_{-\half}(u)}{du}.
\eea

Even though a uniform current results in an even more simple solution, the discontinuous current profile near the filament boundary results in large numerical errors in various physical quantities. In order to eliminate this discontinuity one can consider the current distribution given by Eq.~(\ref{eq:imanufactured1}) with piece-wise constant coefficients, $c_m$,  combining the features of Eqs.~(\ref{eq:currentuniform}) and (\ref{eq:currentpar}) to ``linearly'' decrease the current density over a narrow interval of $u_\snull-\varepsilon<u<u_\snull+\varepsilon$, $\varepsilon\ll1$. Specifically, we define the boundaries, $(u^-_\snull,u^+_\snull)$ of this interval with the equation, $\kappa^\prime(u_\snull^\mp)=\sqrt{1\pm2\varepsilon}\kappa^\prime_\snull)$. This leads to the following expressions:
\bea\label{eq:currentsmeared}
\tilde{j}&=&\left\{\begin{array}{ll} 
\frac{\coth u^-_\snull-\coth u}
{\left[\coth(u^-_\snull)-\coth(u^+_\snull)\right]\Sigma N},\,\mbox{if $u^-_\snull<u<u^+_\snull$} \\ \frac1{\Sigma N},\quad\qquad\qquad\mbox{if $u>u^+_\snull$,}\end{array}\right.\nonumber\\
\tilde{j}_E&=&\left\{\begin{array}{ll} 
\frac{
\frac{1}{E^{(0)}}\coth u^-_\snull-\frac{1}{E^{(1)}}\coth u}
{\left[\coth(u^-_\snull)-\coth(u^+_\snull)\right]\Sigma N},\,\mbox{if $u^-_\snull<u<u^+_\snull$} \\ \frac1{E^{(0)}\Sigma N},\quad\qquad\qquad\mbox{if $u>u^+_\snull$,}\end{array}\right.\nonumber\\
\Sigma N&=&N^-+N^+,\,\,\,\, N^-=\frac{N^\text{par}(u^-_\snull,u^-_\snull)}{\coth(u^-_\snull)-\coth(u^+_\snull)},\nonumber\\
N^+&=&-\frac{N^\text{par}(u^-_\snull,u^+_\snull)}
{\coth(u^-_\snull)-\coth(u^+_\snull)}+N^\text{uni}(u^+_\snull).
\eea
In the case of a thin filament these expressions give $\tilde{j} \approx 1/{\pi a^2}$ if $0\le R_-\le (1-\varepsilon)a$, and $\tilde{j} \approx {(1+\varepsilon)a-R_-}/({2\pi\varepsilon a^3)}$ if $(1-\varepsilon) a\le R_-\le (1+\varepsilon)a$.
The normalized current density distributions given by Eqs.~(\ref{eq:currentuniform}-\ref{eq:currentsmeared}) are shown in Fig.~\ref{fig:Inner} (middle panel).

With the help of Eq.~(\ref{eq:wronskian}) one can evaluate the integral in Eq.~(\ref{eq:fluxfunint}) for $n=0$ (similarly to Eq.~\ref{eq:I0u}), to find the reduced flux function and then the field amplitudes:
\bea\label{eq:Formfactored}
\tilde{\psi}&=&\frac{G_\snull\left(u,u_\snull\right)}{\bar{Q}^{-1}_{-\half}\left(u_\snull\right)}+\nonumber\\
&+&\left\{\begin{array}{ll} 
0,\,\qquad\qquad\qquad\qquad\mbox{if $u<u_\snull$} \\ 
\tilde{j}_E(u)-\frac{\tilde{j}_E(u_\snull )\bar{Q}^{-1}_{-\half}(u)}{\bar{Q}^{-1}_{-\half}(u_\snull )},\,\mbox{if $u>u_\snull$,}\end{array}
\right.. 
\end{eqnarray}
This result is easy to verify and interpret: (1) by applying the differential operator on the LHS of Eq.~(\ref{eq:pcalviaj}) to Eq.~(\ref{eq:Formfactored}) and taking into account Eq.~(\ref{eq:imanufactured2}) one can see that Eq.~(\ref{eq:pcalviaj}) is satisfied in smooth regions; (2) the reduced flux function Eq.~(\ref{eq:Formfactored}) is continuous; and (3) the jump in the derivative of the second term at $u=u_\snull$ is cancelled by the controlled jump in the derivative of the Green function (see the discussion above), as it follows from Eqs.~(\ref{eq:identity}) and (\ref{eq:wronskian}). In the special case of constant form-factor 
 given by Eq.~(\ref{eq:currentuniform}) we get:
\be\label{eq:ffu}
    \tilde{j}_E(u) \equiv - \left(\frac{d\bar{Q}^{-1}_{-\half}(u_\snull )}{du_\snull }\right)^{-1},\quad
    \frac{\mathrm{d}\tilde{j}_E(u)}{\mathrm{d}u}
    \equiv0.
\ee

The amplitudes, $B^\text{(p)}(\kappa^\prime(u))$ and $B^\text{(z)}(\kappa^\prime(u))$, of the poloidal (solid curve) and axial (dashed curve) fields are shown in Fig.~\ref{fig:Inner} with black color. Outside the filament at $\kappa^\prime(u)>\kappa^\prime(u_\snull )$ the field does not depend on the current distribution, therefore, the black and blue curves overlap in this region. For the form-factor Eq.~(\ref{eq:currentsmeared}) the integration span in Eqs.~(\ref{eq:In}) and (\ref{eq:fluxfunint}) splits for domains separated by $u^+_\snull$, resulting in different expressions for the fields in these domains:
\bea\label{eq:fieldsmeared}
\tilde{\psi}&=&\frac{N^-}{\Sigma N}\frac{G_\snull\left(u,u^-_\snull\right)}{\bar{Q}^{-1}_{-\half}\left(u^-_\snull\right)}+\frac{N^+}{\Sigma N}\frac{G_\snull\left(u,u^+_\snull\right)}{\bar{Q}^{-1}_{-\half}\left(u^+_\snull\right)}+\nonumber\\
&+&\left\{\begin{array}{ll} 
0,\,\qquad\qquad\qquad\qquad\mbox{if $u<u^-_\snull$} \\ 
\tilde{j}_E(u)-\frac{\tilde{j}_E(u^-_\snull )\bar{Q}^{-1}_{-\half}(u)}{\bar{Q}^{-1}_{-\half}(u^-_\snull )},\,\mbox{if $u^-_\snull <u$,}\end{array}\right.+\nonumber\\
&+&\left\{\begin{array}{ll} 
0,\,\qquad\qquad\qquad\qquad\mbox{if $u<u^+_\snull$} \\ 
\Delta\left[\tilde{j}_E(u^+_\snull)\right]\frac{\bar{Q}^{-1}_{-\half}(u)}{\bar{Q}^{-1}_{-\half}(u^+_\snull )},\,\mbox{if $u^+_\snull <u$,}\end{array}\right.,
\end{eqnarray}
where $\Delta\left[\tilde{j}_E(u^+_\snull)\right]=
\tilde{j}_E(u^+_\snull-0)-\tilde{j}_E(u+_\snull+0)$ is the difference between the left and right limits of discontinuous function, 
$\tilde{j}_E$ at $u\rightarrow u^+_\snull$ (in contrast with the continuous current density 
function, $\tilde{j}$).

Note, that once Eqs.~(\ref{eq:Bn=0} and \ref{eq:BExternal}) are applied in the CME generator, their vector form allows us to calculate the field in any coordinate system without rotating the vector quantities to the system used in derivations presented in this paper. Indeed, these equations, together with Eq.~(\ref{eq:kappaev}), express the magnetic field vector as a linear combination of vectors $\mathbf{e}_z$ and $\mathbf{R}$, therefore the expression is valid in any coordinate system as long as the vectors $\mathbf{e}_z$ and $\mathbf{R}$ are given in the same coordinate system.

Specifically, in an arbitrary Cartesian coordinate system it is convenient to characterize the position of the current filament by the coordinates of its center, $\mathbf{R}_c$, and the unit vector, $\mathbf{n}_c$, directed along its axis of symmetry. Then, the field vector at a point, $\mathbf{R}^\prime$, is given by Eqs.~(\ref{eq:Bn=0}, \ref{eq:kappaev} and \ref{eq:BExternal}) with the following substitution:
\be
\mathbf{R}=\mathbf{R}^\prime-\mathbf{R}_c,\qquad \mathbf{e}_z=\mathbf{n}_c.
\ee
To calculate scalar functions, we also need to express:
\bea
&&z=\left(\mathbf{R}^\prime-\mathbf{R}_c\right)\cdot\mathbf{n}_c,\quad R^2=\left(\mathbf{R}^\prime-\mathbf{R}_c\right)^2,\nonumber\\
&&r=\sqrt{R^2-z^2},\quad R_\pm=\sqrt{(r\pm R_\infty)^2+z^2}.\label{eq:functionargs}
\eea
To calculate the  toroidal special function in Eqs.~(\ref{eq:BExternal} and \ref{eq:Formfactored}) for field amplitudes, one can calculate their arguments $\kappa$ and $\kappa^\prime$, using 
Eqs.~(\ref{eq:kappas} and \ref{eq:functionargs}). While the formulae for the magnetic field are repeatedly applied at each point where the magnetic field is needed, the filament parameters, $R_\infty=\sqrt{R_\snull^2-a^2}$, $\kappa^\prime_\snull=a/\left(R_\infty+R_\snull\right)$, $\kappa^2_\snull=1-\left(\kappa^\prime_\snull\right)^2$, and the coefficients $\tilde{j}_E(u_\snull),q_1(u_\snull)$ (Eq.~\ref{eq:ffu}) determining the field amplitudes are calculated only once in terms of the major and minor radii, $R_\snull, a$.
\section{Equilibrium Conditions for the Zeroth Harmonic}
\subsection{Full Grad-Shafranov Equation in Cylindrical Coordinates}
\label{sec:GS}
The magnetohydrodynamic (MHD) equilibrium theory of toroidal plasma configurations introduces the key concept of  \textit{magnetic surfaces}, where the flux function, $\Psi$, is constant. To apply this concept, let us start by expressing Eqs.~\ref{eq:BviaPsi} and \ref{eq:jphi1} in cylindrical coordinates ($u\equiv z,\,v\equiv r,\, H_z=H_r\equiv1$):
\be\label{eq:BviaPsirz}
    \mathbf{B}_2 = \frac{\nabla_2\Psi}{r}\times\mathbf{e}_\varphi, \qquad  
    \nabla_2\Psi = \frac{\partial\Psi}{\partial z} \mathbf{e}_z + \frac{\partial\Psi}{\partial r} \mathbf{e}_r,
\ee
\be\label{eq:jphirz}
    \mu_\snull  J_\varphi = -\frac{1}{r} \frac{\partial^2\Psi}{\partial^2 z} -\frac{\partial}{\partial r} \left(\frac{1}{r} \frac{\partial \Psi}{\partial r}\right).
\ee
Since the $\nabla_2\Psi$ vector is orthogonal to the surface of constant $\Psi$, the poloidal magnetic field, $\propto \nabla_2 \Psi \times \bf{e}_\varphi$, is parallel to the magnetic surface everywhere, while ``j-toroidal-cross-B-poloidal'' force, 
\be\label{eq:jtorcrossBpol}
J_\varphi\mathbf{e}_\varphi\times\mathbf{B}_2=\frac{J_\varphi}r\nabla_2\Psi,
\ee
is perpendicular to the magnetic surface (\ie aligned with $\nabla_2\Psi$). This force tends to contract the current filament over the minor radius (\textit{\ie pinch effect}). This contraction may be prevented by the excessive plasma gas-kinetic pressure, $P$, which tends to expand the filament. To balance the force, descibed by Eq.~(\ref{eq:jtorcrossBpol}) that is aligned with $\nabla_2\Psi$, the plasma pressure gradient needs to be aligned with $\nabla_2\Psi$ too. The alignment condition, $\nabla_2\Psi\times\nabla_2P=0$ can be identically rewritten in terms of the Jacobian, $D(\Psi,P)/D(z,r)\equiv\frac{\partial\Psi}{\partial z}\frac{\partial P}{\partial r}-\frac{\partial\Psi}{\partial r}\frac{\partial P}{\partial z}$. This Jacobian vanishes identically if, and only if, $P$ is only a function of $\Psi$, \ie it is constant at each magnetic surface, so that: 
\be\label{eq:gradP}
-\nabla_2P=-\frac{dP(\Psi)}{d\Psi}\nabla_2\Psi.
\ee
In a low-$\beta$ plasma, the pinch-effect is mainly prevented by the counter-action of the toroidal magnetic field, $B_\varphi$, for which the poloidal current density, $\mathbf{J}_2$, can be expressed in terms of a \textit{current function}, $rB_\varphi$:
\be\label{eq:jrz}
    \mu_\snull \mathbf{J}_2 = \frac{\nabla_2\left(rB_\varphi\right)}{r} \times\mathbf{e}_\varphi.
\ee
Again, the current function is required to be constant on magnetic surfaces, and therefore it can be expressed as a function of $\Psi$ only. In this case the poloidal electric current, $\mathbf{J}_2=\frac1{\mu_\snull r} \frac{d(rB_\varphi)}{d\Psi} \nabla_2\Psi \times \bf{e}_\varphi$ is everywhere parallel to the magnetic surface, while the ``j-poloidal-cross-B-toroidal'' force, 
\be\label{eq:jpolcrossBtor}
\mathbf{J}_2\times B_\varphi\mathbf{e}_\varphi=
-\frac{B_\varphi}{\mu_\snull r}
\frac{d\left(rB_\varphi\right)}{d\Psi}\nabla_2\Psi,
\ee
is perpendicular to the magnetic surface (\ie aligned with $\nabla_2\Psi$). Summing up, Eqs.~(\ref{eq:jtorcrossBpol}, \ref{eq:gradP}, and \ref{eq:jpolcrossBtor}) reduce the LHS of the \textit{vector} equilibrium condition, $J_\varphi\mathbf{e}_\varphi\mathbf{ B}_2+\mathbf{J}_2 \times B_\varphi\mathbf{e}_\varphi - \nabla P=0$, to a linear combination of aligned vectors:   
\be\label{eq:GradShafranovVect}
\left[
    \frac{J_\varphi}r -\frac{\left(rB_\varphi\right)}{\mu_\snull r^2} \frac{d\left(rB_\varphi\right)}{d\Psi} -\frac{dP}{d\Psi}\right] \nabla_2\Psi=0.
\ee
For Eq.~(\ref{eq:GradShafranovVect}) to hold everywhere, the expression in the square bracket must vanish.  This condition yields the \textit{scalar} GS equation:
\be\label{eq:GradShafranov}
   \frac{J_\varphi}r =\frac{\left(rB_\varphi\right)}{\mu_\snull r^2} \frac{d\left(rB_\varphi\right)}{d\Psi} +\frac{dP}{d\Psi}.
\ee
The LHS of this equation is often expressed using the RHS of Eq.~(\ref{eq:jphirz}), but in the present derivation this step is not needed. In equilibrium, the toroidal plasma filament boundary must coincide with a magnetic surface. 
\subsection{Shafranov's Virial Theorem and its Consequences for a Uniform Strapping Field}
\label{sec:ShafTheorem}

In Sections~\ref{sec:fourier} and \ref{sec:n=0} we considered only the magnetic field, $\mathbf{B}_2$, induced by the current flowing inside the filament. However, when discussing the MHD equilibrium of a circular current filament, one must also consider the \citet{shafranov66} virial theorem \cite[see also][]{Faddeev:2002} that states that the magnetic field of the current and the internal plasma pressure of the filament are not sufficient to maintain MHD equilibrium. As pointed out by \citet[][Ch.68]{Landau:1984},  the equilibrium condition, $\mathbf{J}\times\mathbf{B}-\nabla P=0$, can be reformulated in terms of the Maxwell stress tensor, $\mathbf{\Pi}$, with the help of \amp's~ law, $\nabla \times \mathbf{B} = \mu_\snull \mathbf{J}$:
\be\label{eq:divPi}
-\nabla\cdot{\mathbf{\Pi}}=0,\quad \mathbf{\Pi} = \left(P+\frac{B^2}{2\mu_\snull}\right)\bm{\mathcal{I}} - \frac{\mathbf{B}\otimes\mathbf{B}}{\mu_\snull},
\ee
$\bm{\mathcal{I}}$ being the unit tensor. By taking the scalar product of Eq.~(\ref{eq:divPi}) with $\mathbf{R}$ and integrating over the entire volume (the pressure and current density are zero outside the filament, but the magnetic field is not) by parts using the identity, $-\mathbf{R}\cdot\left(\nabla\cdot{\mathbf{\Pi}}\right)^T\mathrm=Tr\left(\mathbf{\Pi}\right)-\nabla\cdot\left(\mathbf{\Pi}\cdot\mathbf{R}\right)$ we obtain that the integral of the LHS of Eq.~(\ref{eq:divPi}) is positive definite:
\bea
\label{eq:SafTheorem}
&&-\int{\mathbf{R}\cdot\left(\nabla\cdot{\mathbf{\Pi}}\right)^T\mathrm{d}V}=E>0,\\
&&\quad E=\int{Tr\left(\mathbf{\Pi}\right)\mathrm{d}V}=\int{\left(3P+\frac{B^2}{2\mu_\snull}\right)\mathrm{d}V}.\nonumber
\eea
Since the RHS of Eq.~(\ref{eq:divPi}) is zero therefore it cannot be equal to the LHS, proving that any closed loop configuration is out of equilibrium in the absence of an external magnetic field.

Shafranov's theorem (Eq.~\ref{eq:SafTheorem}) in effect states that the \textit{hoop force} results from the interaction between the loop current and its self-generated magnetic field. 
Even though the integral of this force density ($\mathbf{f}^{(\text{hoop})} = - \nabla\cdot \mathbf{\Pi}$) vanishes over the entire volume ($\int{ \mathbf{f}^{(\text{hoop})} \mathrm{d}V} = -\int{ \nabla\cdot \mathbf{\Pi} \mathrm{d}V}=0$), it has a positive average projection to the radial direction ($\int{\mathbf{R} \cdot \mathbf{f}^{(\text{hoop})} \mathrm{d}V}=E>0$).
It is known from experiments \cite[see][]{Yee:2000} that the hoop force tends to expand the current loop outwards and in the absence of external fields this expansion is approximately self-similar. By approximating the velocity of this self-similar expansion as $\mathbf{v} = \frac{\mathbf{R}}{ R_\infty} \frac{\mathrm{d}R_\infty}{\mathrm{d}t}$, we see that Eq.~(\ref{eq:SafTheorem}) confirms the development of an expanding flow, since
the growth rate of the kinetic energy, $\mathrm{d} E^{(\text{k})} /\mathrm{d}t$ is positive:
\be
\label{eq:kinenergy1}
\frac{\mathrm{d}E^{(\text{k})}}{\mathrm{d}t} = \int{\mathbf{v} \cdot \mathbf{f}^{(\text{hoop})} \mathrm{d}V} = \frac{E}{R_\infty} \frac{\mathrm{d}R_\infty}{\mathrm{d}t}>0.
\ee

A more traditional derivation of the hoop force can be carried out using the energy principle, assuming that the pressure adiabatically scales with volume as $P\propto V^{-\gamma}$, and considering a particular choice of the polytropic index, $\gamma=4/3$. Consider a conformal expansion where each point, $\mathbf{R}$, maps to $\left(\delta R_\infty / R_\infty + 1\right) \mathbf{R}$. In this case the infinitesimal \textit{virtual displacement} is equal to $\delta\mathbf{R} = (\delta R_\infty/R_\infty) \,\mathbf{R}$. For a frozen-in magnetic field  the local magnetic field scales as $\propto(\delta R_\infty + R_\infty)^{-2}$, while the pressure adiabatically scales as $\propto[(\delta R_\infty + R_\infty)^3]^{-4/3}\propto(\delta R_\infty + R_\infty)^{-4}$, similarly to the magnetic pressure that scales as $B^2\propto(\delta R_\infty + R_\infty)^{-4}$ (this is why $\gamma=4/3$ was chosen). According to  general principles, the work done by local forces during the virtual displacement, $\int{\mathbf{f}^{(\text{hoop})}\cdot\delta\mathbf{R}\mathrm{d}V}$, equals the negative of the variation in the energy integral, $-\delta E=-\delta R_\infty\frac{\mathrm{d}E}{\mathrm{d}R_\infty}$:
\bea\label{eq:hoopviaenergy}
&&\frac1{R_\infty}
\int{\mathbf{R}\cdot\mathbf{f}^{(\text{hoop})}\mathrm{d}V}=-\frac{\mathrm{d}E}{\mathrm{d}R_\infty},\\
&&E=\int{\left(\frac{P}{\gamma-1}+\frac{B^2}{2\mu_\snull}\right)\mathrm{d}V}=\int{\left(3P+\frac{B^2}{2\mu_\snull}\right)\mathrm{d}V},\nonumber
\eea
where
$\frac{\mathrm{d}E}{\mathrm{d}R_\infty}=-\frac{E}{R_\infty}$, because the total energy scales as  $\propto(\delta R_\infty + R_\infty)^{-1}$. Even though the energy principle approach does not go beyond the already derived Eq.~(\ref{eq:SafTheorem}), it allows us to evaluate the energy of the motion driven by the hoop force. Combining Eqs.~(\ref{eq:kinenergy1} and \ref{eq:hoopviaenergy}) we find that:
\be\label{eq:kinenergy}
\frac{\mathrm{d}E^{(\text{k})}}{\mathrm{d}t}=-\frac{\mathrm{d}R_\infty}{\mathrm{d}t}\frac{\mathrm{d}E}{\mathrm{d}R_\infty}=-\frac{\mathrm{d}E}{\mathrm{d}t},\ee
hence, $\mathrm{d}\left(E^{(\text{k})}+E\right)/\mathrm{d}t=0$ and $E^{(\text{k})}(t\rightarrow\infty)=E(t=0)$.

For a thin circular current filament, one can approximate $|\mathbf{R}|\approx R_\infty$ in the integrand in Eq.~(\ref{eq:hoopviaenergy}), providing an estimate for the hoop force per unit toroidal angle:
\be\label{eq:hoopperangle}
\frac{\mathrm{d}F^{(\text{hoop})}}{\mathrm{d}\varphi}\approx\frac1{2\pi}\frac{E}{R_\infty}.
\ee
Eq.~(\ref{eq:hoopperangle})
follows from  Eq.~(\ref{eq:SafTheorem}) and is always valid, however, for $\gamma\ne4/3$ $E$ on the RHS is not the energy. This approximation connects our approach to the formalism used to describe the hoop force in the literature, (see Eq.~5 in \citet{titov99} and Eq.~2 in \cite{Toro06}). The main distinction between earlier work and our approach is that our model allows finite $\beta$ values (see details in Section~\ref{Section:Energy} below).

In  application to a CME generator, a current filament can be superposed on top of the model of an active region, so that a ``\textit{strapping magnetic field}'' of the active region maintains the equilibrium if it matches the filament geometry and parameters. In the case when the strapping field at the loop location, $\mathbf{B}^{(\text{s})}$, is uniform, a slight reformulation of the Shafranov theorem provides an estimate for the strapping field in terms of the filament parameters (or vice versa). Indeed, the integration of the modified equilibrium equation, $\left[-\nabla\cdot{\mathbf{\Pi}}+\mathbf{J}\times\mathbf{B^{(\text{s})}}\right]\cdot\mathbf{R}=0$, gives:
\be\label{eq:SafTheorem1}
E+2\mathbf{B}_2^{(\text{s})}\cdot\bm{\mathcal{M}} =0.
\ee
where 
\be\label{eq:magmoment1}
\bm{\mathcal{M}}=\frac12\int{\mathbf{R}\times\mathbf{J}\mathrm{d}V}
\ee
is the magnetic moment that has already been introduced earlier (see Eq.~\ref{eq:magmoment}). Eq.~(\ref{eq:SafTheorem1}) unambiguously determines the intensity of the uniform strapping field in terms of two integral parameters of the configuration. The direction of the strapping field must be aligned with the magnetic moment, otherwise a torque, $\bm{\mathcal{M}} \times \mathbf{B}_2^{(\text{s})} \ne0$ would act on the loop \cite[see][]{jackson1999} breaking the equilibrium. For axisymmetric current configurations 
the magnetic moment is parallel to the axis of symmetry,
\be\label{eq:magmoment2}
\bm{\mathcal{M}} = \mathcal{M}\mathbf{e}_z,
\qquad
\mathcal{M}=\frac12\int{rJ_\varphi\mathrm{d}V}.
\ee
It follows that the strapping field, $B^{(\text{s})}\mathbf{e}_z$,  must also be parallel to the axis of symmetry. For an axisymmetric configuration the square of the magnetic field can be decomposed to contributions from poloidal and toroidal fields, $B^2=B^2_\varphi+[\nabla\times(A_\varphi\mathbf{e}_\varphi]^2$, and thus reduce Eq.~(\ref{eq:SafTheorem1}) to the following form: 
\be\label{eq:SafTheorem2}
\int{\left(\frac{J_\varphi A_\varphi}{2} +\frac{B_\varphi^2}{2\mu_\snull}+3P\right)\mathrm{d}V}+2B^{(\text{s})}\mathcal{M} =0,
\ee
where the integrand is non-zero only inside the filament, simplifying the integration.

Identifying a circular arc inside an active region at which the magnetic field is uniform and orthogonal to the plane of the arc, choosing the loop parameters depending on thus determined strapping field and inserting this current loop along this arc is at the heart of the \cite{titov14} CME generator.  Our approach allows us to generalize the \cite{titov14} model and to extend it to finite $\beta$ current loops. This will be achieved by considering a detailed derivation of Eq.~(\ref{eq:SafTheorem1}) from the local equilibrium condition specified for a particular class of $n=0$ harmonic field
as discussed below in Sections~\ref{sec:ReducedGS} and \ref{subsec:strapping}.
\subsection{Reduced Grad-Shafranov Equation in Toroidal Coordinates}
\label{sec:ReducedGS}

In general, to find the conditions under which a plasma in the magnetic field of the $n=0$ harmonic is in force equilibrium, one needs to solve the Grad-Sharfranov equation in toroidal coordinates. Some of these solutions are known \cite[see, e.g.][]{Zakharov:1986}, however, they include infinite series 
of harmonics and require highly complicated strapping fields. While in application to laboratory plasmas such specially designed confining magnetic fields are not unusual, Nature does not implement such special analytic solutions.

Here, we use a more simple approach and reduce the GS equation in toroidal coordinates by assuming that, rather than finding magnetic surfaces where the true flux function, $\Psi$, is constant, their role in the formalism can be partly substituted by considering ``constant $\psi$-surfaces'', where the reduced flux function, $\psi(u,v)$, is constant.  In the particular case of the $n=0$ harmonic field $\psi(u,v)$ is a function of $u$ only, hence, these surfaces are toroidal coordinate surfaces of constant $u$. 

Note, that constant $\psi$-surfaces are not magnetic surfaces, since the true flux function, $\Psi=\mu_\snull\sqrt{R_\infty r}\psi$, (see Eq.~\ref{eq:tildepsi}) is not constant at constant $\psi$ and $\nabla_2\Psi$ is not orthogonal to constant $\psi$-surfaces, because of the -explicit dependence of $\Psi$ on $r$. In addition to the magnetic field generated by the filament current and characterized by $\psi$ function, the effect of the strapping field, $\mathbf{B}_2^{(\text{s})}=B^{(\text{s})}\mathbf{e}_z$, should be explicitly added to the net force balance, $\mathbf{J}_2 \times (\mathbf{B}_2 + B^{(\text{s})} \mathbf{e}_z) - \nabla P =0$. Now, we can use Eq.~(\ref{eq:Bfield}) for the magnetic field and Eq.~(\ref{eq:jphi1}) for the toroidal current density to describe the pinch-effect force in Eq.~(\ref{eq:jtorcrossBpol}): 
\bea\label{eq:GradShafranovreduced1}
    &&J_\varphi\mathbf{ e}_\varphi\times \left(\mathbf{ B}+B^{(\text{s})}\mathbf{e}_z\right)=
    \frac{j(u,v)R_\infty}{r^4}\times\nonumber\\
    &&\,\,\times\left[\mu_\snull r\nabla_2\psi +\left(
    \frac{\mu_\snull \psi}{2}+\frac{r^{\frac32}}{R^{\half}_\infty}B^{(\rm s)}\right)\mathbf{e}_r\right].
\eea 
The dominant contribution to the force in Eq.~(\ref{eq:GradShafranovreduced1}) is directed along $\nabla_2\psi$, hence, orthogonal to the constant $\psi$-surface. Following the basic idea of the GS equation, we parameterize the toroidal field, $B_\varphi$, and gas-kinetic pressure, $P$, in terms of the representative functions of $\psi$, 
$B_\varphi^2 = b^2 (\psi) (R_\infty/r)^3$ and $P = p (\psi) (R_\infty/r)^3$. 
The total force produced by $B_\varphi$ and $P$ is given by the sum of Eqs.~(\ref{eq:gradP}) and (\ref{eq:jpolcrossBtor}):
\bea\label{eq:GradShafranovreduced2}
&&\frac{\left[\nabla_2\left(rB_\varphi\right)\times\mathbf{ e}_\varphi\right] \times B_\varphi \mathbf{ e}_\varphi}{\mu_\snull r} - \nabla_2 P = \frac{R^3_\infty}{r^4} \times 
\nonumber\\
&& \times \left[-r\frac{\mathrm{d} p^\text{tot}}{\mathrm{d}\psi}\nabla_2\psi  +  \left(p^\text{tot} (\psi)+2p(\psi)\right) \mathbf{ e}_r\right].
\eea
We note that the gradient of the total pressure, $p^\text{tot} = p + b^2/(2\mu_\snull )$ is orthogonal to 
constant $\psi$-surfaces in Eq.~(\ref{eq:GradShafranovreduced2}). 
Similarly to Eq.~(\ref{eq:GradShafranovVect}), the total force, given by the sum of Eqs.~(\ref{eq:GradShafranovreduced1} and \ref{eq:GradShafranovreduced2}) 
vanishes in equilibrium if the following equation holds: 
\bea\label{eq:GradShafranovReduced}
&& \frac{R_\infty}{r^4}\left(\mu_\snull j -R_\infty^2 \frac{\mathrm{d} p^\text{tot}} {\mathrm{d}\psi}\right)r\nabla_2\psi +
\\
&&
+ \left[\frac{J_\varphi A_\varphi}{2} +\frac{B^2_\varphi}{2\mu_\snull} + 3P + rJ_\varphi B^{(\text{s})}\right]\frac{\mathbf{e}_r}r
=0.\nonumber
\eea
In Eq.~(\ref{eq:GradShafranovReduced}) the dominant force comes from the pinch-effect and its opposing pressure gradient. This term  is proportional to $\nabla_2\psi$ (i.e., it is normal to  $\psi=\mathrm{const}$ 
surfaces). This dominant force vanishes identically if the \textit{reduced} version of the GS equation(Eq.~\ref{eq:GradShafranov}) holds:
\be\label{eq:Grad}
j=\frac{R^2_\infty} {\mu_\snull }
\frac{\mathrm{d}p^\text{tot}}
{\mathrm{d}\psi}.
\ee
The reduced GS equation ensures equilibrium against the pinch-effect, similarly to the full equation (see Eq.~\ref{eq:GradShafranov}). However, because of the combined effect of the strapping field and of the $r$-dependent factors in the definitions of the $\psi$-function and the representative functions, $p(\psi)$, $b(\psi)$, there is also a force directed along $\mathbf{e}_r$ in Eq.~(\ref{eq:GradShafranovReduced}). The whole point of the proposed approach is that while balancing the pinch-effect from the analytically known current and magnetic field can be exactly solved with the help of the reduced GS equation as demonstrated in the present subsection, the condition for the radially directed force in Eq.~(\ref{eq:GradShafranovReduced}) to vanish,
\be\label{eq:localhoop}
\left[\frac{J_\varphi A_\varphi}{2} +\frac{B^2_\varphi}{2\mu_\snull} + 3P + rJ_\varphi B^{(\text{s})} \right]\frac{\mathbf{e}_r}r=0,
\ee
cannot be satisfied \textit{locally} with any physically admissible (divergence-free and curl-free) strapping field including the uniform strapping field considered here. In CME generation the situation is even more complicated, because the \textit{local} values of realistic strapping fields are not known. However, a \textit{global} balance for an integral radial force (which is in effect the integrand of Eq.~(\ref{eq:SafTheorem2}) can be achieved if the strapping field satisfies Eq.~(\ref{eq:SafTheorem2}) as we will discuss in section~\ref{subsec:strapping} below.

\begin{figure}[tbh]
\centering
\includegraphics[width=1\linewidth]{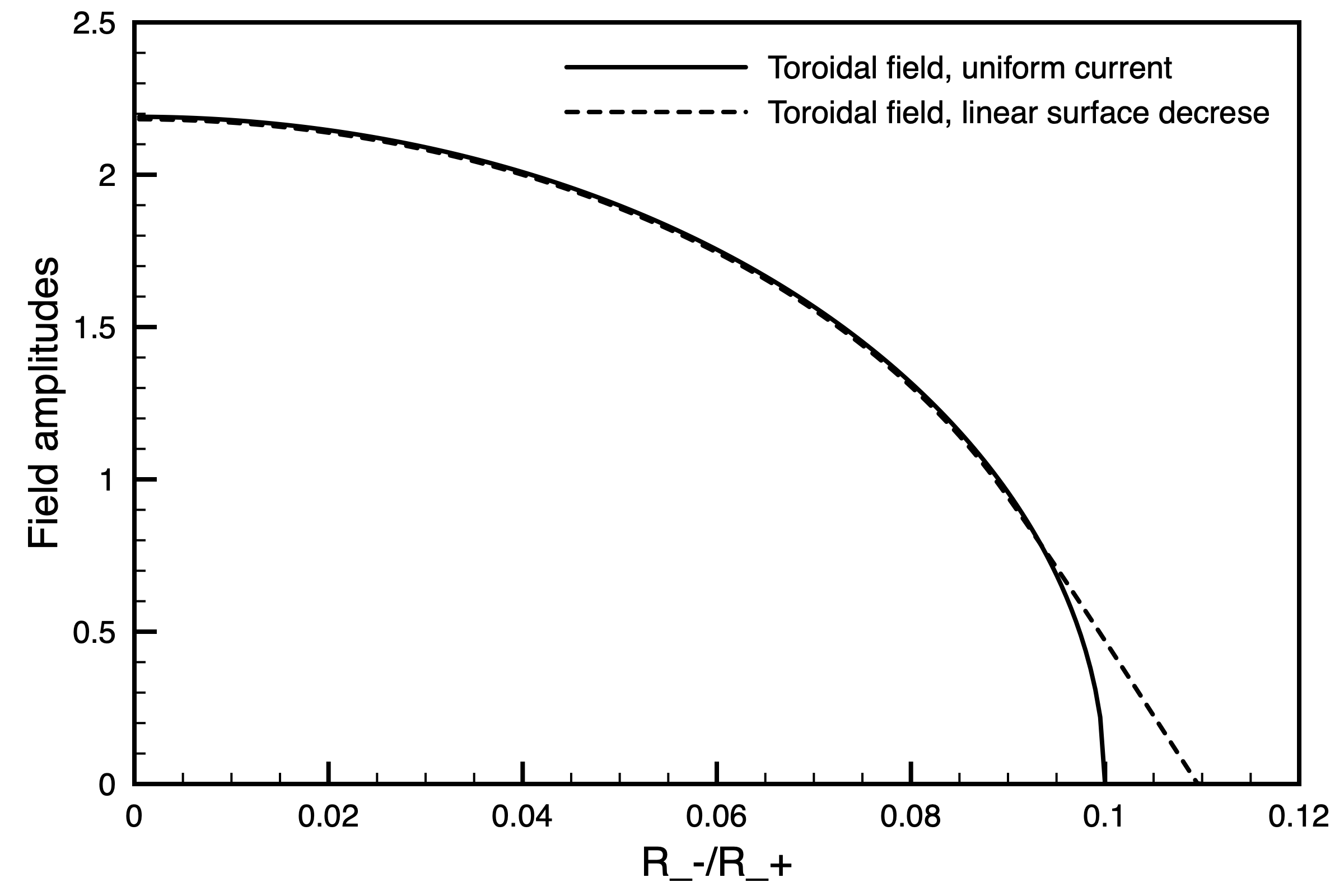}
\caption{Amplitudes of toroidal field component for uniform current (solid line) and for linear surface decrease current (dashed line), for the same filament and in the same coordinate as used in Fig.~\ref{fig:Inner}.}
\label{fig:toroidal}
\end{figure}

Next, we consider the solution of the reduced GS equation for the  current distribution described in Sect.~\ref{sec:n=0} that depends only on $u$,  $j(u,v) = j_\snull (u)$. Eq.~(\ref{eq:Grad}) can be expressed in terms of the normalized quantities, $\tilde{\psi}(u)$, $\tilde{j}(u)$, and the characteristic field, $B_c$, (see Eqs.~(\ref{eq:normalized}) and (\ref{eq:Bc})) and then integrated over $u$:
\be\label{eq:pressuretot}
p^\text{tot}(u) = 8\frac{B_c^2}{2\mu_\snull }\int\limits_{u_\snull }^u\tilde{j} (u_1) \frac{d\tilde{\psi}}{du_1} du_1.
\ee
Here, we note that in the absence of an external toroidal field and pressure the quantity,  $p^\text{tot}(u_\snull)$, vanishes. An \textit{important feature} of our approach is that the plasma parameter $\beta$,
\be\label{eq:defbeta}
\beta=\frac{p(u)}{b^2(u)/(2\mu_\snull)}=\mathrm{const},
\ee
is  assumed to be constant, but finite, so that the toriodal field and gas-kinetic pressure can be expressed in terms of the total pressure:
\be\label{eq:constantbeta}
\frac{b^2}{2\mu_\snull }=\frac{p^\text{tot}(u)}{1+\beta},\quad p(u)=\frac{\beta p^\text{tot}(u)}{1+\beta}.
\ee
Using Eq.~(
\ref{eq:pressuretot}) this can be expressed in terms of the dimensionless toroidal field amplitude, $B^\text{(tor)}$:
\bea\label{eq:torfu}
  p^\text{tot}(u)&=&\frac{\left[B^\text{(tor)}(u)B_c\right]^2}{2\mu_\snull},\nonumber\\
B^\text{(tor)}(u)&=&\sqrt{8\int\limits_{u_\snull }^u\tilde{j} (u_1) \frac{d\tilde{\psi}}{du_1} du_1}.
\eea
For the current and reduced flux functions given by 
Eqs.~(\ref{eq:imanufactured1} and
\ref{eq:Formfactored}) the integral in 
Eq.~(\ref{eq:torfu}) can be carried out by parts using 
Eq.~(\ref{eq:intim}):
\bea\label{eq:torgen}
&\left[B^\text{(tor)}(u)\right]^2=8\tilde{j}(u_1)\left[\tilde{\psi}(u_1)-\tilde{j}_E(u_1)\right]|_{u_\snull}^u+\nonumber\\
&\quad\frac8{E^{(0)}}\frac{\mathrm{d}\tilde{\psi}(u)}{\mathrm{d}\coth u}
\frac{\mathrm{d}}{\mathrm{d}u_1}
\left[\tilde{\psi}(u_1)-\tilde{j}_E(u_1)\right]|_{u_\snull}^u+\nonumber\\&\qquad+\frac{4\tilde{j}^2(u_1)}{E^{(1)}}|_{u_\snull}^u.
\eea
For a uniform current form factor, as in Eq.~(\ref{eq:currentuniform}), using Eqs.~(\ref{eq:q_0}) and (\ref{eq:Formfactored}) one obtains the following:
\bea\label{eq:torfu1}
&&B^\text{(tor)}(u>u_\snull ) = \sqrt{\ell^{(\text{tor})}(u_\snull ) \left[\bar{Q}^{-1}_{-\half}(u_\snull ) - \bar{Q}^{-1}_{-\half}(u)\right]},
\nonumber\\
&& \ell^{(\text{tor})}(u_\snull) =8\left[\frac{\tilde{j}_E(u_\snull)}{\bar{Q}^{-1}_{-\half}(u_\snull )}-\ell^{(\text{s})}(u_\snull )
\right]\tilde{j}(u_\snull),
\eea
where $\ell^{(\text{tor})}$ is another induction coefficient which is discussed below (see Section~\ref{Section:Energy}) to characterize the energy of toroidal magnetic field, expressed in terms of the current density $\tilde{j}(u_\snull)=\frac34\tilde{j}_E(u_\snull)$ and $\tilde{j}_E(u_\snull)$ given by Eq.~(\ref{eq:ffu}).  {The}  toroidal field amplitude {for uniform current} form factor {given by Eq.~(\ref{eq:torfu1})} is shown in  Fig.~\ref{fig:toroidal} with solid line. For the current form factor with linear surface decrease given by Eq.~(\ref{eq:currentsmeared}) the toroidal field near the filament boundary can be calculated by applying the general formula in Eq.~(\ref{eq:torgen}) to the reduced flux function given by Eq.~(\ref{eq:fieldsmeared}). Thus calculated toroidal field
amplitude is shown in Fig.~\ref{fig:toroidal} with dashed line. A comparison of the curves in Fig.~\ref{fig:toroidal} shows how the singularity in the toroidal field near the boundary (infinite spatial derivative of the solid line as $\kappa^\prime\to\kappa^\prime_\snull$) is eliminated by using a linearly decreasing current near the surface (dashed line).    

\begin{figure*}[htb]
\centering
\includegraphics[width=0.95\textwidth]{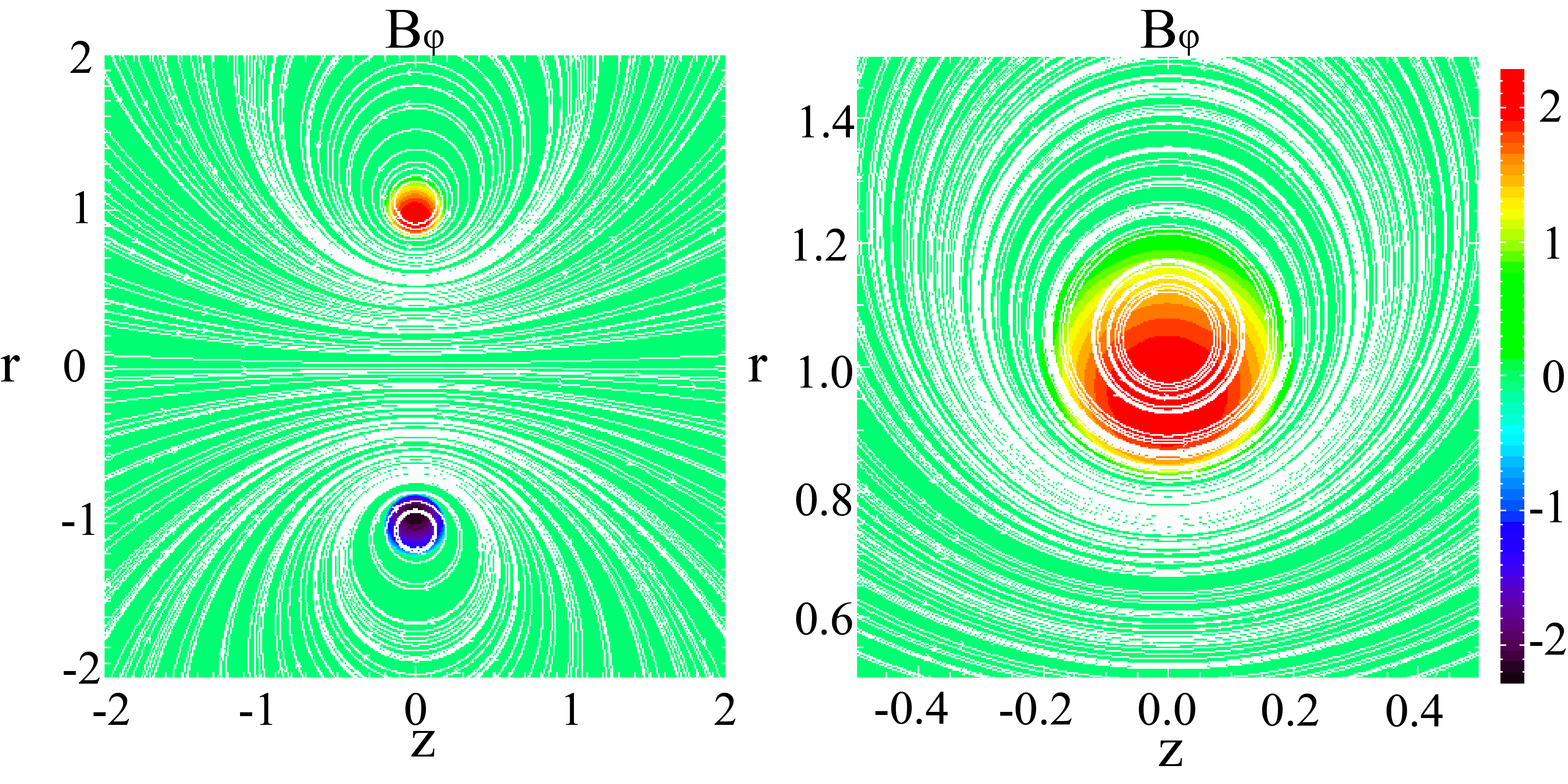}
\caption{\textbf{Left panel:} Solution of the reduced GS equation, in the meridional cross-section ($z,r$  plane), for $R_\infty=1,\,\kappa^\prime_\snull =0.1, B_c=1$ and positive helicity. White lines: field lines of the poloidal field $(B_z,B_r)$. Color: toroidal magnetic field,  $B_\varphi$, perpendicular to the image plane, outgoing field being positive, incoming one being negative. The figure corresponds to positive helicity, otherwise, the blue and red circles would swap.  \textbf{Right panel:} Close-up image of the cross-section of the plasma filament.}
\label{fig:NoStrapField} 
\end{figure*}

Eq.~(\ref{eq:torfu}) allows us to express the total magnetic field (including the toroidal component) that satisfies the reduced GS equation:
\bea
\label{eq:BFull}
&&\mathbf{B}_\snull  = \frac{R_\infty^{\frac32}}{r^{\frac32}} B_c \times\\
&&\times\left[B^\text{(z)}(u) \mathbf{e}_z \pm \frac{B^\text{(tor)}(u)}
{\sqrt{1+\beta}} \mathbf{e}_\varphi -
\left[\frac{B^\text{(p)}(u)}{\kappa^\prime}\right] 
\left(\kappa^\prime\mathbf{e}_v\right)\right],\nonumber
\eea
where $\mathbf{e}_\varphi=\mathbf{e}_z\times\mathbf{R}/r$. Since the toroidal field amplitude, $B^\text{(tor)}(u)$, is positive as is the toroidal current density, $j_\snull (u)$, the choice of plus or minus sign in Eq.~(\ref{eq:BFull}) corresponds to the positive or negative \textit{helicity}, $\mathrm{sign}(B_\varphi/J_\varphi)$. The magnetic field calculated using Eq.~(\ref{eq:BExternal}) for $\kappa\le\kappa_\snull $ and with Eq.~(\ref{eq:BFull}) for $\kappa\ge\kappa_\snull $ and with field amplitudes obtained assuming uniform current form factor is shown in Fig.~\ref{fig:NoStrapField} for $R_\infty=1,\,\kappa^\prime_\snull =0.1$.

This magnetic field produced by an azimuthal current  (white magnetic field lines) and magnetic field inside the toroidal filament (red and blue color) satisfies the reduced GS equation.  However, the right (zoomed) panel demonstrates that the equilibrium is not yet complete. As emphasized in Sect.~\ref{sec:GS}, under equilibrium conditions the (yellow) boundary of the filament, where the total pressure turns to zero (hence it is constant), must coincide with a magnetic surface. Inspection of Fig.~\ref{fig:NoStrapField} shows that magnetic surfaces defined by closed (poloidal) magnetic field lines (white lines) intersect the plasma boundary (yellow circle) at  multiple locations, indicating the absence of true equilibrium.

For constant $\beta$ one can find the corresponding gas-kinetic pressure inside the filament using 
Eq.~(\ref{eq:torfu}):
\be\label{eq:pressure}
P(u\ge  u_\snull ) = \frac{\beta}{1+\beta} \frac{R_\infty^3}{r^3}\frac{\left[ B^\text{(tor)}(u)B_c\right]^2}{2\mu_\snull },
\ee
Assuming constant electron and ion temperatures inside the  filament, $T_e$ and  $T_i$,  one can also derive the distribution of plasma density that will form the ejecta:
\be\label{eq:rho}
\rho(u\ge  u_\snull )=\frac{\beta }{1+\beta}\frac{R_\infty^3}{r^3}\frac{\left[B^\text{(tor)}(u)B_c\right]^2m_i}{2\mu_\snull k_\sB\left(T_i+Z_iT_e\right)},
\ee
where $m_i$ and $Z_i$ are the average mass and charge state of ions, and $k_\sB$ is the Boltzmann constant. An equation for the \alf-wave speed in the $\varphi$ direction,
\be\label{eq:alf}
V^2_{{\text{A}},\varphi}=\frac{B_\varphi^2}{\mu_\snull\rho}=\frac{2 k_\sB\left(T_i+Z_iT_e\right)}{\beta m_i},
\ee
directly follows from Eq.~(\ref{eq:rho}). This useful parameter is constant as long as $\beta,T_i,T_e$ are assumed to be constant.  

The total ejected mass can be expressed in terms of the total pressure integral over the filament volume:
\be\label{eq:Mass}
M=\int{\rho\mathrm{d}V}
=\frac{2\int{\left(P+\frac{B_\varphi^2}{2\mu_\snull }\right)\mathrm{d}V} }{\left(1+\beta\right)V^2_{\text{A},\varphi}},
\ee
which is calculated and discussed in Sect.~\ref{Section:Energy}.
\subsection{Balancing the hoop force by a strapping field}
\label{subsec:strapping}
We demonstrated, that the reduced GS equation 
(Eq.~\ref{eq:Grad}) ensures the cancelation of the dominant pinching force 
at \textit{each point} inside the filament. However, in the full force balance, there is an unbalanced radial force described by Eq.~(\ref{eq:localhoop}) that does not vanish \textit{locally}. The first three terms in Eq.~(\ref{eq:localhoop}) describe the density of the \textit{hoop force} directed radially outwards (compare them with the integrand in Eq.~\ref{eq:SafTheorem2}). This force is fully determined by the parameters of the plasma configuration. The last term in Eq.~(\ref{eq:localhoop}) describes the effect of the strapping field on the toroidal current which may oppose the hoop force if the strapping field is negative (\ie antiparallel to the magnetic moment).

\begin{figure*}[htb]
\centering
\includegraphics[width=0.95\textwidth]{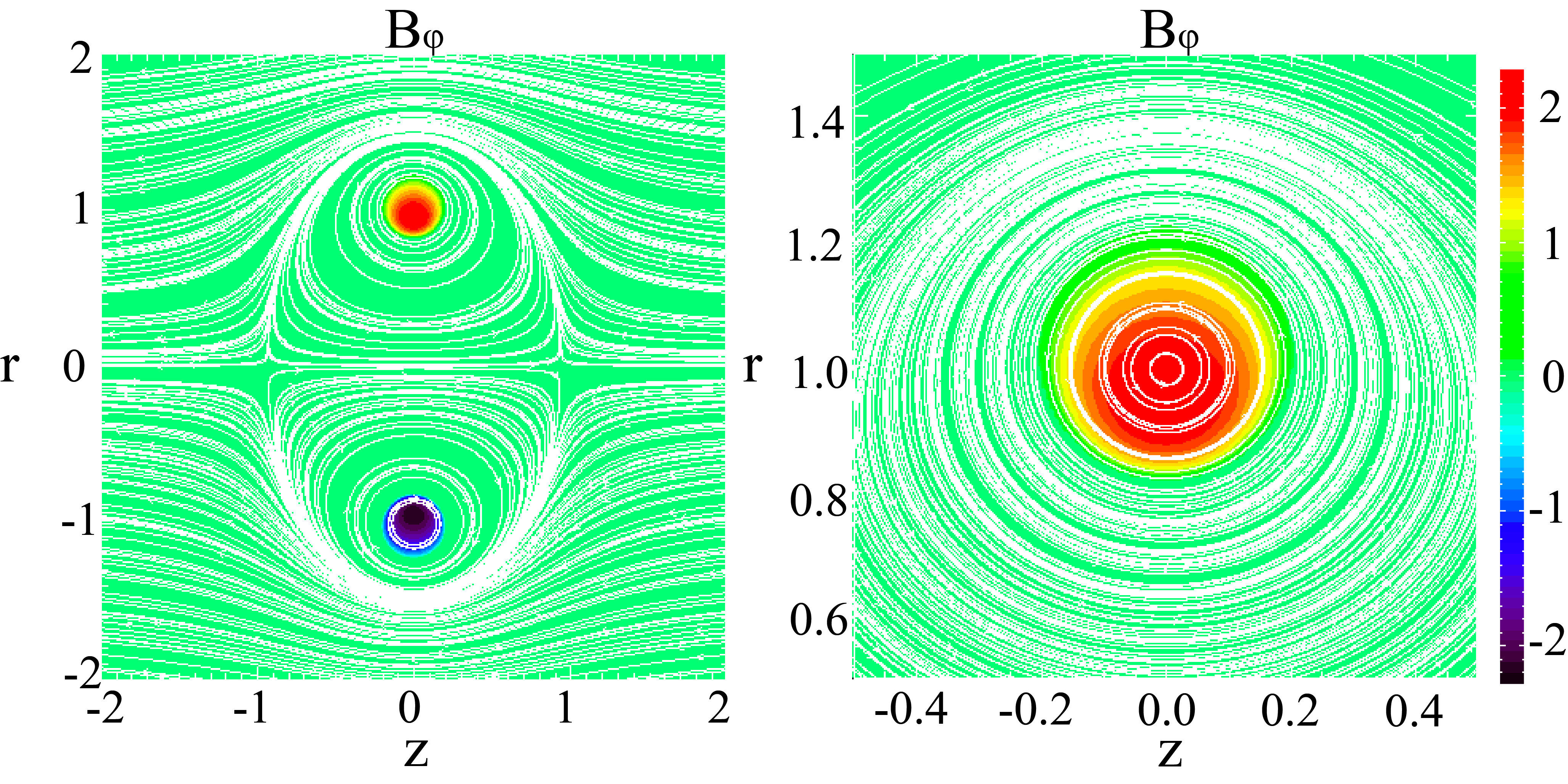}
\caption{
\textbf{Left panel:} Same as in Fig.~\ref{fig:NoStrapField}, with superposed uniform strapping field, $B^\mathrm{(s)}= - L_\snull B_c/(2\pi \mu_\snull R_\infty)$, where the inductance for $n=0$ harmonics is given by Eq.~(\ref{eq:Esnull}) below. 
White lines: field lines of the poloidal field $(B_z,B_r)$. 
Color: toroidal magnetic field,  $B_\varphi$, perpendicular to the image plane, outgoing field being positive, incoming one being negative. The figure corresponds to positive helicity, otherwise, the blue and red circles would swap.  \textbf{Right panel:} Close-up image of the cross-section of the plasma filament. Coincidence of the filament boundary (yellow color) with the magnetic surface (white line) demonstrates that the equilibrium condition is satisfied.}
\label{fig:StrapField} 
\end{figure*}
Since the the hoop force density and the current density are different functions of the coordinates, the local forces cannot be balanced by a uniform strapping field. Alternatively, if we express the strapping magnetic field from Eq.~(\ref{eq:localhoop}), both the divergence and the curl of this field would not vanish.   This situation can be rectified by ensuring that \textit{global equilibrium}, holds, i.e., requiring that the  integrated radial force  (given by Eq.~\ref{eq:localhoop}) vanishes  over the plasma volume. This means that the integrated hoop force is balanced by the overall effect of the adjusted strapping field. 
This can be achieved by taking the scalar product of Eq.~(\ref{eq:localhoop}) and $\mathbf{R}$ and integrating the resulting scalar equation over the entire plasma volume. This way we obtain Eq.~(\ref{eq:SafTheorem2}) as an integral radial force balance equation, unambiguously determining the strapping field:
\be
\label{eq:NetForce}
    \int\left(\frac{J_\varphi A_\varphi}{2} + \frac{B_\varphi^2}{2\mu_\snull } + 3P\right) 
    \text{d}V +2\pi R^2_\infty I^\mathrm{tot}
    B^\text{(s)}=0,
\ee
where in Eq.~(\ref{eq:SafTheorem2}) we substituted Eq.~(\ref{eq:magmoment}) for the magnetic moment, $\mathcal{M}$. As we discussed in Section~\ref{sec:ShafTheorem}, in the absence of a strapping field, $B^\text{(s)}=0$,  the radial hoop force (parameterized by the volume integral of a function that is positive definite everywhere) would disrupt the current filament over the major radius. However, 
the \amp~force from the strapping field, $\left(J_\varphi\mathbf{e}_\varphi\right)\times\left(B^\text{(s)}\mathbf{e}_z\right)=J_\varphi B^\text{(s)}\mathbf{e}_r$, tends to contract the filament in case $B^\text{(s)}<0$ and it may balance the hoop force. The condition for the force balance can be parameterized in terms of the inductance, $L_\snull$, since the volume integral evaluating the hoop force in Eq.~(\ref{eq:NetForce}) is very close to the magnetic free energy 
(exactly coincides with that for $\gamma=4/3$ -- see Section~\ref{sec:ShafTheorem} for more detail):
\bea\label{eq:Bstrap}
    B^{\rm(s)}&=& - \frac{L_\snull  I^\mathrm{tot}}{4\pi R_\infty^2} = -\frac{1}{2\pi} \frac{L_\snull }{\mu_\snull  R_\infty} B_c,
    \\
    \frac{L_\snull \left(I^\mathrm{tot}\right)^2}2 &=& E_\snull=\int \left(\frac{J_\varphi A_\varphi}{2}+\frac{B_\varphi^2}{2\mu_\snull }+3P\right)\text{d}V\nonumber.
\eea
The strapping field is anti-parallel to the $\mathbf{B}_c$ field and its magnitude can be derived from the inductance. Eq.~(\ref{eq:Bstrap}) shows, that in order to derive the strapping field that is needed for obtaining full equilibrium solutions, one has calculate the inductance for the given current density profile. This derivation is discussed next (Sect.~\ref{Section:Energy}). 

In Fig.~\ref{fig:StrapField}  the we consider the same configuration as in Fig.~\ref{fig:NoStrapField}  ($R_\infty=1,\kappa^\prime_\snull=0.1$), but with a superposed uniform strapping field given by Eq.~(\ref{eq:Bstrap}) (the inductance for $n=0$ is determined by Eq.~(\ref{eq:Esnull}) discussed below). The left panel shows a drastically changed topology compared to the no-strapping-field configuration in Fig.~\ref{fig:NoStrapField}. The separator surface separates the external region of strapping field from the region of the field generated by the filament current. In the right panel the coincidence of the filament boundary (yellow color) with a magnetic surface (white line) demonstrates that in the presence of a strapping field this equilibrium condition is satisfied, while in the configuration with no strapping field (see Fig.~\ref{fig:NoStrapField}) this condition is not met.

\begin{figure*}[htb]
\centering
\includegraphics[width=0.496\textwidth]{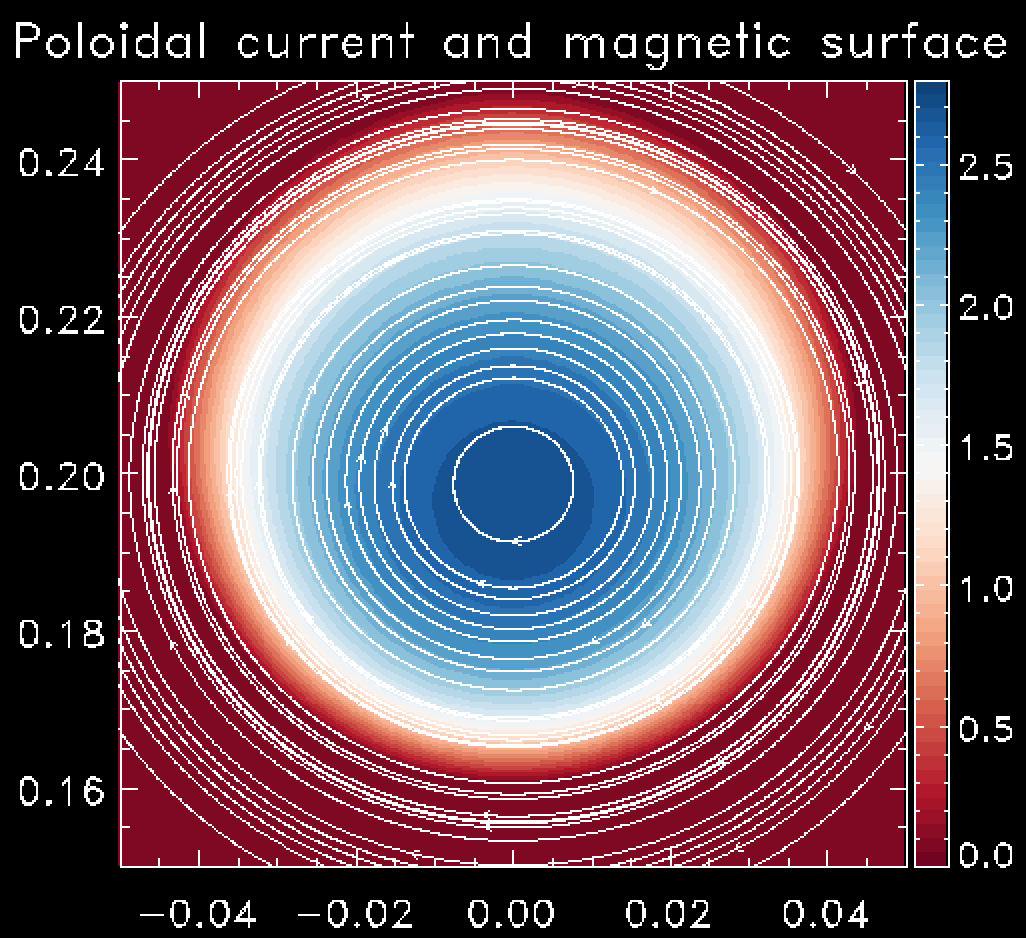}
\includegraphics[width=0.496\textwidth]{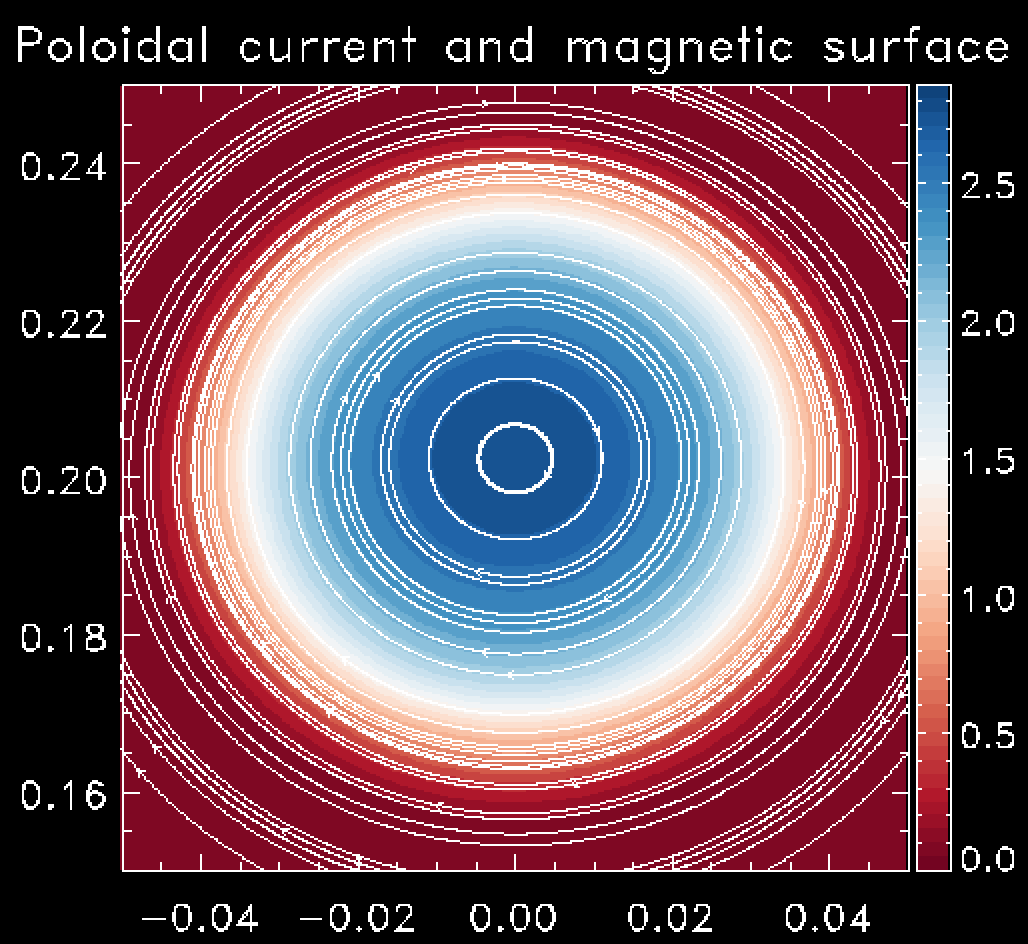}
\caption{
Close-up image of the meridional cross-section 
of the plasma filament. The $z,r$ coordinates are measured in units of $R_\sSun$. White curves: closed field lines of the poloidal field $(B_z,B_r)$ (=meridional cross-sections of the magnetic surfaces). Color: level contours of the poloidal current function, $rB_\varphi$, in G$\times R_\sSun$. \textbf{Left panel:} Initial, close to equilibrium field distribution given by Eq.~(\ref{eq:BFull}). Slight misalignment between the levels of the current function and magnetic surfaces demonstrate imperfectness of estimated equilibrium. \textbf{Right panel:} same quantities are visualized after relaxation to ``true'' equilibrium after 6,000 seconds simulation with the R-MHD equations. Perfect alignment of the constant current surfaces with magnetic surfaces demonstrate that the configuration reached the equilibrium state, which is reasonably close to the estimated one.}
\label{fig:Relaxation} 
\end{figure*}

Alternatively, the equilibrium condition can be verified if the near-equilibrium magnetic field (given by Eq.~\ref{eq:BFull}), gas-kinetic pressure (Eq.~\ref{eq:pressure}) and density (Eq.~\ref{eq:rho}) distributions are used as the initial condition for the \textit{Relaxation} MagnetoHydroDynamics (R-MHD) model. In this model an artificial friction force density, $-\rho\mathbf{U}/\tau$, is added to the momentum equation,  that is oppositely directed than the plasma velocity vector, $\mathbf{U}$. The friction force relaxes the residual plasma motions with a characteristic time of $\tau=\mathrm{const}$, thus damping the possible oscillations around the equilibrium state.

The result of such a simulation is presented in Fig.~\ref{fig:Relaxation}. The initial condition corresponds to a current filament with major and minor radii of $R_\snull=0.202/0.99R_\sSun$, $a=0.04/0.99R_\sSun$, so that $R_\infty=0.1R_\sSun$ and $\kappa^\prime_\snull=0.1$. The horizontal ($z$) and vertical ($r$) coordinates are also measured in units of solar radii, $R_\sSun$. The uniform strapping field is chosen to be $B^{\rm(s)}=2.7$ G, while the current is $I^\mathrm{tot}\approx1.5\times10^{11}$ A, is expressed in terms of the strapping field using the equilibrium condition, Eq.~(\ref{eq:Bstrap}). The other parameters are $T_e=T_i=5\times10^5$ K and $\beta=0.1$. As we recommend for any application, the uniform current form factor is used with linear surface decrease in a narrow region ($\varepsilon=0.1$), however, the inductance characterizing equilibrium strapping field is calculated for purely uniform current ($\varepsilon=0$).  

The meridional cross-section of the initial field distribution is shown in the left panel of Fig.~\ref{fig:Relaxation}. The white circles show the closed magnetic field lines of the poloidal field (=the meridional cross-sections of the magnetic surfaces), with the color scale showing the levels of $rB_\varphi$ $[G\cdot R_\sSun]$ (=constant levels of the poloidal current function). While the exact Grad-Shafranov equation requires the poloidal current function to be constant at magnetic surfaces (see Section~\ref{sec:GS} for more detail), in the initial plasma configuration there is a slightly visible misalignment between the level contours of the current function and the magnetic surfaces.

Using this initial condition we integrate the axi-symmetric R-MHD equations with a relaxation time of $\tau=1\times10^3$ s, on a grid of $1000\times500$ cells covering the coordinate range of $-0.5R_\sSun\le z\le 0.5R_\sSun$, $0\le r\le 0.5R_\sSun$, to evolve the initial distribution for $t=6\times10^3$ s. A background plasma of negligible pressure but finite density is added outside the filament to limit the characteristic speeds of the MHD perturbations and avoid too small time steps. The result of the numerical simulation is presented in the right panel of Fig.~\ref{fig:Relaxation}. Perfect alignment of the current function levels with the magnetic surfaces demonstrates that the plasma filament reached equilibrium. The video-file in the online version of the journal shows that the relaxation to equilibrium proceeds via damping of small-amplitude internal oscillations, with no collapse by the pinch-effect (prevented by the counter-pressure of the toroidal field) and no disruption by the hoop force (prevented by the strapping field).

\subsection{Magnetic Energy and Inductance}
\label{Section:Energy}
\begin{figure}[tbh]
\centering
\includegraphics[width=1\linewidth]{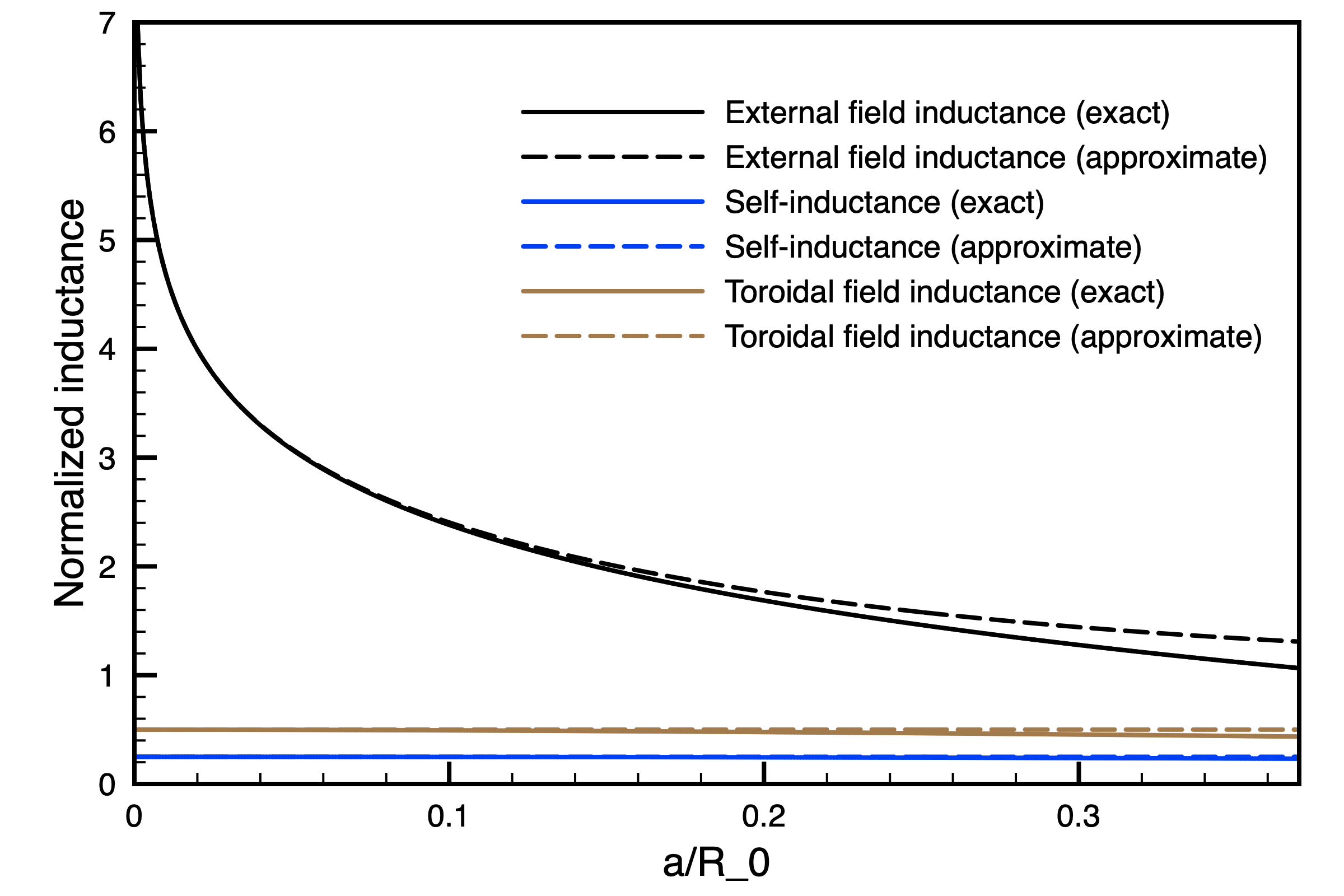}
\caption{External field inductance, $L^{(\text{ext})}_\snull$ (Eq.~(\ref{eq:Lext0}), black line), self-inductance, $L^{(\text{int})}_\snull$ (Eq.~(\ref{eq:Lint1}), blue line), and toroidal field inductance, $L^{(\text{tor})}_\snull$  (Eq.~(\ref{eq:LTor1}), brown line), for the $n=0$ harmonic. The normalized inductance coefficients are  related to $\mu_\snull R_\infty$ and are presented as functions of the ${a}/{R_\snull } = {2\kappa^\prime_\snull} / (1 + \kappa^{\prime^2}_\snull)$ ratio. The functions and their arguments are all calculated in terms of the value of $\kappa^\prime_\snull$ at the toroidal surface. For comparison, approximate solutions given by Eqs.~(\ref{eq:LextApprox},\ref{eq:LintUniform},
\ref{eq:ETor})  are shown by dashed lines.}
\label{Fig:Inductance}
\end{figure}

A contribution to the integral in Eq.~(\ref{eq:Bstrap}) determining the strapping field from the free energy of the poloidal magnetic field, $E^\mathrm{(p)} = \half\int{J_\varphi A_\varphi (2\pi r) H_v} H_u\,dv\,du$, can be derived from Eqs.~(\ref{eq:lame}), (\ref{eq:pcalseries}) and (\ref{eq:jphiseries}):
\bea
\label{eq:Eint}
    E^\mathrm{(p)} &=& \pi{R_\infty\mu_\snull } \int{\frac{j(u,v)\psi(u,v)} {\sinh^2u}\,\mathrm{d}u \mathrm{d}v} = \nonumber\\
    &=& 2\pi^2{R_\infty\mu_\snull } \sum_{n=-\infty}^\infty {\int_{u_\snull }^\infty {\frac{i^*_n(u) \psi_n(u)} {\sinh^2u} \, \mathrm{d}u}}
    =\nonumber\\
   &=&\half\sum_{n=-\infty}^\infty \left(L^{(\text{ext})}_n  +L^{(\text{int})}_n\right) |I_{n_\snull }|^2,
\eea
where the external field inductance,
\be\label{eq:Lext}
\frac{L^{(\text{ext})}_n}{\mu_\snull  R_\infty } =\frac{4\pi^2\tilde{\psi} (u_\snull )} {\bar{Q}^{-1}_{n-\half}(u_\snull )}= 4\pi^2\frac{\left(\oneeigths - \frac{n^2}{2}\right) \bar{P}^{-1}_{n-\half} (u_\snull )} {\bar{Q}^{-1}_{n-\half}(u_\snull )},
\ee
(see Eq.~\ref{eq:fluxfuncext})~quantifies the energy of the magnetic field produced by surface currents concentrated on the filament boundary. In the particular case of $n=0$ harmonic the external inductance, 
\be\label{eq:Lext0}
\frac{L^{(\text{ext})}_\snull}{\mu_\snull  R_\infty }=4\pi^2\ell^{(\text{s})}(u_\snull)=\frac{\pi^2}2\frac{\bar{P}^{-1}_{-\half}(u_\snull )}{\bar{Q}^{-1}_{-\half}(u_\snull )}, 
\ee
(see Eqs.~\ref{eq:q_0} and \ref{eq:Lext}) is shown in Fig.~\ref{Fig:Inductance} (solid black curve). For a thin filament ($\kappa^\prime_\snull\to0$) the toroidal functions in Eq.~(\ref{eq:Lext0}) can be approximated with the help of Eqs.~(\ref{eq:PTransformed}) and (\ref{eq:qminus1}):
\bea\label{eq:LextApprox}
\frac{L^{(\text{ext})}_\snull}{\mu_\snull R_\infty} &\approx& \log\left(\frac4{\kappa_\snull^\prime}\right) -2,\\
\frac{L^{(\text{ext})}_{\pm1}}{\mu_\snull R_\infty} &\approx& \frac{2}{\left(\kappa_\snull^\prime\right)^2},\nonumber
\eea
(see the dashed black curve in Fig.~\ref{Fig:Inductance}). Comparison of the green and magenta curves in Fig.~\ref{Fig:Inductance}) shows that the accuracy of this approximation for a thin filaments is good enough to make it attractive for CME modeling.

Another contribution to the poloidal field energy is characterized by the positive definite self-induction coefficient, which we calculate only for the $n=0$ harmonic:
\be\label{eq:Lint0}
   \frac{ L^{(\rm int)}_\snull}{\mu_\snull  R_\infty }  = 
   4\pi^2\int_{u_\snull }^\infty {\frac{\tilde{j}(u) \tilde{\psi}(u)} {\sinh^2u} \, \mathrm{d}u}-\frac{L^{(\text{ext})}_\snull}{\mu_\snull  R_\infty }.
\ee
For the manufactured current profile given by Eq.~(\ref{eq:imanufactured1}) and the reduced flux function from Eq.~(\ref{eq:Formfactored}), the integration in Eq.~(\ref{eq:Lint0}) can be done using Eq.~(\ref{eq:intim}):
\be\label{eq:Lint}
\frac{L^{(\text{int})}_\snull }{\mu_\snull  R_\infty} = 4\pi^2 \left[\int\limits_{u_\snull }^\infty{\frac{\tilde{j}(u) \tilde{j}_E(u) \mathrm{d}u} {\sinh^2 u}} - \frac{\tilde{j}_E(u_\snull )} {\bar{Q}^{-1}_{-\half}(u_\snull )}\right]. 
\ee
In the case of uniform form-factor (see Eq.~\ref{eq:ffu})
the limiting value of the self-inductance coefficient for thin filaments (\ie $u_\snull\to\infty$ and $\kappa^\prime_\snull\to0$)  can be obtained if we de-normalize the currents the following way:
\bea\label{eq:Lint1}
&&\frac{L^{(\text{int})}_\snull }{\mu_\snull  R_\infty}=\frac{4\pi^2 }{\bar{Q}^{-1}_{-\half}(u_\snull )}\times\\
&&\times\frac{\tilde{j}(u_\snull) \tilde{j}_E(u_\snull)\bar{Q}^{-1}_{-\half}(u_\snull )(\coth u_\snull-1) - \tilde{j}_E(u_\snull )I(u_\snull)}{I^2(u_\snull)},\nonumber
\eea
and then apply L'H\^opital's rule to the second fraction. By differentiating both numerator and denominator over $\mathrm{d}u_\snull$ at constant $\tilde{j}$ and $\tilde{j}_E$, and by using Eq.~(\ref{eq:In}) to derive $\mathrm{d}I(u_\snull)/\mathrm{d}u_\snull$ and Eq.~(\ref{eq:imanufactured2}) to express $\mathrm{d}\bar{Q}^{-1}_{-\half}(u_\snull)/\mathrm{d}u_\snull=-I(u_\snull)/j_E(u_\snull)$, one finds:
\bea\label{eq:LintUniform}
    && \lim_{u_\snull \to \infty} \frac{L^{(\text{int})}_\snull }{\mu_\snull  R_\infty} =\lim_{u_\snull \to \infty}\frac{4\pi^2 }{\bar{Q}^{-1}_{-\half}(u_\snull )}\times\nonumber\\
    &&\qquad\times\frac1
    {2(\coth u_\snull+1)\bar{Q}^{-1}_{-\half}(u_\snull )}=\frac{1}{4},
\eea
since according to Eq.~(\ref{eq:qminus1}) $\bar{Q}^{-1}_{-\half}(u_\snull\to \infty)=2\pi$.

The rest of the integral, $E_\snull$, determining the strapping  field in Eq.~(\ref{eq:Bstrap}) for constant plasma $\beta$, can be expressed in terms of the integral of total pressure:
\bea\label{eq:PTotInt}
&&E_\snull=\half\left(L^{(\rm ext)}_\snull+L^{(\rm int)}_\snull+\frac{1+3\beta}{1+\beta}L^{(\rm tor)}_\snull\right)\left(I^\mathrm{tot}\right)^2,\nonumber\\
&&\int{
 \left[\frac{B_\varphi^2}{2\mu_\snull } + P\right]dV}=\half L^{(\rm tor)}_\snull\left(I^\mathrm{tot}\right)^2.
\eea
The integrand in Eq.~(\ref{eq:PTotInt}), describing the magnetic free energy density due to the toroidal field, can be calculated by multiplying the representative function, $p^\text{tot}(u)$ for the total pressure (see Eq.~\ref{eq:torfu}) by $(R_\infty/r)^3$ while the volume element equals $\mathrm{d}V=H_u H_v 2\pi r\mathrm{d}u\,\mathrm{d}v$. Integrating over $\mathrm{d}u\,\mathrm{d}v$ using Eqs.~(\ref{eq:Bc} and \ref{eq:torfu}) results in the following: 
\be
\frac{L^{(\text{tor})}_\snull}{\mu_\snull R_\infty}= \pi^2 \int\limits_{u_\snull }^\infty {\frac{\left[B^\text{(tor)}(u)\right]^2} {\sinh^2 u} \mathrm{d}u}.
\ee
In the special case of uniform current form-factor this expression can be rewritten using Eq.~(\ref{eq:torfu1}):
\be
    \frac{L^{(\text{tor})}_\snull}{\mu_\snull R_\infty}=
    \pi^2\ell^{(\text{tor})}(u_\snull )\int\limits_{u_\snull }^\infty{\frac{\bar{Q}^{-1}_{-\half}(u_\snull )-\bar{Q}^{-1}_{-\half}(u)}{\sinh^2 u}\mathrm{d}u},
\ee
or, by simplifying Eq.~(\ref{eq:torfu1}) using Eqs.~(\ref{eq:q_0}, \ref{eq:ffu}, \ref{eq:wronskian}):
\bea\label{eq:LTor1}
&&\frac{L^{(\text{tor})}_\snull }{\mu_\snull  R_\infty}=\pi^2 \frac{\mathrm{d}\bar{P}^{-1}_{-\half}(u_\snull )}{\mathrm{d}u_\snull}\times\\
&&\times\frac{\tilde{j}(u_\snull) \tilde{j}_E(u_\snull)\bar{Q}^{-1}_{-\half}(u_\snull )(\coth u_\snull-1) - \tilde{j}_E(u_\snull )I(u_\snull)}{I^2(u_\snull)}.\nonumber
\eea
In the approximation of thin filament, we have
\be\label{eq:ETor}
\lim_{u_\snull \to \infty}\frac{\mathrm{d}\bar{P}^{-1}_{-\half}(u_\snull)}{\mathrm{d}u_\snull}=\frac4\pi,\quad\lim_{u_\snull \to \infty} \frac{L^{(\text{tor})}_\snull }{\mu_\snull  R_\infty} =\frac{1}{2},
\ee
according to Eqs.~(\ref{eq:dP0du}, \ref{eq:PTransformed}) and analogous derivations for Eq.~(\ref{eq:Lint1}) above. Comparison of the exact (solid lines) and approximate (dashed lines) expressions for the self-inductance and toroidal field inductance in Fig.~\ref{Fig:Inductance} shows that the differences between the exact solutions and the approximate ones are hardly visible, therefore it is fully adequate to use the approximate solutions.  With these simplifications, the inductance of the $n=0$ harmonic field, determining the magnitude of hoop force, strapping field, and, for a specific adiabatic index, also a magnetic free energy can be obtained with the help of Eqs.~(\ref{eq:Lext0}, \ref{eq:LintUniform}, \ref{eq:PTotInt}, and \ref{eq:ETor}):
\bea
\label{eq:Esnull}
    E_\snull &=& \half L_\snull\left(I^\text{tot}\right)^2,\nonumber\\
    L_\snull&=&L^{(\text{ext})}_\snull + c_{\threeforth}\mu_\snull R_\infty,
\eea
where
\be\label{eq:c34}
c_{\threeforth}=\frac34+\frac\beta{1+\beta},
\ee
is a frequently used constant, turning to $3/4$ as $\beta\to0$. The expression for the hoop force, Eq.~(\ref{eq:hoopperangle}) with the energy integral given by Eq.~(\ref{eq:Esnull}) can be compared with that found in literature (see Eq.~5 in \citet{titov99} and Eq.~2 in \cite{Toro06}). The (inessential) difference of our approach is in the use of the exact Eq.~(\ref{eq:q_0}) for the external field inductance instead of the approximate Eq.~(\ref{eq:LextApprox}) and in the term allowing for, if desired, the contribution from the gas-kinetic pressure. However, the difference is small, which justifies our model.

\subsection{Improved Equilibrium Theory for a Thin Filament.}
For a thin filament the equilibrium condition can be simplified allowing us to improve the accuracy of the solution and get it much closer to a real equilibrium. To achieve this, we reevaluate the condition under which the sum of Eqs.~(\ref{eq:GradShafranovreduced1} and \ref{eq:GradShafranovreduced2}) vanishes,
\bea\label{eq:thinequil}
&&\sinh u\left[\left(\tilde{j}\frac{\mathrm{d}\tilde{\psi}}{\mathrm{d}u}-\frac{\partial \tilde{p}^\mathrm{tot}}{\partial u}\right)\mathbf{e}_u-
\frac{\partial \tilde{p}^\mathrm{tot}}{\partial v}\mathbf{e}_v\right]+\\
&&\quad+\left(
    \frac{\tilde{j} \tilde{\psi}-\tilde{p}^\text{tot}}{2}+\frac{r^\frac32}{R^\frac32_\infty}\frac{\tilde{j}B^{(\rm s)}}{2B_c}+\frac{3\tilde{b}^2}4+\frac{7\tilde{p}}2\right)\mathbf{e}_r=0,\nonumber
\eea
where: (1) similar to the dimensionless representative functions for current and flux, $\tilde{j},\tilde{\psi}$, we introduced analogous functions for the pressures and magnetic field:
\be\label{eq:tildeptot}
\tilde{p}^\text{tot}=\frac{\tilde{b}^2}2+\tilde{p}, \quad \tilde{p}=\frac{R^2_\infty p}{\mu_\snull(I^\mathrm{tot})^2}, \quad\tilde{b}=\frac{R_\infty b}{\mu_\snull I^\text{tot}};
\ee
(2) we divided the sum of Eqs.~((\ref{eq:GradShafranovreduced1} and \ref{eq:GradShafranovreduced2}) by a common factor, $\mu_\snull\left(I^\mathrm{tot}\right)^2R_\infty/r^4$; (3) we used Eqs.~(\ref{eq:nabla2uv} and \ref{eq:lame}) to express the $\nabla_2$ operator; and (4) we used Eq.~(\ref{eq:Bc}) to relate the total current to $B_c$, $\mu_\snull I^\mathrm{tot}=2B_c R_\infty$. However, we do not use the assumption of $p^\mathrm{tot}=p^\mathrm{tot}(\psi)$ any longer. To the contrary, while $\psi$ is a function of $u$, $p^\mathrm{tot}$ is now assumed to be a function of both $u$ and $v$. 

For a thin current filament, \ie for $\sinh u\gg1$, the term that is proportional to $\sinh u$ is dominant in Eq.~(\ref{eq:thinequil}) and the following simplifications are possible. First, by keeping only the $n=0$ term in the expansion given by Eq.~(\ref{eq:q1series}), the geometric factor multiplying the strapping field becomes $(r/R_\infty)^{\frac32} \approx \bar{Q}^{-1}_{-\half}(u) /{2\pi} \approx \bar{Q}^{-1}_{-\half}(u_\snull)/ {2\pi}$. Second, the strapping field itself is approximated using Eqs.~(\ref{eq:Bstrap}, \ref{eq:q_0} and \ref{eq:Lext0}): 
\bea\label{eq:approxBStrap}
\frac{B^{(\rm s)}}{2B_c}&=&-\frac{L_\snull}{4\pi\mu_\snull R_\infty}\approx-\frac{L^{(\text{ext})}_\snull}{4\pi\mu_\snull R_\infty}=\nonumber\\
&=&-\frac\pi8\frac{\bar{P}^{-1}_{-\half}(u_\snull)}{\bar{Q}^{-1}_{-\half}(u_\snull)}=-\frac{\pi\tilde{\psi}(u_\snull)}{\bar{Q}^{-1}_{-\half}(u_\snull)},
\eea
since for a thin filament the inductance of the external field dominates (see Fig.~\ref{Fig:Inductance}). Third, we use Eq.~(\ref{eq:evapprox}) to approximate the radial unit vector, $\mathbf{e}_r \approx -\sin v\,\mathbf{e}_v-\cos v\,\mathbf{e}_u$.  With these approximations,  one can rewrite Eq.~(\ref{eq:thinequil}) keeping only the leading terms in the factors multiplying  $\mathbf{e}_v,\mathbf{e}_u$:
\bea\label{eq:thinequil2}
&&\left(\tilde{j}\frac{\mathrm{d}\tilde{\psi}}{\mathrm{d}u}-\frac{\partial \tilde{p}^\mathrm{tot}}{\partial u}\right)\mathbf{e}_u-\left(\frac{\partial \tilde{p}^\mathrm{tot}}{\partial v}+\frac{\sin v}{\sinh u}\times\right.\\
&&\,\,\times\left.\left\{
    \frac{\tilde{j} \left[\tilde{\psi}-\tilde{\psi}(u_\snull)\right]-\tilde{p}^\text{tot}}{2}+\frac{3\tilde{b}^2}4+\frac{7\tilde{p}}2\right\}\right)\mathbf{e}_v=0.\nonumber
\eea
In the zeroth order approximation for small $\frac{\sin v}{\sinh u}$, one gets $\partial \tilde{p}^\mathrm{tot}/\partial v=0$ and the condition for the coefficient of $\mathbf{e}_u$ to vanish results in the reduced GS equation (Eq.~\ref{eq:Grad}), giving:
\bea\label{eq:0thorder}
\left(\frac{\tilde{b}^2}2\right)^{(0)}&=&\frac1{1+\beta}\int\limits_{u_\snull}^u{\tilde{j}(u_1)
\frac{\mathrm{d}\tilde{\psi}}{\mathrm{d}u_1}\mathrm{d}u_1},\nonumber\\
\tilde{p}^{(0)}&=&\frac\beta{1+\beta}\int\limits_{u_\snull}^u{\tilde{j}(u_1)
\frac{\mathrm{d}\tilde{\psi}}{\mathrm{d}u_1}\mathrm{d}u_1},
\eea
where the superscript ``(0)'' denotes the zeroth order approximation. To get the first order approximation, Eqs.~(\ref{eq:0thorder}) are used to evaluate the expression in braces in Eq.~(\ref{eq:thinequil2}). Particularly, for uniform current form, $\left(\tilde{p}^{(\mathrm{tot})}\right)^{(0)}=\tilde{j} \left[\tilde{\psi}-\tilde{\psi}(u_\snull)\right]$, so that the first term inside the braces vanishes. In the first order approximation, corrections that are $\propto\frac{\cos v}{\sinh u}$ should be added to the magnetic and gas-kinetic pressures, to get the factor multiplying $\mathbf{e}_v$ vanish:
\bea\label{eq:1storder}
\left(\frac{\tilde{b}^2}2\right)^{(1)}&=&\frac{1+\frac32\frac{\cos v}{\sinh u}}{1+\beta}\int\limits_{u_\snull}^u{\tilde{j}(u_1)
\frac{\mathrm{d}\tilde{\psi}}{\mathrm{d}u_1}\mathrm{d}u_1},\\
\tilde{p}^{(1)}&=&\frac{\beta\left(1+\frac72\frac{\cos v}{\sinh u}\right)}{1+\beta}\int\limits_{u_\snull}^u{\tilde{j}(u_1)
\frac{\mathrm{d}\tilde{\psi}}{\mathrm{d}u_1}\mathrm{d}u_1},\nonumber
\eea
where, according to Eqs.~(\ref{eq:lame} and \ref{eq:toroidalcoords}),
\be\label{eq:cosoversinh}
\frac{\cos v}{\sinh u}=\frac{R^2-R^2_\infty}{2rR_\infty}.
\ee
The corrections given by Eqs.~(\ref{eq:1storder} and \ref{eq:cosoversinh}) have two remarkable properties. First, the corrections do not modify the integral $E_\snull$ in Eq.~(\ref{eq:Bstrap}), hence, the estimate for the strapping field. Indeed, $E_\snull$ reduces to integrals of $b^2(u,v)$ and $p(u,v)$ over $\mathrm{d}u\mathrm{d}v$, therefore, the contributions to the integrand, which are proportional to $\cos v$, vanish once integrated over $\mathrm{d}v$. 

Now, we use Eqs.~(\ref{eq:1storder} and \ref{eq:cosoversinh}) as well as Eqs.~(\ref{eq:torfu}, \ref{eq:BFull} and \ref{eq:pressure}) to derive the first order approximation for the dimensional quantities:
\bea\label{eq:1storderphys}
B^{(1)}_\varphi&=&\sqrt{\frac{1+\frac32\frac{R^2-R^2_\infty}{2rR_\infty}}{1+\beta}}B_c\left(\frac{R_\infty}r\right)^\frac32B^\text{(tor)}(u),\\
P^{(1)}&=&\frac{\beta\left(1+\frac72\frac{R^2-R^2_\infty}{2rR_\infty}\right)}{1+\beta}\left(\frac{R_\infty}r\right)^3\frac{\left[B_cB^\text{(tor)}(u)\right]^2}{2\mu_\snull},\nonumber
\eea
We note that 
$$
B^\text{(tor)}\propto\left[\int\limits_{u_\snull}^u{\tilde{j}(u_1)
\frac{\mathrm{d}\tilde{\psi}}{\mathrm{d}u_1}\mathrm{d}u_1}\right]^\half\propto\left[\tilde{\psi}(u)-\tilde{\psi}(u_\snull)\right]^\half$$
for a uniform current form factor. Within the adopted accuracy, we can approximate $(R^2-R^2_\infty)/2rR_\infty \approx (r-R_\infty)/R_\infty$, so that $[1+\threeforth(R^2-R^2_\infty)/rR_\infty]^\half \approx \left(r/R_\infty\right)^\threeforth$ and $1+\frac74(R^2-R^2_\infty)/rR_\infty \approx\left(r/R_\infty\right)^\frac72$. Finally, we arrive at the following scaling for the current function (see Section~\ref{sec:GS} for more detail): $rB^{(1)}_\varphi \approx \mathrm{const}\times r^\oneforth [\tilde{\psi}-\tilde{\psi}(u_\snull)]^\half$ as well as for pressure: $P^{(1)} \approx \mathrm{const}\times\,r^\half [\tilde{\psi}-\tilde{\psi}(u_\snull)]$. The second remarkable property of the first order approximation is that these quantities only depend on the function, $r^\half[\tilde{\psi}-\tilde{\psi}(u_\snull)]$, which can be expressed in terms of the total flux function, $\Psi^\mathrm{tot}=\Psi+B^{(\rm s)}r^2/2$, including the contribution from the uniform strapping field, as we demonstrate next.

To express the function, $r^\half [\tilde{\psi}(u)-\tilde{\psi}(u_\snull)]$, that vanishes as $u\to u_\snull$, we re-define the total flux function by adding a constant, equal to $\frac32B^{(\rm s)}R^2_\infty$, so that as $u\to u_\snull$, the total flux function vanishes, $\Psi^\mathrm{tot}\to0$. Using Eqs.~(\ref{eq:Bc}, \ref{eq:approxBStrap} and \ref{eq:sumseries}) the redefined flux function can be transformed as follows:
\bea
&&\Psi^\mathrm{tot}=\Psi+B^{(\rm s)}\left(\half r^2+\frac32R^2_\infty\right)=\mu_\snull\sqrt{r R_\infty} I^\text{tot}\times\nonumber\\
&&\quad\times\left\{\tilde{\psi}+\frac{B^{(\rm s)}}{2B_c}\left[\half \left(\frac{r}{R_\infty}\right)^\frac32+\frac32\left(\frac{R_\infty}{r}\right)^\half \right]\right\}\approx\nonumber\\
&&\quad\approx\mu_\snull\sqrt{r R_\infty} I^\text{tot}\left[\tilde{\psi}-\frac{\pi\tilde{\psi}(u_\snull)}{\bar{Q}^{-1}_{-\half}(u_\snull)}\frac{\bar{Q}^{-1}_{-\half}(u)}\pi\right]\approx\nonumber\\
&&\quad\approx\left(\mu_\snull R^\half_\infty I^\text{tot}\right)r^\half
\left[\tilde{\psi}(u)-\tilde{\psi}(u_\snull)\right].
\eea
Thus, the pressure and current functions both depend on the function, $r^\half
[\tilde{\psi}(u)-\tilde{\psi}(u_\snull)]$, which differs only by a constant factor from the flux function, $\Psi^\mathrm{tot}$. In order to eliminate the extra contributions to the force in Eq.~(\ref{eq:thinequil}), which are aligned with $\mathbf{e}_u$ and are proportional to $\cos v$, one needs to replace the $\tilde{j}^{(0)}=\tilde{j}_\snull(u)\approx\mathrm{const}$ approximation of the current density with:
\bea\label{eq:1stordercurrent}
&&j^{(1)}=I^{\mathrm{tot}}\tilde{j}_\snull(u)\left(1+2C_\frac34\frac{\cos v}{\sinh u}\right)\approx\\
&&\approx I^{\mathrm{tot}}\tilde{j}_\snull(u)\left[1+\left(\frac32+\frac{2\beta}{\beta+1}\right)\left(\frac{r}{R_\infty}-1\right)\right].\nonumber
\eea
This modification satisfies the exact Grad-Shafranov equation (Eq.~\ref{eq:GradShafranov}), requiring that 
\be
J_\varphi=
\frac1r
\frac{\mathrm{d}}
{\mathrm{d}\Psi}
\frac{\left(rB_\varphi\right)^2}{2\mu_\snull}+r\frac{\mathrm{d}P}
{\mathrm{d}\Psi},
\ee 
hence, 
\be
j^{(1)}=I^{\mathrm{tot}}\frac{\tilde{j}_\snull(u)}{\beta+1}\left[\left(\frac{r}{R_\infty}\right)^\frac32+\beta\left(\frac{r}{R_\infty}\right)^\frac72\right]
\ee
This becomes Eq.~(\ref{eq:1stordercurrent}) for $|r-R_\infty|\ll R_\infty$. The extra current harmonics, $j_{\pm1}\cos v=C_\frac34I^{(\mathrm{tot})}\frac{\tilde{j}_\snull(u)}{\sinh u}\cos v$, in Eq.~(\ref{eq:1stordercurrent}) result in: 
(1) the generation of the first harmonics of the reduced flux function, $\psi_{\pm1}(u)\cos v\propto C_\frac34 $; (2) the modification of the external field given by Eq.~(\ref{eq:extfieldseries}); (3) an extra requirement on the magnitude of the strapping field, which, in addition to the zeroth-order approximation of $\propto L^{(\mathrm{ext})}t_\snull$ given by Eq.~(\ref{eq:approxBStrap}), also gives  a contribution of $\propto C_\frac34$, in accordance with Eqs.~(\ref{eq:Bstrap} and \ref{eq:Esnull}). However, to satisfy the exact equilibrium condition, the strapping field must satisfy a more  
restrictive condition for not only its ``average'' magnitude, but also for the particular distribution over the current filament cross-section \cite[see details in][including the shapes of strapping field for different filament parameters]{Zakharov:1986}, to separately balance the force on three current harmonics (for $n=0,\pm1$).

Based on these considerations we arrive to an important conclusion. Although solving the \textit{reduced} GS equation is sufficient to find a configuration sufficiently close to equilibrium, this approach may look misaligned within the general framework of the full GS, since the current function and pressure are not directly expressed via the flux function. however, this contradiction is resolved with the improved approximation described here, since within the accuracy of the approximation the functional dependencies become $rB^{(1)}_\varphi\approx\mathrm{const}\times\sqrt{\Psi^\mathrm{tot}}$ and $P^{(1)}\approx\mathrm{const}\times\Psi^\mathrm{tot}$, in compliance with the full GS equation. Despite formally being more accurate and consistent, the improved equilibrium solution is more laborious and difficult to compute, and, which is even more problematic, poses more severe restriction on the shape of the strapping field. In a realistic magnetic field which hardly satisfies these requirements, the ``improved'' solution may appear to be even farther from equilibrium than the simple and easy-to-compute single-harmonic solution for $n=0$. Solving the R-MHD equations with the simple $n=0$ harmonic solution as initial condition seems to be a more practical, and therefore preferred approach. This way both the magnetic configuration and its external field automatically adjust to the realistic strapping field.

\subsection{CME-Generator Based on Finite-Beta Zeroth Harmonic 
Solution}
\begin{figure*}[htb]
\centering
\includegraphics[width=1\textwidth]{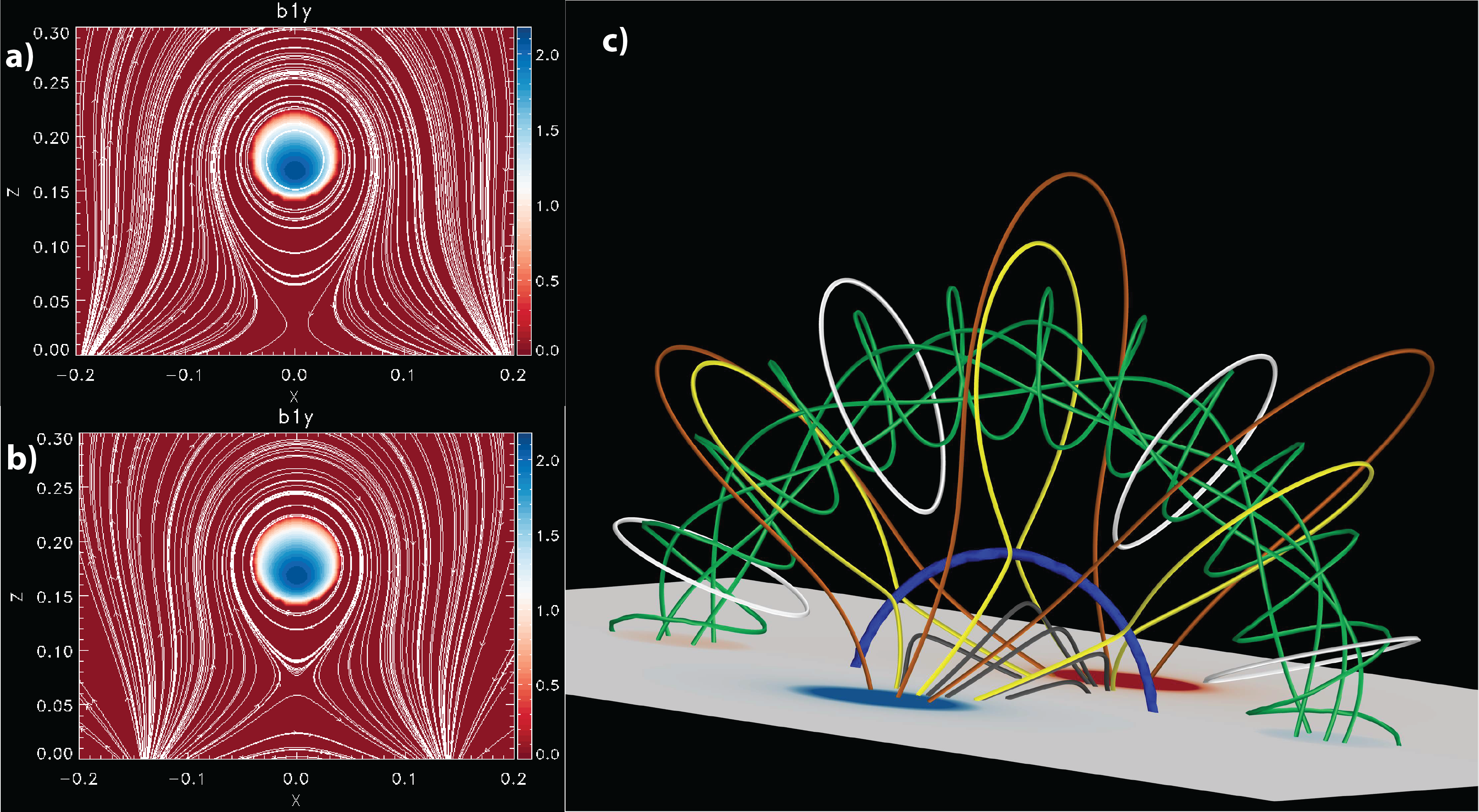}
\caption{Magnetic field lines from the current filament characterized by  $R_\infty=0.2R_\sSun$, $\kappa^\prime_\snull=0.1$, $B_c=1$, strapped by the field from a pair of positive and negative magnetic charges. The configuration center is at the depth of $d=0.025R_\sSun$ below the solar surface, the distance, $2D$, between the charges is $D=R_\infty$ for panel a) and $D=0.7R_\infty$ for panels b) and c). Null points can be seen in panels a)  and b), which show meridional cross-sections of the configuration similarly to Figs.~\ref{fig:NoStrapField} and \ref{fig:StrapField}. In the 3-D topology, shown in panel c), the null-line is marked with a thick blue line.}
\label{fig:StrapCharge} 
\end{figure*}
In actual numerical simulations of CMEs an important distinction from idealized configurations is that only a part of the toroidal filament rises above the solar surface, with the center of configuration located at a depth, $d$, below the surface. From simple geometric considerations one can determine the angular size of this circular arc above the solar surface:
\be\label{eq:angularsize}
\Delta\alpha=2\arccos\left({\frac{2R_\sSun d-d^2-R_\infty^2}{2(R_\sSun - d)R_\infty}}\right).
\ee
where $R_\sSun$ is the solar radius. For small values of $d$ the configuration gets close to an idealized situation when the highly conducting solar surface cuts the circular ring of the filament to two half circles. In this case the ``hidden'' part of the filament (that is under the solar surface) can be considered as an ``image'' current below the surface. For such idealized situations one gets $\Delta \alpha\approx\pi$. As long as in such model the CME is driven by the hoop force, the work done by this force in the course of expansion (according to Eq.~\ref{eq:kinenergy}) can be expressed via the change in the total magnetic free energy, which for the described circular arc can be obtained from Eq.~(\ref{eq:Esnull}):
\be\label{eq:ECME}
E_\mathrm{CME}=\frac{\Delta\alpha}{4\pi}\left(L^{(\text{ext})}_\snull+c_{\frac34}\mu_\snull R_\infty\right)
\left(I^\mathrm{tot}\right)^2.
\ee
The mass of the ejecta is obtained from Eqs.~(\ref{eq:Mass}, \ref{eq:PTotInt} and \ref{eq:ETor}):
\be\label{eq:MCME}
M_\mathrm{CME}=\frac{\Delta\alpha}{4\pi}\mu_\snull R_\infty
\frac{\left(I^\mathrm{tot}\right)^2}{V^2_{A\varphi}},
\ee
where we assumed a low $\beta$  plasma, so that  $1+\beta\approx1$, and the \alf\  speed inside the filament, $V_{\text{A},\varphi}$, has been defined in Eq.~(\ref{eq:alf}). 
Assuming that in the CME the available free energy is fully converted to the kinetic energy of ejecta, $E_\mathrm{CME}= \frac12M_\mathrm{CME}V^2_\mathrm{CME}$, \ie by neglecting the interaction of the strapping field with the starting to expand flux rope, we can estimate the asymptotic CME speed, that is independent of the current, $I^\mathrm{tot}$, and the angular extent of the erupting arc, $\Delta\alpha$:
\be\label{eq:VCME}
V_\mathrm{CME}=\sqrt{2\left(\frac{L^{(\text{ext})}_\snull}{\mu_\snull R_\infty}+c_{\frac34}\right)}\,V_{\text{A},\varphi}\sim2V_{\text{A},\varphi}.
\ee
It can be seen that the normalized external inductance, $L^{(\text{ext})}_\snull/(\mu_\snull R_\infty)$, controls the physically important speed ratio, $V_\mathrm{CME}/V_{A_\varphi}$. According to Eq.~(\ref{eq:Eint}) this inductance is a function of $u_\snull$, but it also can be parameterized with $\kappa^\prime_\snull$ or with the $a/R_\snull$ ratio (see Fig.~\ref{Fig:Inductance}). For a thin filament this coefficient is about $1.2-2.5$, indicating that the CME speed can exceed the \alf\ speed in the initial filament configuration by a factor of two (see Eq.~\ref{eq:VCME}). Another potentially important contribution to the energy budget is due to gravity. 
With an account of negative potential energy, the energy conservation law, $E_\mathrm{CME} - \frac{GM_\sSun}{R_\sSun}M_\mathrm{CME}= \frac12M_\mathrm{CME}V^2_\mathrm{CME}$, gives:
\be
V^2_\mathrm{CME}
=
2\left(\frac{L^{(\text{ext})}_\snull}{\mu_\snull R_\infty}+c_{\frac34}\right)V^2_{\text{A},\varphi}-V^2_{G},
\ee
where $G$ is a gravitation constant, $M_\sSun$ is a solar mass, and
\be
V_{G}=\sqrt{\frac{2GM_\sSun}{R_\sSun}}\approx615\,[\mathrm{km/s}]
\ee
is an escape velocity.

The fact that our model can produce super-Alfv{\'e}nic CMEs raises several interrelated questions, such as what is the mechanism of energy conversion from magnetic free energy to kinetic energy of the ejecta? How fast is the energy conversion?

Under these circumstances, an essential element of the CME initiation scenario is magnetic reconnection. In addition to fast removal of the field tying the current filament to the active region and subsequent acceleration of the CME to super-\alf{ic} speeds, the reconnection can also explain the X-ray flare accompanying the CME \cite[see, e.g.][]{forbes00}, as well as the accelerated particle release \citep{Masson:2013}.

\begin{figure*}[htb]
\centering
\includegraphics[width=0.505\textwidth]{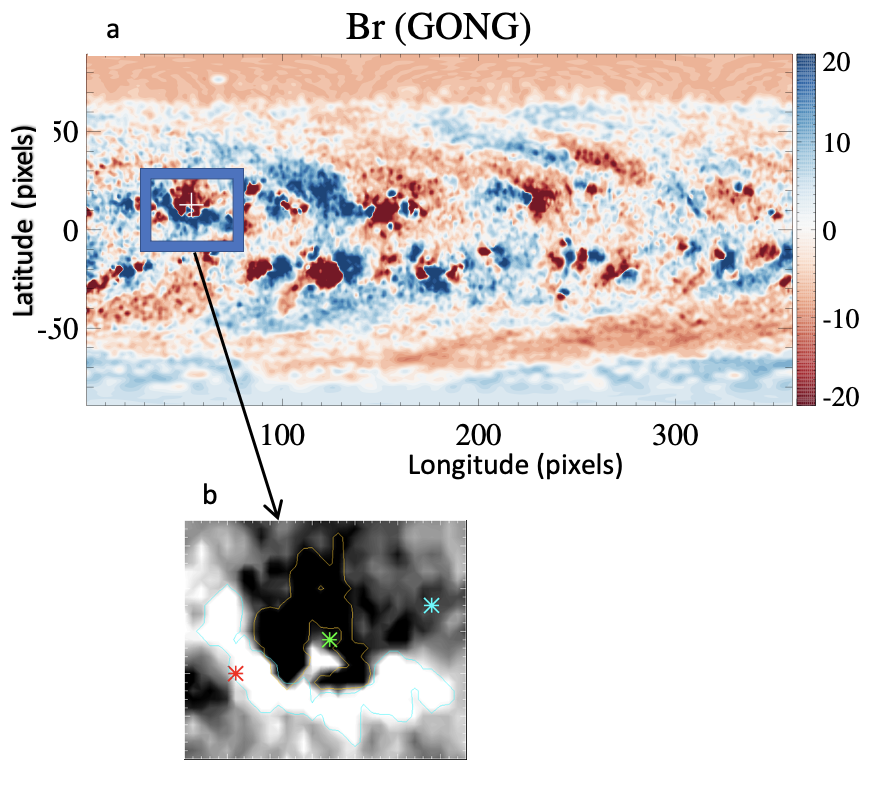}
\includegraphics[width=0.485\textwidth]{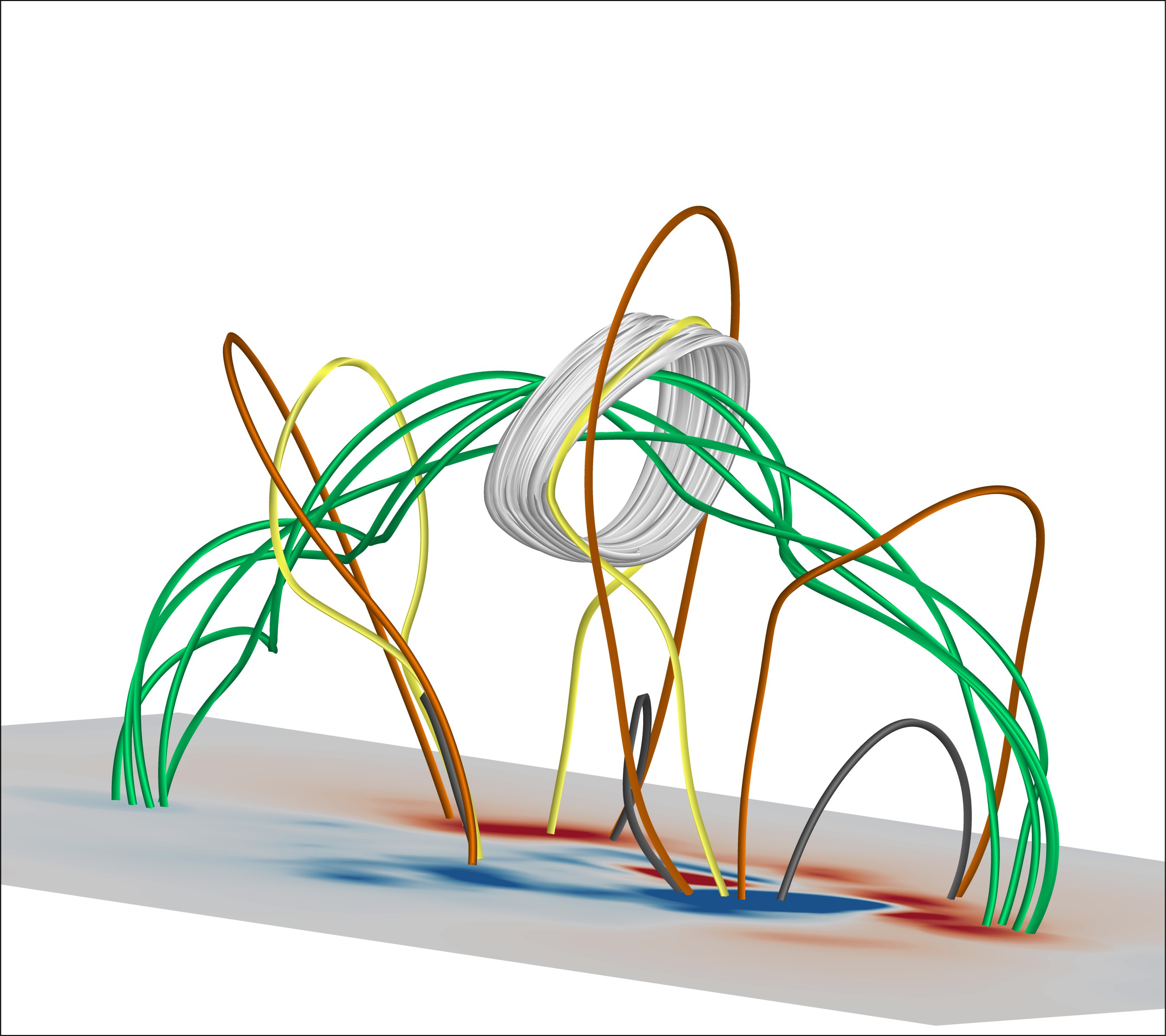}
\caption{\textbf{Top left:} GONG magnetogram 
as of April, 11 2013 with intensified weak field. \textbf{Bottom left:} Zoomed in AR field with the chosen locations for the current filament footpoints (red and blue asterisks) and the center of configuration (green asterisk).   \textbf{Right panel:} The magnetic configuration superposed with realistic magnetic field of the active region adjusted to simulate the CME event of April, 11 2013.}
\label{Fig:7}
\end{figure*}

This new scenario is demonstrated in Fig.~\ref{fig:StrapCharge}. 
Previously (see section \ref{sec:ReducedGS}) we considered a uniform horizontal strapping field, that at the center of the current loop was oppositely oriented to the magnetic field of the current filament, $B_c\mathbf{n}_c$. However, the magnitude of this strapping field was smaller than $B_c$, and therefore, the superposed field, $\left(B_c+B^\mathrm{(s)}\right)\mathbf{n}_c$, did not change direction (see Fig.~\ref{fig:StrapField} and compare it to the case of no strapping field, depicted in Fig.~\ref{fig:NoStrapField}). In contrast with the uniform field the new scenario involves an altitude dependent overarching strapping field. This strapping field originates from the active region and it is anchored to the solar surface. The strapping field balances the hoop force at the apex and it can be sufficiently strong near the solar surface to flip the direction of the superposed field. This flip results in the formation of \textit{null-points} that are the seeds of future reconnection.

A still idealized, but more realistic, case of a strapping field created by a pair of positive and negative magnetic charges at the axis of symmetry of the configuration,   which mimic positive and negative magnetic spots of a bipolar active region \cite[see][]{titov99} is illustrated in Fig.~\ref{fig:StrapCharge}. On the left (panels a and b) we demonstrate how the field topology depends on the distance, $2D$, between the charges. Here we used current filament parameters, $R_\infty=0.2R_\sSun$, $\kappa^\prime_\snull=0.1$, $B_c=1$, while the depth of the configuration center was $d=0.025R_\sSun$. The magnitude of the charges was chosen in a way that the field at the current filament location is sufficient for strapping. For $D\gg R_\infty$ (not shown) the strapping field is almost uniform, the only distinction from Fig.~\ref{fig:StrapField} is that at large distances the field lines connect to the solar surface. In the intermediate case when $D = R_\infty$ (see Fig.~\ref{fig:StrapCharge}a) the field of the current filament near the solar surface is balanced by the strapping field, therefore the null point forms near the origin. When the strapping field is even more non-uniform, $D=0.7R_\infty$ (see Fig.~\ref{fig:StrapCharge}b) the null point raises and gets closer to the filament.

Panel Fig.~\ref{fig:StrapCharge}c  shows the 3-D topology of the field for the $D=0.7R_\infty$ case. There are five families of topologically different magnetic field lines: 
\begin{outline}[enumerate]
\1 
Twisted magnetic field lines inside the filament (green lines),
\1
Circular field lines looping around the filament generated by its current (white circles), 
\1 
Arcade-type strapping magnetic field lines originating from the magnetic charges (brown lines). Their tension balances the hoop force and maintains the equilibrium,
\1 
Below the null line (marked by blue) there are black lines connecting the positive and negative magnetic charges. These field lines are completely disconnected from the filament and its own field,
\1 
Stretched magnetic field lines with null-points (yellow lines). These are separators: the upper loop separates the brown strapping field lines from the field lines looping around the filament. The bottom loop separates the strapping (brown) field lines from the black field lines closed below the null line. 
\end{outline}

Using numerical simulations for a nearly identical configuration, \cite{roussev03a} demonstrated that reconnection at the null line naturally results in loss of equilibrium (note the similarity between our Fig.~\ref{fig:StrapCharge}c and Fig.~1 in \cite{roussev03a}). This loss of equilibrium is due to the fact that the strapping field (brown lines) partially reconnects and its strapping effect decreases. Note, that the strapping field is introduced to ensure equilibrium, while the height dependence of the realistic strapping field in the active region naturally results in the appearance of null-points and null-lines, which make the configuration prone to magnetic reconnection, thus potentially breaking the equilibrium.

It is important that when applying this methodology to realistic CME simulations, it is not enough to choose a location and appropriate model parameters to obtain an equilibrium configuration of the underlying active region \textit{together} with the superposed filament model (as described by \cite{titov14}). In addition, one must find a configuration that is ripe for spontaneous eruption due to magnetic reconnection. Note, that reconnection by itself does not have to be spontaneous \cite[although it can be - see][]{roussev03a}. Another possible mechanism to enforce reconnection is horizontal motion of photospheric plasma together with the frozen-in footpoints of strapping field lines converging toward the polarity inversion line, \textit{flux cancellation} \cite[see, e.g.][]{Linker:2003}. Such motion builds up the current along the null loci below the flux rope ending up with reconnection and further eruption.

In Fig.~\ref{Fig:7} we present such a configuration created to simulate the CME event of April 11, 2013. The GONG magnetogram as of April,11 2013 is shown in top left panel. Because of the limitations of the observed 
geometry there is significant uncertainty of the radial magnetic field measurements in the polar regions. In order to reduce this uncertainty and achieve better agreement of global simulation results with observations it is customary to modify the photospheric radial magnetic field in the polar regions. Specifically, the observed radial field, $B^{(obs)}_R$, used as the boundary condition at $R=R_\sSun$, is intensified in weak field regions: 
\bea
&&B_R|_{R=R_\sSun}=\mathrm{sign}\left(B^{(obs)}_R\right)\times\nonumber\\
&\times&\min\left(3.5\left|B^{(obs)}_R\right|,\left|B^{(obs)}_R\right|+5\,\text{Gs}\right).
\eea
To get a 3-D distribution of the strapping field, the Potential Field Source Surface Model (PFSSM) is applied by expressing the intensified field as a series of spherical harmonics to the order of 180.

Once the 3-D PFSSM field of the active region and the approximate location of the CME source are obtained, we iterate the locations of the two filament footpoints near the polarity inversion line and analyze the PFSSM field along the filament passing through these footpoints and the topology of the 
total (superposed) field. The iterated locations are shown with red and blue asterisks in the left bottom panel of  Fig.~\ref{Fig:7} displaying a zoomed fragment of the magnetogram. The best choice for the center of configuration in heliographic coordinates are $(80^\circ,13^\circ)$ as shown with the green asterisk in the left bottom panel of  Fig.~\ref{Fig:7}) and the depth is $d=0.03R_\sSun$. The major and minor radii of the current filament are $0.21R_\sSun$ and $0.04R_\sSun$, with the horizontal axis of symmetry rotated $290^\circ$ counter-clockwise from the local direction of heliographic parallel. The helicity sign is negative. 

Under these conditions, the  strapping field along the filament is approximately uniform and perpendicular to the plane of filament. The magnitude of the strapping field, $B^{(\text{s})}\approx-2.7$ Gauss, determines the current according Eq.~(\ref{eq:Bstrap}), thus balancing the hoop force in equilibrium.  On the other hand, the topology of superposed field of the current filament on top of the active region (presented in the right panel of Fig.~\ref{Fig:7}) shows  null points below the filament, which make the configuration prone to reconnection, and thus eruption. As we described in this paper, one must chose the model parameters in a way that the resulting CME matches the total mass and kinetic energy of the observed eruption. With these choices our proposed eruption generator will automatically match a significant number of observational constraints.  

\section{Discussion and Summary}

In this paper we described the relations between the current, the poloidal field it produces and the toroidal field preventing the pinch-effect by accurate analytical expressions that allow for finite thermal pressure. However, we only provide an integral approximation for  the strapping field. This is still very useful, because in coronal mass ejection simulations the strapping field is quite uncertain: it is  non-uniform and even if we were able to describe an exact equilibrium of ideally shaped ring with the prescribed current would not describe a realistic scenario. On the other hand, the accurately described filament in which the pinch-effect is prevented is capable to self-adjust its height and curvature radius to create an equilibrium configuration.  

It is important that the direction of the strapping field is opposite to that of $B_c$, and its magnitude for thin filament (of large inductance and large stored magnetic free energy) can exceed the field at the axis.

In summary, this paper presents a mathematically rigorous extension of the \cite{titov99, titov14} CME generator based on the \cite{grad58} -- \cite{shafranov66} equation. The main new features of the proposed model are:
\begin{outline}
\1
The filament is filled with plasma thus the model describes a finite $\beta$ initial configuration with finite mass and energy,
\1
The model describes an equilibrium solution that will spontaneously erupt due to magnetic reconnection of the strapping magnetic field arcade,
\1
There are analytic expressions connecting the model parameters to the asymptotic velocity and total mass of the resulting CME, providing a way to connect the simulated CME properties to multipoint coronograph observations.
\end{outline}
\section{Acknowledgments}
We are grateful to Drs. V. S. Titov, J. Linker, J. Karpen, and S. Antiochos for useful discussions and to Dr. Lulu Zhao for her kind help in visualization. This work was supported by a NASA LWS Strategic Capability (SCEPTER) project at the University of Michigan under NASA grant 80NSSC22K0892, and by NSF ANSWERS grant GEO-2149771.
\appendix
\section{Superconducting Ring with Current}\label{Section:SCR}
\subsection{Magnetic Field Produced by Superconducting Ring with Current}
The flux function of the magnetic field, produced by a superconducting ring with current, has a constant value, $\Psi(u=u_\snull ) = \frac{\Phi_\snull }{2\pi}$, at the surface, $\Phi_\snull $ being the magnetic flux through the ring. The flux function can be continued into the ring interior, which gives: $\psi(u \ge u_\snull ) \equiv \frac{\Phi_\snull }{2\pi\mu_\snull \sqrt{R_\infty r}}$, or (see Eq.~\ref{eq:qminus1series})

\be\label{eq:scrpsi}
    \psi(u\ge u_\snull )=\sum_{n=-\infty}^\infty{ \psi_ n(u) e^{\text{i}nv}},\qquad\psi_ n(u) = \frac{\Phi_\snull \bar{Q}^{-1}_{n-\half} (u)}{4\pi^2\mu_\snull R_\infty} .
\ee
The continuity with the external field as in Eq.~(\ref{eq:fluxfuncext}) gives equations for the  current amplitudes: 
\be\label{eq:scrcurrents}
    I_{n_\snull } = \frac{\psi_n(u_\snull)}{\left(\frac18-\frac{n^2}2\right)\bar{P}^{-1}_{n-\half} (u_\snull)}=\frac{\Phi_\snull }{L^{(\text{ext})}_n}, 
\ee
the definition of the external inductances has been given in Eq.~(\ref{eq:Lext}).  The total current is:
\be\label{eq:scrcurrents1}
    I^\mathrm{tot} = \sum\limits_{n=-\infty}^\infty I_{n_\snull } = \Phi_\snull  \sum\limits_{n=-\infty}^\infty \frac{1}{L^{(\text{ext})}_n}.
\ee
From here, one can express the inductance of a superconducting circular loop \cite[see][]{Fock32,Malm65,Bhad68,Bele83}:
\be\label{eq:scrcurrents2}
    L^{(\text{s.c.r})} = \frac{\Phi_\snull }{I^\mathrm{tot}} = \left(\sum_{n=-\infty}^\infty\frac{1}{L^{(\text{ext})}_n}\right)^{-1}.
\ee
This expression looks similar to Eq.~(3-62) for the electric capacitance of a ring conductor given by \cite{Ioss81}. Interestingly, \cite{Buck65} pointed out that this solution for the external field has been known since the 19th century describing the flow of a perfect fluid circulating around a solid toroid.

Note an interesting minimum principle that is the consequence of Eq.~(\ref{eq:scrcurrents}). Using the Cauchy inequality in a space of the vectors, $\left\{I_{n_\snull }\right\}$ with a scalar product based on the matrix, $\text{diag}(
L^{(\text{ext})}_n)$, we find that, for a given total current, $I^\mathrm{ext}$, its distribution over harmonics (as in Eq.~\ref{eq:scrcurrents1}) \textit{minimizes} the magnetic energy, $E^{(\text{ext})}$. Indeed, within this framework, Eq.~(\ref{eq:In0}) for the total current can be written as an equation for projecting of the vector of current  amplitudes, 
$\left\{I_{n_\snull}\right\}$, on the direction of $\left\{1/{L^\mathrm{(ext)}_n}\right\}$:
\be
\left\{\frac{1}{L^\mathrm{(ext)}_n}
\right\}\cdot\left\{I_{n_\snull}\right\}=I^\mathrm{tot},\quad\text{where}\,\{a_n\}\cdot\{b_n\}=\sum\limits_{n=-\infty}^\infty {a_nL^{(\text{ext})}_nb_n}.
\ee

With this constrained projection, the $L_2$ norm of the current amplitude vector, $\left\|\left\{I_{n_\snull}\right\}\right\|^2=\{I_{n_\snull}\}\cdot\{I_{n_\snull}\}=\sum{L^\mathrm{(ext)}_nI^2_{n_\snull}}=2E^\mathrm{(ext)}$ (see Eq.~\ref{eq:Eint}) satisfies the Cauchy inequality: 
\be
\left\|\{I_{n_\snull}\}\right\|^2\ge\frac{\left(I^\mathrm{tot}\right)^2}{\left\|\{\frac1{L^\mathrm{(ext)}_n}\}\right\|^2} \equiv L^{(\text{s.c.r})}\left(I^\mathrm{tot}\right)^2.
\ee
If the vector of current amplitudes (not harmonics), 
$\left\{I_{n_\snull}\right\}$, is parallel to  $\left\{1/{L^\mathrm{(ext)}_n}\right\}$, the Cauchy inequality becomes an equality and the magnetic energy is minimized : 
\be
I_{n_\snull} = \frac{L^{(\text{s.c.r})}I^\mathrm{tot}}{L^\mathrm{(ext)}_n} \qquad E^\mathrm{(ext)}=\half L^{(\text{s.c.r})}\left(I^\mathrm{tot}\right)^2.
\ee

\begin{figure}[tbh]
\centering
\includegraphics[width=0.7\linewidth]{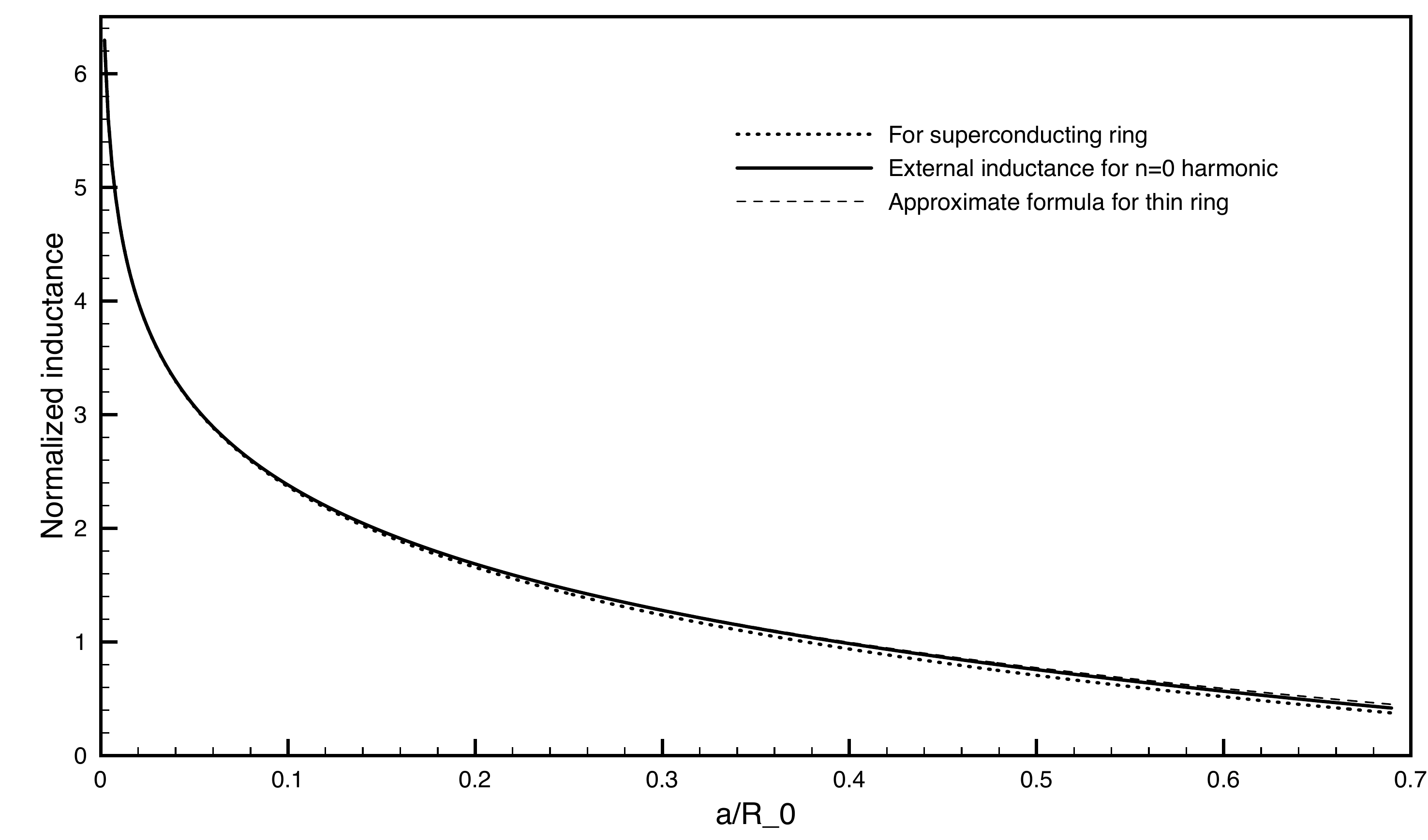}
\caption{Inductance of superconducting field (dotted line)  in comparison with the external field inductance, $L^{(\text{ext})}_\snull$, of $n=0$ harmonic (Eq.~(\ref{eq:Lext0}), solid line), and the approximate solution for thin ring given by Eq.~(\ref{eq:LextApprox}  (dashed line). All inductances are normalized per $\mu_\snull R_\infty$.}
\label{Fig:InductanceMod}
\end{figure}

For $a\ll R_\snull $ (thin toroid) the $n=0$ inductance is $L_\snull ^{(\text{ext})} \approx \mu_\snull R_\infty \left[\log\left(8R_\infty/a\right) - 2\right]$, while the higher order inverse inductances are small: $1/L_n^{(\text{ext})} \propto (a/R_\snull )^{2n}$, and therefore negligible (see Fig.~\ref{Fig:InductanceMod}). However, the distribution of magnetic field over harmonics is not so simple even in this limiting case. Specifically, the total magnetic field inside the superconductor, of course vanishes, while direct derivation of the $n=0$ field (using Eqs.~\ref{eq:Bfield}, \ref{eq:scrpsi}, and \ref{eq:QTransformed}) gives a non-zero limiting value at $u\to\infty$: 
\be\label{eq:B0z}
\lim\limits_{u\to\infty}\mathbf{B}_\snull (u) = \frac{\Phi_\snull }{4\pi R^2_\infty} \mathbf{e}_z = \frac{I_{0_\snull}  L^{(\text{ext})}_\snull }{4\pi R^2_\infty} \mathbf{e}_z
\ee
This field is cancelled by  those from the $n=\pm1$ harmonics: $\mathbf{B}_{\pm1}(u\to\infty) = - \frac{\Phi_\snull } {8\pi R^2_\infty} \mathbf{e}_z$. Contributions from harmonics $n=0,\pm1$ to the magnetic field outside the ring are of the same order and they are non-zero. We conclude that even for $a/R_\snull \ll1$ the field \textit{is not} dominated by the $n=0$ harmonic.       
\subsection{Superconducting Ring in an External Magnetic Field}
Next, we consider a superconducting ring with no current placed into a uniform magnetic field, $B^{(\text{s})}\mathbf{e}_z$. In this scenario a current is induced in the ring in a way that the flux functions of this induced current,  $\Psi(u\ge u_\snull ) \equiv\mu_\snull \sqrt{R_\infty r} \, \psi(u\ge u_\snull )$ (see Eq.~\ref{eq:tildepsi}), and that of the external field, $B^{(\text{s})}r^2/2$, cancel each other to keep total flux through the ring zero (same as prior to bringing the ring into the field). According to Eq.~(\ref{eq:q1series}),  the total flux inside the super-conductor vanishes if:
$$
\psi(u\ge u_\snull ) = - \frac{B^{(\text{s})}R_\infty}{2\mu_\snull } \left(\frac{r}{R_\infty}\right)^{\frac32} = \frac{B^{(\text{s})}R_\infty}{\pi\mu_\snull } \sum_{n=-\infty}^\infty \left(n^2 - \frac14\right) \bar{Q}^{-1}_{n-\half} (u) e^{\text{i}nv}
$$
Just as before, the continuity with the external field (as in Eq.~\ref{eq:fluxfuncext}) gives equations for the currents: 
\begin{equation}\label{eq:scrinfield}
I_{n_\snull } = -\frac{2B^{(\text{s})} R_\infty \bar{Q}^{-1}_{n-\half}(u_\snull )} {\pi\mu_\snull  \bar{P}^{-1}_{n-\half}(u_\snull )} = \frac{\left(4n^2-1\right)\pi R^2_\infty B^{(\text{str})}} {L^{(\text{ext})}_n}.
\end{equation}
Again, the axial field harmonics, $n=0,\pm1$, are all of the same order as the external field.   
\subsection{Cancellation of the Axial Field by the Strapping Field}
\label{sec:scr}
We found that the field from a thin superconducting ring is different from that of the $n=0$ harmonic. One of the reasons is that the field of the $n=0$ harmonic has a non-vanishing axial component, (see Eq.~\ref{eq:B0z}). However, in the particular choice of the \textit{external} field cancelling the magnetic field of the $n=0$ harmonics inside superconducting ring (see Eq.~\ref{eq:B0z})
\be\label{eq:bzcancel}
    B^{(\text{s})} = -\frac{I_{{\snull_\snull} }L^{(\text{ext})}_\snull } {4\pi R^2_\infty},
\ee
we can assume that the solution in a form of single $n=0$ harmonic is applicable. Indeed, in this case the boundary condition,
$$
    \mu_\snull \sqrt{R_\infty r} \psi(u=u_\snull ) + B^{(\text{s})} r^2/2 \approx \Phi_\snull /(2\pi)
$$
is second order accurate, since its derivative over $r$ vanishes: 
$$
    \frac{\mu_\snull}2\sqrt{\frac{R_\infty}{r}} \psi(u=u_\snull ) + B^{(\text{s})}r \approx \frac{I_{0_\snull}  L^{(\text{ext})}_\snull } {4\pi R_\infty} +B^{(\text{s})} R_\infty = 0.
$$
In other words, the constant value of $\psi(u=u_\snull )$ with our specific external magnetic field makes the flux function to be (almost) constant. 

For a more formal derivation of Eq.~(\ref{eq:bzcancel}), one can take a superposition of solutions, Eqs.~(\ref{eq:scrcurrents}) and (\ref{eq:scrinfield}): 
\be\label{eq:B11}
I_{n_\snull}=\frac{\Phi_\snull +\left(4n^2-1\right)\pi R^2_\infty B^{(\text{s}) }}
{L^{(\text{ext})}_n},
\ee
and require that the quantities, $\Phi_\snull $ and $B^{(\text{s})}$, are related in a way that $\Phi_\snull  \approx  -3\pi R^2_\infty B^{(\text{s})}$. This choice cancels the currents, $I_{\pm1_\snull}$, so that their reduced fields  are no longer comparable with the now dominant field $\mathbf{B}_\snull $ from harmonic $n=0$. In this case Eq.~(\ref{eq:B11}) reduces to  Eq.~(\ref{eq:bzcancel})  for $n=0$.

When the current ring is thin the expressions for the external (or strapping) field via the ring (or filament) current in Eqs.~(\ref{eq:Bstrap} and \ref{eq:bzcancel}) are close to each other, so that  $L^\mathrm{(ext)}_\snull\approx L_\snull$. This means that the requirement that the axial field of the $n=0$ harmonic is cancelled by the external field approximately coincides with the condition for balancing the hoop force with the Amp\'ere force, $\mathbf{j}\times \mathbf{B}^\mathrm{(s)}$. This observation allows an alternative view: the hoop force may be interpreted as an Amp\'ere force, $\mathbf{j}\times\mathbf{B}_z$, so that the role of the strapping field, $\mathbf{B}^\mathrm{(s)}=-\mathbf{B}_z$, is to cancel the axial field, $\mathbf{B}_z$, of the $n=0$ harmonic.

\section{Toroidal Functions}
\label{app:toroidal}
\subsection{Definition and Expressions via Hypergeometric Functions}
The toroidal functions \cite[see definition in][Ch. 3.13]{Bate53}  used in the present paper are, 
\be
\bar{Q}^{-m}_{n-\half}(u) = \sqrt{2\sinh u} \, Q^{-m}_{n-\half}(\cosh u),\quad \bar{P}^{-m}_{n-\half}(u) = \sqrt{2\sinh u} \,P^{-m}_{n-\half}(\cosh u), \quad m=0,1.
\ee
They differ from the usually introduced associated Legendre functions of semi-integer index, $Q^{-m}_{n-\half}(\cosh u)$, $P^{-m}_{n-\half}(\cosh u)$,  by a factor of $\sqrt{2\sinh u}$. The associated Legendre function of the first kind is expressed in terms of the hypergeometric series, $F(a,b;c;z)={_2F_1}(a,b;c;z)$ (see  Eq.~8.852(2) in \cite{GR14} and also Eq.~(5) in Ch.3.13 of \cite{Bate53}):
\be\label{eq:Pminus1}
    \bar{P}^{-1}_{n-\half}( u) = \frac{\kappa^3}4 \left(\kappa^\prime\right)^n F\left(\frac32,n+\frac32;3;\kappa^2\right).
\ee
According to Eq.~(8.752(3)) in \cite{GR14}: $P^{-1}_\nu(\cosh u) = \frac1{\sinh u} \int_1^{\cosh u}P_\nu(z)dz$, so that:
\be\label{eq:Pdiff}
    \frac{d}{du} \left[\bar{P}^{-1}_{n-\half}( u)\right] = \bar{P}_{n-\half}(u) - \frac{2-\kappa^2}{2\kappa^2} \bar{P}^{-1}_{n-\half}(u)
\ee
where
\be\label{eq:P}
    \bar{P}_{n-\half}( u) = \kappa\left(\kappa^\prime\right)^n F\left(\half,n+\half;1;\kappa^2\right).
\ee
In the particular case of $n=0$ the  difference of the two functions in Eq.~(\ref{eq:Pdiff}) can be expressed through a single toridal function \cite[see][Eq.~(8) in Ch.3.8]{Bate53}:
\be\label{eq:dP0du}
    \frac{d}{du} \left[\bar{P}^{-1}_{-\half}(u)\right] = \frac{3\kappa^\prime}{\kappa^2} \bar{P}^{-1}_{\half}(u).
\ee
For $\kappa\approx1$ the original hypergeometric series in Eq.~(\ref{eq:Pminus1}) converges slowly and it is worthwhile to transform the series to one based on the variable, $1-\kappa^2$ \cite[see Eqs.~(15.1.2, 15.8.10, and 15.8.12) in][]{DLMF}. Specifically, at $\kappa^\prime\to0$ one gets
\be\label{eq:PTransformed}
    \bar{P}^{-1}_{-\half}(u) \approx \frac4\pi \left(\log\frac4{\kappa^\prime}-2\right) ,\qquad
    \bar{P}^{-1}_{\half}(u) \approx \frac4{3\pi\kappa^\prime}.
\ee

The toroidal function of the second kind is given by Eqs.~(8.736(4) and 8.852(1)) in \cite{GR14}:
\be\label{eq:qminus1}
    \bar{Q}^{-1}_{n-\half}( u) = -\frac{\Gamma(n-\half) \sqrt{\pi}}{\Gamma(n+1)} \kappa^3 \left(\kappa^\prime\right)^n F\left(\frac32,n+\frac32;n+1;(\kappa^\prime)^2\right)
\ee
According to Eq.~8.752(5) in \cite{GR14}, $Q^{-1}_\nu(\cosh u) = -\frac1{\sinh u} \int_{\cosh u}^\infty Q_\nu(z)dz$, and 
\be\label{eq:Qdiff}
\frac{d}{du}\left[\bar{Q}^{-1}_{n-\half}( u)\right]=\bar{Q}_{n-\half}(u) - \frac{2-\kappa^2}{2\kappa^2}\bar{Q}^{-1}_{n-\half}( u)
\ee
where \cite[see Eq.~8.852 in][]{GR14}:
\be\label{eq:q}
    \bar{Q}_{n-\half}( u) = \frac{\Gamma(n+\half) \sqrt{\pi}}{\Gamma(n+1)} \kappa \left(\kappa^\prime\right)^n F\left(\half,n+\half;n+1;(\kappa^\prime)^2\right)
\ee
In the particular case of $n=0$ Eq.~(\ref{eq:Qdiff}) reduces to a small difference of two separate hypergeometric functions which both are near unity. More practical way to calculate this is to express in Eq.~(\ref{eq:Qdiff}) via a single function using Eq.~8.734(2) in \cite{GR14}:
\be\label{eq:dQ0du}
    \frac{d}{du} \left[\bar{Q}^{-1}_{-\half}( u)\right] = \frac{3\kappa^\prime}{\kappa^2} \bar{Q}^{-1}_{\half}(u) = -3\pi \kappa(\kappa^\prime)^2 F\left(\frac32,\frac52;2;(\kappa^\prime)^2\right)
\ee
For $\kappa^\prime\to0$ one gets
\be\label{eq:QTransformed}
    \bar{Q}^{-1}_{-\half}(u) \approx 2\pi,\quad
    \bar{Q}^{-1}_{\half}(u) \approx -\pi\kappa^\prime,\quad\frac{d}{du} \left[\bar{Q}^{-1}_{-\half}( u)\right] \approx -3\pi (\kappa^\prime)^2, \quad\frac{d}{du} \left[\bar{Q}^{-1}_{\half}( u)\right] \approx\pi \kappa^\prime\approx-\bar{Q}^{-1}_{\half}( u).
\ee
\subsection{Wronskian of Toroidal Functions}
The Wronskian of the Legendre functions may be found in \cite[Eq.~(13) in Ch.3.2]{Bate53}:  
\be\label{eq:wronskian}
    \bar{Q}^{-1}_{n-\half}(u) \frac{d \bar{P}^{-1}_{n-\half}(u)}{du} -\bar{P}^{-1}_{n-\half}(u) \frac{d \bar{Q}^{-1}_{n-\half}(u)}{du} = \frac{2}{(\frac14-n^2)},
\ee
\subsection{Series of Toroidal Functions}
The Fourier series for semi-integer powers of $r/R_\infty = \sinh u/(\cosh u-\cos v)$, can be obtained from the following equation \cite[\cf][]{Shus97}:
\be\label{eq:qseries}
    \frac1{\sqrt{2(\cosh u-cosv)}} = \frac1\pi \sum_{n=-\infty}^\infty{Q_{n-\half} \, (\cosh u) e^{\text{i}nv}}.
\ee
Once Eq.~(\ref{eq:qseries}) is multiplied by $(\cosh u-\cos v)\sqrt{2/\sinh u}$, its LHS equals $\sqrt{R_\snull /r}$. In the RHS one can express $\cos v=\frac12\left(e^{\text{i}nv}+e^{-\text{i}nv}\right)$ and partial sum of the multipliers by $e^{\text{i}nv}$ 
using Eqs.~(8.734(3-4)) in \cite{GR14} reduces to $\frac1{2\pi}\bar{Q}^{-1}_{n-\half}(u)$. Therefore:
\be\label{eq:qminus1series}
    \sqrt{\frac{R_\infty}{r}} =\frac{1}{2\pi} \sum_{n=-\infty}^\infty {\bar{Q}^{-1}_{n-\half}(u) e^{\text{i} n v}}.
\ee
Another series can be obtained by differentiating Eq.~(\ref{eq:qseries}) over $u$ and using Eqs.~(8.736(4) and 8.752(4)) in \cite{GR14}, which show that $dQ_{n-\half}(\cosh u)/du =\left(n^2-\frac14\right) Q^{-1}_{n-\half}(\cosh u)$:
\be\label{eq:q1series}
    \left(\frac{r}{R_\infty}\right)^{\frac32} =\frac{2}\pi \sum_{n=-\infty}^\infty {\left(\frac14  -n^2\right) \bar{Q}^{-1}_{n-\half}(u) e^{\text{i} n v}}.
\ee  
A particular linear combination of Eqs.~(\ref{eq:qminus1series} and \ref{eq:q1series}) has the following remarkable property:
\be\label{eq:sumseries}
\frac32\sqrt{\frac{R_\infty}{r}}+\frac12\left(\frac{r}{R_\infty}\right)^{\frac32}=\frac{1}\pi \sum_{n=-\infty}^\infty {\left(1  -n^2\right) \bar{Q}^{-1}_{n-\half}(u) e^{\text{i} n v}}=\frac{1}\pi\bar{Q}^{-1}_{-\half}(u)+O[(\kappa^\prime)^2],
\ee
since the terms for $n=\pm1$ vanish.
\subsection{Some Integrals of the Modified Toroidal Functions}
\label{sec:integral}
The current form factor functions, $j_n^{(m)}(u)$, 
utilized in this paper to approximate the profile of the toroidal current, are eigenfunctions of the equation, 
\be
    \left[\sinh^2u\left(-\frac{d^2}{du^2}+n^2\right) +\frac34 \right]j_n^{(m)} = E^{(m)}j_n^{(m)}.
\ee
While $\bar{Q}^{-1}_{n-\half}(\cosh u)$ is the eigenfunction for $E=0$, we note that 
$\bar{Q}_{n-\half}^{m+\half}(\cosh u)$ is the eigenfunction for the eigenvalue of
$E^{(m)}=1-(m+\half)^2.$ These eigenfunctions can be transformed to Legendre \textit{polynomials} of argument $\coth u$ using  Eq.~(8.739) in \cite{GR14}. For the particular case of $n=0$ we introduce the following definition (the subscript ``0'' denoting the $n=0$ harmonic is omitted herewith) : 
$$
    j^{(m)}(u) = j^{(m)}_{\infty} P_m(\coth u),
$$
where 
$$
    j^{(m)}_{\infty} =\lim_{u\to\infty} j^{(m)}(u).
$$
Specifically, $j^{(0)}(u)\equiv i^{(0)}_{0\infty}$, $E^{(0)}=\frac34$ and $j^{(1)}(u) =i^{(1)}_{0\infty}\coth u$, $E^{(1)}=-\frac54$.

For such current profiles, Eq.~(\ref{eq:In}) can be integrated analytically. Upon integrating by parts and using the equation, $\left[-\frac{d^2}{du^2} +\frac{3}{4\sinh^2u}\right] \bar{Q}^{-1}_{-\half}(u)=0$ we get the following expression for $n=0$:
\bea\label{eq:intim}
    I(u) &=& \int\limits_u^\infty 
    \frac{j^{(m)} (u_1) \bar{Q}^{-1}_{-\half}(u_1)du_1} {\sinh^2u_1} = \int\limits_u^\infty \bar{Q}^{-1}_{-\half}(u_1) \left[-\frac{d^2}{du^2_1} + \frac{3}{4\sinh^2u_1}\right] 
    \frac{j^{(m)} (u_1)}{E^{(m)}}du_1 =
    \nonumber\\
    &=& \frac{1}{E^{(m)}} \left[\frac{dj^{(m)}(u)}{du} \bar{Q}^{-1}_{-\half}(u) - j^{(m)}(u) \frac{d\bar{Q}^{-1}_{-\half}(u)}{du}\right].
\eea
\bibliographystyle{aasjournal}

\end{document}